\documentclass[%
 reprint,
superscriptaddress,
 amsmath,amssymb,
 aps,
]{revtex4-2}

\usepackage{graphicx}
\usepackage{float}
\usepackage{ulem}
\usepackage{xcolor}
\usepackage{dcolumn}
\usepackage{bm}
\usepackage[T1]{fontenc}
\usepackage[utf8]{inputenc}
\usepackage{bbm}

\newcommand\numberthis{\addtocounter{equation}{1}\tag{\theequation}}
\DeclareUnicodeCharacter{202F}{\,}
\begin{document}

\preprint{APS/123-QED}

\title{Lattice thermal conductivity in the  anharmonic overdamped regime}

	\author{{\DJ}or{\dj}e Dangi{\'c}}
	\email{dorde.dangic@ehu.es}
	\affiliation{Fisika Aplikatua Saila, Gipuzkoako Ingeniaritza Eskola, University of the Basque Country (UPV/EHU), 
		Europa Plaza 1, 20018 Donostia/San Sebasti{\'a}n, Spain}
	\affiliation{Centro de F{\'i}sica de Materiales (CSIC-UPV/EHU), 
		Manuel de Lardizabal Pasealekua 5, 20018 Donostia/San Sebasti{\'a}n, Spain}	
        \author{Giovanni Caldarelli}
        \affiliation{Dipartimento di Fisica, Universit{\'a} di Roma La Sapienza, Piazzale Aldo Moro 5, 00185 Roma, Italy}
	\author{Raffaello Bianco}
    \affiliation{Ru\dj er Bo\v{s}kovi\'c Institute, 10000 Zagreb, Croatia}

    \affiliation{Dipartimento di Scienze Fisiche, Informatiche e Matematiche, Universit\`a di Modena e Reggio Emilia, Via Campi 213/a I-41125 Modena, Italy}

    \affiliation{Centro S3, Istituto Nanoscienze-CNR, Via Campi 213/a, I-41125 Modena, Italy}
    \author{Ivana Savi{\'c}}
    \affiliation{Department of Physics, King’s College London, The Strand, London WC2R 2LS, United Kingdom}
	\author{Ion Errea}%
	\affiliation{Fisika Aplikatua Saila, Gipuzkoako Ingeniaritza Eskola, University of the Basque Country (UPV/EHU), 
		Europa Plaza 1, 20018 Donostia/San Sebasti{\'a}n, Spain}
	\affiliation{Centro de F{\'i}sica de Materiales (CSIC-UPV/EHU),
		Manuel de Lardizabal Pasealekua 5, 20018 Donostia/San Sebasti{\'a}n, Spain}
	\affiliation{Donostia International Physics Center (DIPC),
		Manuel de Lardizabal Pasealekua 4, 20018 Donostia/San Sebasti{\'a}n, Spain}

\date{\today}

\begin{abstract}

In crystalline materials, low lattice thermal conductivity is often associated with strong anharmonicity, which can cause significant deviations from the expected Lorentzian lineshape of phonon spectral functions. These deviations, occurring in an overdamped regime, raise questions about the applicability of the Boltzmann transport equation. Furthermore, strong anharmonicity can trigger structural phase transitions with temperature, which cannot be adequately described by the standard harmonic approximation. To address these challenges, we propose a novel approach for computing the lattice thermal conductivity. Our method combines the Green-Kubo linear response theory with the stochastic self-consistent harmonic approximation. The latter allows us to describe the temperature-dependent evolution of the crystal structure, including first- and second-order phase transitions, as well as the vibrational properties in highly anharmonic materials. The Green-Kubo method considers the entire lineshapes of phonon spectral functions in the calculation of the lattice thermal conductivity, thus eliminating the questionable use of phonon lifetimes in the overdamped regime, as well as naturally including coherent transport effects. Additionally, we extend our theory to model complex dynamical lattice thermal conductivity, enhancing our understanding of time-dependent thermoreflectance experiments. As a practical application, we employ this approach to calculate the lattice thermal conductivity of CsPbBr$_3$, a complex crystal known for its anomalous thermal transport behavior with a complex phase diagram. Our method is able to determine the thermal conductivity across different phases in good agreement with experiments.       
\end{abstract}

\maketitle


	\section{Introduction}

The thermal conductivity coefficient measures how efficiently heat is conducted through a material under a temperature gradient. It has two main contributions, one from the lattice and another from electrons. Materials with low lattice thermal conductivity find widespread technological applications. They can serve as thermal insulators, for example, in internal combustion engines where they retain heat to enhance efficiency~\cite{Thermalinsulation1,Thermalinsulation2}. Lowering the lattice thermal conductivity also improves the efficiency of thermoelectric materials~\cite{GeTe,IonSnSe,PbTe,PbGeTe}. This efficiency is quantified using the thermoelectric figure of merit $ZT$, which is inversely proportional to the material's thermal conductivity. As a result, cutting-edge thermoelectric devices utilize materials with exceptionally low lattice thermal conductivity. For these reasons understanding the microscopic origin of the lattice thermal conductivity is essential in order to further improve and optimize functional materials.

In insulating materials where lattice thermal conductivity is the dominant contribution, most of the heat is conducted by phonons. Phonons are quasiparticles that are convenient representations of the atomic vibrations and are well-defined only in perfect crystals with harmonic interaction between atoms. In these hypothetical systems, they are found as solutions to atomic equations of motion. This means that their definition depends heavily on how we represent the interaction between atoms. For example, we can expand the Born-Oppenheimer energy surface (BOES), which governs lattice dynamics, into a Taylor series with respect to atomic displacements and truncate it at the second order. Phonons that we get by solving equations of motion parametrized in this way are usually called harmonic phonons. However, when thermal and quantum atomic displacements are larger than the region in which this truncation is reasonable, these harmonic phonons are not useful representations of the atomic vibrations, and a different basis set is needed. There are different ways to obtain these alternative basis sets, such as the stochastic self-consistent harmonic approximation (SSCHA)~\cite{SSCHA1, SSCHA2, SSCHA3}  or temperature-dependent effective potential TDEP~\cite{TDEP} methods. The relationship between these different basis sets is discussed in detail in Ref.~\cite{monacelli2024simulatinganharmoniccrystalslights}. In any of these basis sets (harmonic, SSCHA, TDEP), the physical observables (scattering cross-sections, thermal conductivity, etc.) should be the same; however, the better the basis set the easier it is to obtain them. That is, the better the basis set, the lower-level theory is needed to obtain the correct physical observables. For example, in strongly anharmonic materials one would need to use second-, third- and fourth-order perturbative force constants to obtain correct phonon frequencies, while in the SSCHA and TDEP only effective second-order force constants are sufficient. 

The phonon scattering determines the thermal conductivity of insulating materials. It can arise from crystalline imperfections like vacancies~\cite{vacancies}, impurities~\cite{isotope}, and boundaries~\cite{DWs}, or from phonon-phonon interactions due to the anharmonicity of the Born-Oppenheimer energy surface (BOES)~\cite{BoronArsenide,Broido_review,Fugallo_2018,Anharmonicity_Knoop}. In perfect crystals, in the absence of phonon anharmonicity, phonons behave as infinitely long-lived quasiparticles, exhibiting infinitely narrow spectral lines in scattering experiments. However, phonon-phonon interaction, which arises due to the anharmonicity of the BOES, broadens these spectral lines, limiting phonon lifetimes. This broadened lineshape can often be well-modeled with a Lorentzian, with the linewidth inversely proportional to the phonon lifetime~\cite{Si_lifetimes}. Increased anharmonicity leads to larger phonon linewidths (lower phonon lifetimes) and consequently lower lattice thermal conductivities.

The main method for calculating the lattice thermal conductivity from first principles is the Boltzmann transport equation~\cite{Fugallo_2018, Broido_review,phonopy-phono3py-JPSJ,Alamode,ShengBTE}, which treats phonons as a weakly interacting gas of particles diffusing due to a temperature gradient while scattering from each other. To calculate the lattice thermal conductivity using this method, phonon frequencies, group velocities, and lifetimes are necessary. These calculations have become standard in the scientific community and have successfully been applied to various systems~\cite{PbGeTe,PbTe,GeTe,IonSnSe}.

In the presence of strong anharmonicity two important challenges arise. When anharmonicity is low, the thermal expansion is linear with temperature and this phenomenon is usually well captured by the quasiharmonic theory~\cite{GeTeTexp,QHarm}. On the other hand, there are many systems that undergo displacive structural phase transitions with increasing temperature, like charge-density wave~\cite{CDW} or ferroelectric transitions~\cite{GeTe}, so that the experimentally observed phases are not minima of the BOES, but are stabilized by quantum or thermal lattice fluctuations. This also implies that the harmonic phonons calculated for the high-temperature phases from the Hessian of the BOES may be imaginary, impeding any calculation of the thermal conductivity with the standard method. Interestingly, many good thermoelectric materials with low thermal conductivity fall into this category~\cite{PbGeTe,PbTe,GeTe,IonSnSe}. In recent years several first principles approaches have been developed to correctly model these strongly anharmonic situations~\cite{TDEP, QSCAILD, SSCHA1, SDM}. One of these alternatives is the so-called stochastic self-consistent harmonic approximation (SSCHA) method~\cite{SSCHA1, SSCHA2, SSCHA3, SSCHA4, SSCHA5}, which has been successfully applied to correctly model a number of first and second-order phase transitions in strongly anharmonic materials~\cite{IonSnSe, Monacelli2021, KTa03}.

The second challenge is related to the treatment of phonon lifetimes. In cases with strong phonon-phonon interactions, phonon spectral functions can deviate substantially from the Lorentzian lineshape leading to overdamped dynamics of phonon modes~\cite{overdamped_theory}. For instance, in the vicinity of structural phase transitions where anharmonicity is the strongest, phonon spectral functions can develop satellite peaks, as observed in materials like PbTe or SnTe~\cite{SnTe_PbTe}. In some other cases, anharmonicity is so strong that it leads to complete damping of the phonon mode, without any noticeable peak structure in the spectral signature~\cite{dampedCsPbBr3,Niedziela2019,IonSnSe}. In such instances, it becomes challenging to assign meaningful values to either phonon lifetimes or energies. This raises an open question: can the Boltzmann transport equation be used to model the lattice thermal conductivity in these and similar systems?

A recent work by some of us addressed this issue for the calculation of the lattice thermal conductivity in germanium telluride~\cite{GeTe}. GeTe undergoes a second-order phase transition at around 700 K. This phase transition is marked by the anomalous increase in the lattice thermal conductivity and complete softening of the zone center optical phonon mode. To account for the non-Lorentzian lineshapes and softening of phonon modes, a Green-Kubo approach in combination with the temperature-dependent effective potential method for phonon properties~\cite{TDEP} was implemented. The difference between lattice thermal conductivities calculated with the Green-Kubo method and Boltzmann transport equation was relatively low, only around 10\% even at the phase transition.

Here we combine the mentioned Green-Kubo linear response theory~\cite{GeTe} and the SSCHA~\cite{SSCHA1, SSCHA2,SSCHA3,SSCHA4,SSCHA5}, which is capable of tracking temperature-induced structural changes and vibrational properties at the same time. By using the Green-Kubo method, we retain the dependence of the lattice thermal conductivity on the entire lineshape of the phonon spectral function, not solely on the phonon lifetime. Consequently, this method remains applicable even in the regime of strong anharmonicity, where the phonon quasiparticle picture fails and phonon energies and lifetimes are not well-defined. Moreover, this method automatically incorporates the effects of coherent phonon transport~\cite{Simoncelli2019,Simoncelli2022,Simoncelli_glasses,Isaeva2019,Fiorentino,fiorentino2023hydrodynamic}, which has proven crucial in amorphous materials and complex crystals. Finally, we propose a way to calculate the complex dynamical lattice thermal conductivity based on the Green-Kubo method. We apply this method to study the lattice thermal conductivity of the strongly anharmonic CsPbBr$_3$ compound across several phase transitions, obtaining a good agreement with experiments.

\section{Method}

\subsection{The stochastic self-consistent harmonic approximation}

The SSCHA is a variational approach for determining the structural and vibrational properties of materials at a finite temperature accounting for the quantum nature of the ions. At its core, it minimizes the trial free energy $\mathcal{F}[\tilde{\rho}]$, which is a functional of the trial density matrix $\tilde{\rho}$. The Gibbs–Bogoliubov variational principle demonstrates that the minimum of this trial free energy $\mathcal{F}[\tilde{\rho}]$ is an upper bound for the true free energy of the system $F$:
\begin{equation}
\mathcal{F}[\tilde{\rho}] = \langle K + V\rangle_{\tilde{\rho}} + \frac{1}{\beta} \langle \ln \tilde{\rho} \rangle_{\tilde{\rho}}
\geq F.
\end{equation}
Here the first term is the energy, calculated as the quantum statistical average of the ionic kinetic energy $K$ plus the Born-Oppenheimer potential $V$, and the second is the entropic contribution to the free energy given by  $\tilde{\rho}$. As usual $\beta=1/(k_BT)$, where $k_B$ is Boltzmann's constant and $T$ the temperature, and the quantum statistical average of an operator $O$ means $\langle O \rangle_{\tilde{\rho}}=\mathrm{tr} [O\tilde{\rho}]$, which, in case it only depends on the ionic positions $\mathbf{R}$, can be written as 
\begin{equation}
\langle O \rangle _{\tilde{\rho}} = \int d\mathbf{R} O (\mathbf{R}) \tilde{\rho}(\mathbf{R}).
\end{equation}

$\tilde{\rho}(\mathbf{R})$ is the ionic probability distribution function defined by the trial density matrix $\tilde{\rho}$.

This problem is not well defined since there are many different choices for the trial density matrix $\tilde{\rho}$. An especially useful approximation that the SSCHA employs is to restrict the probability distribution defined by the trial density matrix to be a Gaussian. In this approximation, we define an auxiliary harmonic Hamiltonian 
\begin{equation}
\mathcal{H}_{\boldsymbol{\mathcal{R}},\boldsymbol{\Phi}} = K + \frac{1}{2}\sum _{a,b}(R_{a} - \mathcal{R}_{a})\Phi _{ab}(R_{b} - \mathcal{R}_{b}),
\end{equation}
parametrized by centroid positions $\boldsymbol{\mathcal{R}}$ and auxiliary force constants $\boldsymbol{\Phi}$,
%
whose associated probability distribution function is the Gaussian
\begin{eqnarray}
&& \tilde{\rho}_{\boldsymbol{\mathcal{R}},\boldsymbol{\Phi}}(\mathbf{R})=\sqrt{\textrm{det}(\mathbf{\Psi}^{-1}/2\pi)} \nonumber \\ && \ \ \ \times \exp\left[-\frac{1}    {2}\sum_{a,b}(R_a-\mathcal{R}_a)\Psi ^{-1}_{ab}(R_b-\mathcal{R}_b)\right].
\end{eqnarray}
Here $\mathbf{\Psi}$ is the displacement-displacement correlation matrix and in the SSCHA it has the closed analytical form 
\begin{equation}
\Psi _{ab} = \frac{1}{\sqrt{M_aM_b}}\sum _{j}\frac{2n_{j} + 1}{2\omega _{j}}e^{a}_{j}e^{b}_{j},
\end{equation}
where $\omega _{j}^2$ and $\mathbf{e}_j$ are the auxiliary eigenvalues and eigenvectors of the mass-scaled SSCHA auxiliary force constants, $\Phi_{ab}/\sqrt{M_aM_b}$, and $n_j$ is the Bose-Einstein factor for the frequency $\omega _j$. The label $a$ above refers to both an ion index and a Cartesian index. The Gaussian probability distribution function makes the quantum statistical average of the kinetic energy and the entropic term analytical functions of the auxiliary phonon frequencies $\omega _{j}$. 

To obtain the desired crystal structure at a given temperature we need to minimize the free energy $\mathcal{F}[\tilde{\rho}_{\boldsymbol{\mathcal{R}},\boldsymbol{\Phi}}]$ with respect to the two set of parameters: the centroids $\boldsymbol{\mathcal{R}}$ and the auxiliary force constants $\boldsymbol{\Phi}$. After the minimization, the centroid $\mathcal{R}_a$ represents the most probable positions of the ion $a$. The auxiliary force constants $\Phi _{ab}$ define the variances of the Gaussians that determine the ionic probability distribution function and are related to the uncertainties in determining the true average positions of the ions. Also, they can be seen as the renormalized second-order force constants due to quantum, temperature and anharmonic effects. The SSCHA method considers in its optimization of the crystal structure also the relaxation of the lattice parameters by calculating the stress tensor from the derivative of the SSCHA free energy with respect to a strain tensor~\cite{SSCHA5}. Thus the SSCHA offers a complete relaxation of the crystal structure considering quantum/thermal fluctuations and the associated anharmonicity.

It can be shown~\cite{SSCHA4} that for a given position of the centroids $\boldsymbol{\mathcal{R}}$, the SSCHA auxiliary force constants are equal to the quantum statistical average of the second derivatives of the Born-Oppenheimer potential:
\begin{equation}
\Phi _{ab}(\boldsymbol{\mathcal{R}}) = \left\langle\frac{\partial ^2 V}{\partial R_a\partial R_b}\right\rangle _{\tilde{\rho}_{\boldsymbol{\mathcal{R}},\boldsymbol{\Phi}}}. 
\label{eq:def_phi}
\end{equation}
This is the self-consistent equation of the SSCHA.
Since these quantities are positive definite, they can not tell us anything about the dynamical stability of the system. For a particular $\boldsymbol{\mathcal{R}}$, it is the Hessian of the total free energy ($\boldsymbol{\mathcal{D}}^{F}$) with respect to the centroid positions that determines the dynamical stability~\cite{SSCHA4}:
\begin{equation}
\mathbf{\mathcal{D}}^{F}_{ab} = \frac{1}{\sqrt{M_aM_b}}\frac{\partial ^2\mathcal{F}}{\partial \mathcal{R}_a \partial \mathcal{R}_b} = (\stackrel{(2)}{\mathbf{\mathcal{D}}}_{\mathbf{\mathcal{R}}})_{ab} + \Pi_{ab}(0).
\end{equation}
Here, we have defined  $(\stackrel{(2)}{\mathbf{\mathcal{D}}}_{\mathbf{\mathcal{R}}})_{ab}=\Phi_{ab}/\sqrt{M_aM_b}$ and $\boldsymbol{\Pi}$ is the SSCHA phonon self energy~\cite{SSCHA4}, which in compact notation is given by
\begin{equation}
\boldsymbol{\Pi}(\Omega) = \stackrel{(3)}{\boldsymbol{\mathcal{D}}}_{\mathcal{R}}:\boldsymbol{\Lambda}_{\mathcal{R}}(\Omega):\left[ \boldsymbol{\mathbbm{1}} -  \stackrel{(4)}{\boldsymbol{\mathcal{D}}}_{\mathcal{R}}:\boldsymbol{\Lambda}_{\mathcal{R}}(\Omega)\right]^{-1}:\stackrel{(3)}{\boldsymbol{\mathcal{D}}}_{\mathcal{R}}. 
\label{eq:self-energy}
\end{equation}
Here $\stackrel{(3,4)}{\boldsymbol{\mathcal{D}}}_{\mathcal{R}}$ have analogous definitions to the $\stackrel{(2)}{\boldsymbol{\mathcal{D}}}_{\mathbf{\mathcal{R}}}$ (see Eq.~\eqref{eq:def_phi}), meaning they are the ionic mass scaled ensemble averages of the third and fourth derivatives of the BOES. The double-dot product $\mathbf{X}:\mathbf{Y}$ indicates the contraction of the last two indices of $\mathbf{X}$ with the first two indices of $\mathbf{Y}$. Finally, the $\mathbf{\Lambda}$ fourth-order tensor is given in  components as
\begin{widetext}
	\begin{equation}
	\Lambda _{\mathcal{R}}^{abcd}(\Omega) = \sum _{jl}\frac{1}{4\omega _{j}\omega _{l}}\left[\frac{(\omega _{j} - \omega _{l})(n_{j} - n_{l})}{(\omega _{j} - \omega _{l})^2 - \Omega ^2 + i\epsilon} - \frac{(\omega _{j} + \omega _{l})(1 + n_{j} + n_{l})}{(\omega _{j} + \omega _{l})^2-\Omega ^2 + i\epsilon}\right]e^{a}_{j}e^{b}_{l}e^{c}_{j}e^{d}_{l}. 
	\label{eq:lambda}
	\end{equation}
\end{widetext}
If the hessian of the total free energy ($\boldsymbol{\mathcal{D}}^{F}$) has a negative eigenvalue for a particular set of centroid positions, the crystal structure is unstable towards a displacive structural phase transition. 

Using the expression for the SSCHA phonon self-energy, it is possible to define a dynamical extension of the theory~\cite{TDSSCHA,PhysRevResearch.3.L032017}:
\begin{equation}
\mathbf{G}^{uu}(\Omega) = \left(\Omega^2\boldsymbol{\mathbbm{1}} - \stackrel{(2)}{\boldsymbol{\mathcal{D}}}_{\mathbf{\mathcal{R}}} - \boldsymbol{\Pi}(\Omega)\right)^{-1}.
\end{equation}
Here $\mathbf{G}^{uu}(\Omega)$ is the Green function of the mass scaled atomic displacement $\sqrt{M_{a}}(R^{a} - \mathcal{R}^{a}) = u^{a}$. From this Green function, we can define phonon lifetimes and spectral functions.

In the current implementation of the thermal conductivity in the SSCHA code, which will be detailed below, the dynamical properties (phonon spectral functions and lifetimes) are calculated by assuming that $\stackrel{(4)}{\boldsymbol{\mathcal{D}}}_{\mathcal{R}}:\boldsymbol{\Lambda}_{\mathcal{R}}(\Omega) \ll \boldsymbol{\mathbbm{1}}$ and, consequently, neglecting this term. This so-called bubble approximation is usually good enough to calculate phonon spectral functions~\cite{SSCHA4}. 
The inclusion of $\stackrel{(4)}{\boldsymbol{\mathcal{D}}}_{\mathcal{R}}$, namely fourth-order anharmonic vertices in the SSCHA self-energy, is straightforward and will be a focus of future work (see Eq.~\eqref{eq:self-energy}). 

\begin{figure}
	\centering
	\includegraphics[width=0.99\linewidth]{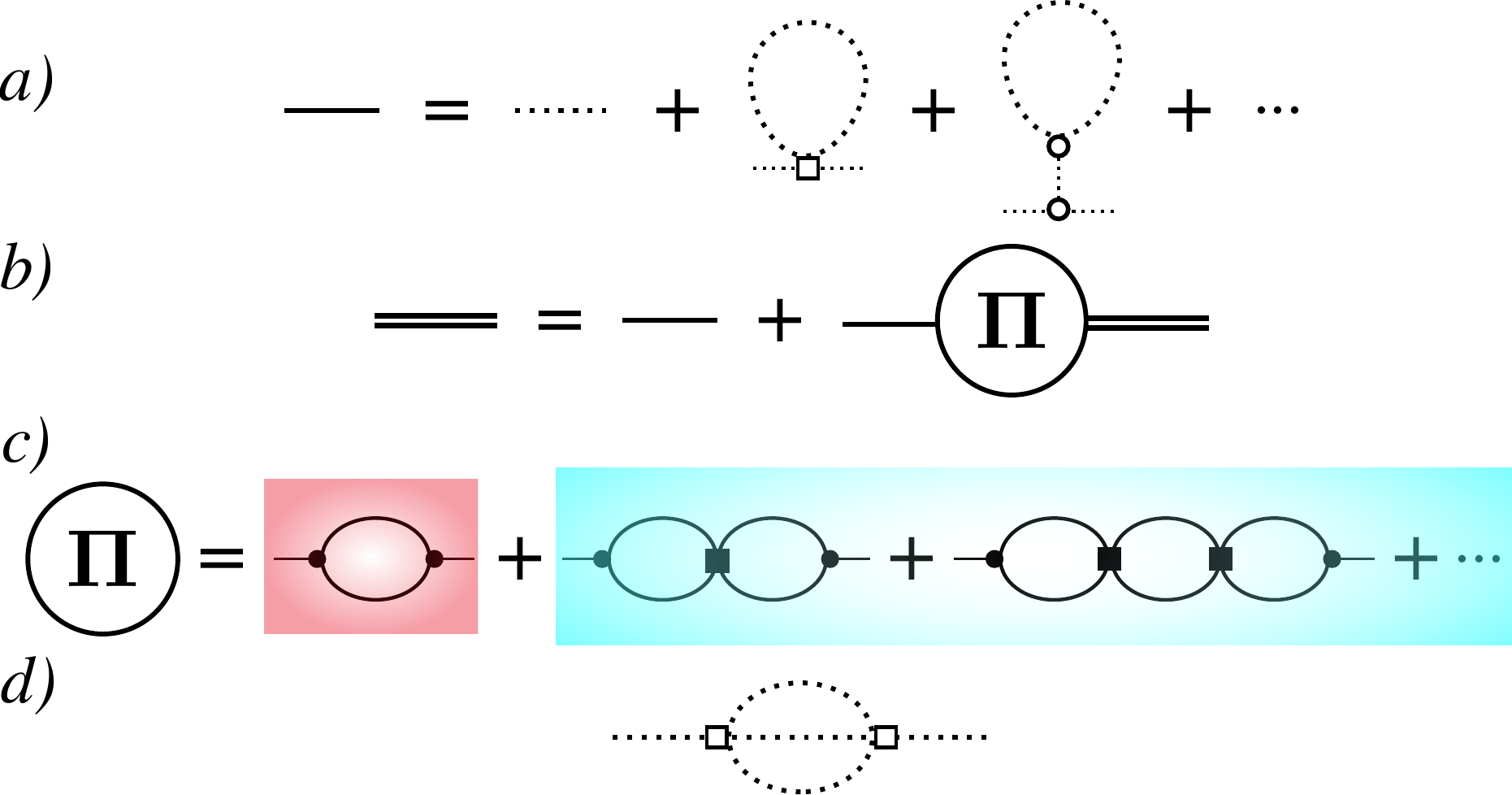}
	\caption{Graphical representation of the Feynman diagrams included in the stochastic self-consistent harmonic approximation (SSCHA). a) The Dyson equation for the SSCHA auxiliary phonon Green function. b) The Dyson equation for the dynamical SSCHA Green function. c) Phonon self-energy expansion in the SSCHA. d) The self-energy term currently implemented in some transport codes for the fourth-order anharmonicity contribution to the phonon lifetimes. Dashed lines represent the harmonic phonon Green functions, full lines SSCHA auxiliary phonon Green functions, double lines dynamical phonon Green functions, empty symbols are perturbative anharmonic vertices in the expansion of BOES and full symbols are SSCHA anharmonic terms. Circles represent third-order vertices and squares fourth-order vertices.}
	\label{fig:self_energy}
\end{figure}

It is pertinent to compare the SSCHA approach with the traditional perturbative one. In Fig.~\eqref{fig:self_energy} we can see a graphical representation of the self-consistent harmonic approximation. In panel a) we show the Dyson equation for an SSCHA phonon auxiliary Green's function. The full single line is the SSCHA auxiliary phonon Green function, while the dashed line is the harmonic one. The SSCHA auxiliary phonon Green function includes the renormalization due to anharmonicity of the so-called tadpole and loop lowest-order self-energy diagrams, among many others. In panel b) we can see the Dyson equation for the interacting SSCHA Green's function. Again the full line is the SSCHA auxiliary Green's function, the double line is the full dynamical Green's function, and the circle is the SSCHA self-energy. This self-energy includes an infinite number of terms as shown in the panel c). However, in the current implementation, we are only including the first term (red box), which only relies on third-order anharmonicity. Equation~\eqref{eq:self-energy} will include all the rest of the terms in the expansion that include additionally fourth-order anharmonic vertices (cyan box). It is important to point out that these terms considered in the SSCHA theory (cyan box) are different from the fourth-order anharmonic terms currently implemented in some other thermal transport codes~\cite{fourth1, fourth2} (see panel d) of Fig.~\eqref{fig:self_energy})~\cite{Siciliano}. It has been argued that, at least in some systems, the self-energy term in panel d) of Fig.~\eqref{fig:self_energy} and those marked in cyan in panel c) of Fig.~\eqref{fig:self_energy} should be of comparable size~\cite{Tripathi1974,Monga1, Monga2}. Benchmarking between these different terms will be a subject of future studies.

\subsection{Green-Kubo formula for the lattice thermal conductivity}

To derive an expression for the lattice thermal conductivity, we start with Kubo's well-known results for the lattice thermal conductivity in the linear response regime~\cite{Green, Kubo}:
\begin{equation}
\kappa ^{xy} (\nu) = \frac{NV}{k_BT^2}\int _{0}^{\infty}\textrm{d}t\frac{e^{i\nu t}}{\beta}\int _{0}^{\beta}\textrm{d}s\langle S^x (0)S^y(t+is)\rangle .\numberthis
\label{eq:heat_current}
\end{equation}
Here $x$ and $y$ are Cartesian directions, $NV$ is the volume of the crystal, and $\langle S^\alpha (0)S^\beta(t+is)\rangle$ is the heat autocorrelation function. We use $\hbar = 1$. $\nu$ is the driving frequency, the frequency of the time modulation of the temperature gradient. The dynamical lattice thermal conductivity ($\nu > 0$) can be calculated and compared, for instance, with time-dependent thermoreflectance experiments~\cite{Thermoreflectance1, Thermoreflectance2, Thermoreflectance3}. The dynamical lattice thermal conductivity $\kappa ^{xy}(\nu)$ is actually a complex-valued function. To obtain a quantity that is measured in the experiment we take the $\nu\rightarrow 0$ limit of the real part of $\kappa ^{xy}$.

First, we need a model for the heat current $S^x(t)$. We will use the one given by Hardy~\cite{Hardy}:
\begin{equation}
S^{x} (t) =\frac{1}{2NV}\sum _{\mathbf{q},j,j'}\omega_{\mathbf{q},j'}v^{x}_{\mathbf{q},j,j'}A_{\mathbf{q},j}(t)B_{\mathbf{q},j'}(t).
\label{heatcurrent}
\end{equation}
$\omega_{\mathbf{q},j}$ is the SSCHA auxiliary frequency of the phonon with wave vector $\mathbf{q}$ and branch $j$. $v^{x}_{\mathbf{q},j,j'}$ is the component of the generalized phonon group velocity matrix in the $x$th direction:
\begin{equation}
\mathbf{v}_{\mathbf{q},j,j'} = \frac{1}{2\sqrt{\omega _{\mathbf{q},j}\omega _{\mathbf{q},j'}}}\sum_{a,b}e^{a*}_{\mathbf{q},j}e^{b}_{\mathbf{q},j'}(\nabla _{\mathbf{q}}\stackrel{(2)}{\mathbf{\mathcal{D}}}_{\mathbf{\mathcal{R}}}(\mathbf{q}))_{ab},
\end{equation}
where $\stackrel{(2)}{\mathbf{\mathcal{D}}}_{\mathbf{\mathcal{R}}}(\mathbf{q})$ is the Fourier transform of $\stackrel{(2)}{\mathbf{\mathcal{D}}}_{\mathbf{\mathcal{R}}}$. In the equation above $a$ and $b$ label now Cartesian and atom indexes of atoms in the unit cell. We will adopt this meaning for Latin indexes for all quantities in Fourier space.
Operators $A(t)$ and $B(t)$ are given in Heisenberg's picture as:
\begin{eqnarray}
A_{\mathbf{q},j}(t) &=&   a^{\dagger}_{\mathbf{q},j}(t) - a_{-\mathbf{q},j}(t) \\ B_{\mathbf{q},j}(t) &=& a_{\mathbf{q},j}(t) + a^{\dagger}_{-\mathbf{q},j}(t),
\end{eqnarray}
where $a_{\mathbf{q},j}(t)$ and $a^{\dagger}_{\mathbf{q},j}(t)$ are phonon annihilation and creation operators. Operators $A$ and $B$ are related to atomic displacements and momenta operators up to some scaling factor. It is easy to show that this expression for the heat current in the harmonic approximation simplifies to the more commonly used Peierls' definition~\cite{Hardy}. However, in our case, adopting this approximation will not simplify the subsequent discussion. Therefore, we will continue using the slightly more general definition, as given in Eq.~\eqref{heatcurrent}. One important note is that the dynamical matrices used to calculate phonon frequencies and group velocities in Eq.~\eqref{heatcurrent} come from the SSCHA auxiliary force constants. Hardy derived Eq.~\eqref{heatcurrent} assuming a general harmonic Hamiltonian. The SSCHA auxiliary Hamiltonian is a better approximation to the full Hamiltonian compared to the harmonic one due to the variational minimization procedure.

When we substitute the heat current expression into Kubo's formula we end up with a four-operator correlation function. In principle, we should be working with this correlation function, but in order to make calculations more tractable we implement Wick's decoupling scheme~\cite{DepBehera}. Giving up on the two phonon Green's function means we are neglecting the vertex corrections to the phonon self-energy and we will not be able to recover the full solution to the Boltzmann transport equation. Hence, we will not be able to model the hydrodynamic phenomena at relatively low temperatures~\cite{Fiorentino}. However, here we are mostly interested in the modeling of the transport properties of highly anharmonic materials at elevated temperatures where hydrodynamic effects are strongly diminished. With the decoupling procedure, we approximate the correlation function as
\begin{eqnarray}
&\langle A_{\mathbf{q},j}(0)B_{\mathbf{q},j'}(0)A_{\mathbf{q}',l}(t)B_{\mathbf{q}',l'}(t)\rangle \approx \nonumber \\
&\langle A_{\mathbf{q},j}(0)B_{\mathbf{q},j'}(0)\rangle \langle A_{\mathbf{q}',l}(t)B_{\mathbf{q}',l'}(t)\rangle + \nonumber \\
&\langle A_{\mathbf{q},j}(0)B_{\mathbf{q}',l'}(t)\rangle \langle B_{\mathbf{q},j'}(0)A_{\mathbf{q}',l}(t)\rangle + \nonumber \\
&\langle A_{\mathbf{q},j}(0)A_{\mathbf{q}',l}(t)\rangle \langle B_{\mathbf{q},j'}(0)B_{\mathbf{q}',l'}(t)\rangle = \nonumber\\
&J^{AB}_{\mathbf{q},j,\mathbf{q}',l'}(t)J^{BA}_{\mathbf{q},j',\mathbf{q}',l}(t) + J^{AA}_{\mathbf{q},j,\mathbf{q}',l}(t)J^{BB}_{\mathbf{q}, j',\mathbf{q}',l'}(t).
\label{eq_corr}
\end{eqnarray}
The first term in this expansion describes the correlations between phonon displacement and momenta at the same time and thus it is zero~\cite{Zwanzig}. After Fourier transformation of the rest of the correlation functions ($J^{AA}_{\mathbf{q},j,\mathbf{q}',l}(t), J^{AB}_{\mathbf{q},j,\mathbf{q}',l'}(t), J^{BA}_{\mathbf{q},j',\mathbf{q}',l}(t), J^{BB}_{\mathbf{q},j',\mathbf{q}',l'}(t)$) we obtain:
\begin{widetext}
	\begin{eqnarray}
	\kappa ^{xy} (\nu) =& \frac{\beta ^2k_{B}}{16\pi^2NV}\sum _{\mathbf{q},j,j'}\sum _{\mathbf{q}',l,l'}\omega _{\mathbf{q},j'}v^{x}_{\mathbf{q},j,j'}\omega _{\mathbf{q}',l'}v^{y}_{\mathbf{q}',l,l'}\frac{1}{\beta}\int_0 ^{\beta}e^{(\Omega _1 - \Omega _2)s}\textrm{d}s\int_0 ^{\infty}e^{-i(\Omega _1-(\Omega _2 + \nu))t}\textrm{d}t\times \nonumber \\
	&\int_{-\infty}^{\infty}\textrm{d}\Omega _1\textrm{d}\Omega _2 \left(J^{AA}_{\textbf{q},j,\mathbf{q}',l}(\Omega _1)J^{BB}_{\textbf{q},j',\mathbf{q}',l'}(\Omega _2) + J^{AB}_{\textbf{q},j,\mathbf{q}',l'}(\Omega _1)J^{BA}_{\textbf{q},j',\mathbf{q}',l}(\Omega _2)\right).
	\end{eqnarray}
\end{widetext}
Using the fact that $i\dot{A}_{\mathbf{q},j}(t) = [A_{\mathbf{q},j},\mathcal{H}] = -\omega _{\mathbf{q},j}B^{\dagger}_{\mathbf{q},j}(t)$ and the properties of the spectral representation of the correlation functions~\cite{Zubarev}, we can represent these four correlation functions only through $J^{AA}$~\cite{PhilAllen} (see Supp. Material~\cite{supp_mat}). The correlation function $J^{AA}$ will exist only for the operators at the same wave vector $\mathbf{q}$ (or rather for a pair ($\mathbf{q},-\mathbf{q}$)) and branch index $j$ (so-called no-mode-mixing approximation). We checked for simple systems that the full (keeping the off-diagonal terms in phonon branches of the phonon spectral function for the same wave vector $\mathbf{q}$) and no-mode-mixing approximation gives almost the same result for the thermal conductivity (see Supplementary Material~\cite{supp_mat}). After evaluating the integrals over $s$ and $t$, the no-mode-mixing approximation simplifies the expression for the real part of the lattice thermal conductivity to
\begin{widetext}
	\begin{equation}
	\kappa ^{xy} (\nu) = \frac{\beta ^2k_{B}}{16NV\pi}\sum _{\mathbf{q},j,j'}v^x_{\mathbf{q},j,j'}v^{y*}_{\mathbf{q},j,j'}\frac{e^{\beta\nu} - 1}{\beta\nu}\int_{-\infty}^{\infty}\textrm{d}\Omega \Omega(2\Omega + \nu)e^{\beta\Omega}J^{AA}_{\mathbf{q},j}(\Omega + \nu)J^{AA}_{\mathbf{q},j'}(\Omega). 
	\label{eq:dyn_kappa}
	\end{equation}
\end{widetext}

Let us now discuss how to obtain the correlation functions $J^{AA}$. These correlation functions can be found as~\cite{Zubarev}
\begin{equation}
J^{AA}_{\mathbf{q},j}(\Omega) = -\frac{2}{\exp(\beta\Omega) - 1}\textrm{Im}G^{AA}_{\mathbf{q},j}(\Omega),
\end{equation}
where $G^{AA}_{\mathbf{q},j}(\Omega)$ is the Green function of the operator $A$. As we have noted in the previous section, in the SSCHA we define Green's functions in terms of mass-scaled phonon displacement operators $u_{\mathbf{q},j}$, which introduces a factor $2\omega_{\mathbf{q},j}$. Additionally, if we define phonon spectral function as $\sigma _{\mathbf{q},j} (\Omega)= -\frac{\Omega}{\pi}\mathrm{Im}G^{uu}_{\mathbf{q},j}(\Omega)$ and take a $\nu\rightarrow 0$ limit, we reach the final expression for the lattice thermal conductivity as implemented in the SSCHA code:
\begin{widetext}
	\begin{equation}
	\kappa ^{xy} = \frac{2\pi\beta ^2k_{B}}{NV}\sum _{\mathbf{q},j,j'}v^{x}_{\mathbf{q},j,j'}v^{y*}_{\mathbf{q},j,j'}\omega _{\mathbf{q},j}\omega _{\mathbf{q},j'}\int_{-\infty}^{\infty}\textrm{d}\Omega \frac{\exp(\beta\Omega)}{\left(\exp(\beta\Omega) - 1\right)^2}\sigma _{\mathbf{q},j}(\Omega)\sigma_{\mathbf{q},j'}(\Omega).\numberthis
	\label{kappa_eq}
	\end{equation}
\end{widetext}
The equation above can be split into two contributions, the diagonal part in the phonon branch index ($j=j'$) and the off-diagonal part ($j\neq j'$). The off-diagonal part of the lattice thermal conductivity describes what was termed as the coherent contribution to transport~\cite{Simoncelli2019}, and these two terms will be used interchangeably in the rest of the paper. To obtain the lattice thermal conductivity, we integrate over the entire phonon spectral function, thereby avoiding the question of whether the phonon lifetimes and energies are well defined.

In the limit of small anharmonicity, where phonon linewidths are much smaller than the phonon frequencies and the phonon shift can be disregarded, the diagonal part of the equation above reduces to the solution of the Boltzmann transport equation in the single relaxation time approximation (SRTA). To show this, we start from the expression for $\sigma _{\mathbf{q},j}(\Omega)$ in the no-mode-mixing approximation~\cite{SSCHA4,SSCHA1}:
\begin{eqnarray}
\sigma _{\mathbf{q},j}(\Omega) = \frac{1}{2\pi}&\Bigg[ \frac{-\mathrm{Im}\mathcal{Z}_{\mathbf{q},j}(\Omega)}{\left(\Omega - \mathrm{Re}\mathcal{Z}_{\mathbf{q},j}(\Omega)\right)^2 + \mathrm{Im}\mathcal{Z}^2_{\mathbf{q},j}(\Omega)}+\nonumber \\
+&\frac{\mathrm{Im}\mathcal{Z}_{\mathbf{q},j}(\Omega)}{\left(\Omega + \mathrm{Re}\mathcal{Z}_{\mathbf{q},j}(\Omega)\right)^2 + \mathrm{Im}\mathcal{Z}^2_{\mathbf{q},j}(\Omega)}\Bigg],
\label{nmmspectralfunction}
\end{eqnarray}
with $\mathcal{Z}_{\mathbf{q},j}(\Omega) = \sqrt{\omega ^2_{\mathbf{q},j} + \Pi _{\mathbf{q},j}(\Omega)}$, where $\Pi_{\mathbf{q},j}(\Omega)$ is the phonon self-energy in the mode basis (see Eq.~\eqref{eq:self-energy}).
The form of Eq.~\eqref{nmmspectralfunction} resembles a Lorentzian function peaked around $\mathrm{Re}\mathcal{Z}_{\mathbf{q},j}(\Omega)$, which in the limit of low anharmonicity is just $\omega _{\mathbf{q},j}$. Since this function has a strong peak in the limit of low anharmonicity, we can approximate one of the spectral functions in Eq.~\eqref{kappa_eq} as a delta function $\delta (\Omega - \omega _{\mathbf{q},j})$, which will annihilate the integral over $\Omega$ and substitute all $\Omega$'s with $\omega _{\mathbf{q},j}$. With this, we obtain:
\begin{equation}
\kappa^{xy} = \frac{1}{NV}\sum _{\mathbf{q},j}v^{x}_{\mathbf{q},j}v^{y}_{\mathbf{q},j}c_{\mathbf{q},j}\frac{-1}{2\mathrm{Im}\mathcal{Z}_{\mathbf{q},j} (\omega_{\mathbf{q},j})}.
\label{kappa_bte}
\end{equation}
Here $c_{\mathbf{q},j} = k_B\beta ^2\omega ^2_{\mathbf{q},j}n_{\mathbf{q},j}(n_{\mathbf{q},j}+1)$ is the heat capacity of the phonon mode $(\mathbf{q},j)$ and $\mathbf{v}_{\mathbf{q},j}=\mathbf{v}_{\mathbf{q},jj}$ is its group velocity. From the equation above we identify $\frac{-1}{2\mathrm{Im}\mathcal{Z}_{\mathbf{q},j}(\omega_{\mathbf{q},j})}$ as the phonon lifetime $\tau _{\mathbf{q},j}$. In previous publications~\cite{SSCHA1}, this approximation was called one-shot approximation. The phonon lifetimes can further be approximated as (as it is done in most other codes) $\tau _{\mathbf{q},j} = -\frac{\omega _{\mathbf{q},j}}{\mathrm{Im}\Pi_{\mathbf{q},j}(\omega_{\mathbf{q},j})}$, which is what we term perturbative approximation. However, the best approximation for the phonon lifetime and  frequency is obtained by solving self-consistently the equation
\begin{equation}
\Omega _{\mathbf{q},j} = \mathrm{Re}\mathcal{Z}_{\mathbf{q},j} (\Omega_{\mathbf{q},j})
\end{equation}
to get the shifted phonon frequency due to anharmonicity $\Omega _{\mathbf{q},j}$ and, then, taking $\tau _{\mathbf{q}, j} = -\frac{1}{2\mathrm{Im}\mathcal{Z}_{\mathbf{q},j} (\Omega_{\mathbf{q},j})}$ for the lifetime.

A similar procedure can be applied in order to get the off-diagonal contribution in the limit of low anharmonicity (see Supp. Material~\cite{supp_mat}). If we assume that $\textrm{Im}\mathcal{Z}_{\mathbf{q},j}(\Omega)= \textrm{Im}\mathcal{Z}_{\mathbf{q},j}(\omega _{\mathbf{q},j}) = \Gamma_{\mathbf{q},j}$, i.e. the lifetime does not depend on $\Omega$, and that $\textrm{Re}\mathcal{Z}_{\mathbf{q},j} (\Omega) = \textrm{Re}\mathcal{Z}_{\mathbf{q},j} (\omega _{\mathbf{q}, j}) = \omega _{\mathbf{q},j}$, we obtain that the lattice thermal conductivity in the perturbative limit is split into two terms, $\kappa^{xy}=\kappa _{R}^{xy}+\kappa_{A}^{xy}$, with 
\begin{align*}
	\kappa _{R}^{xy} = \frac{1}{2NV}&\sum _{\mathbf{q},j,j'}v^{x}_{\mathbf{q},j,j'}v^{y*}_{\mathbf{q},j,j'}\left(\omega _{\mathbf{q},j'}\frac{c_{\mathbf{q},j}}{\omega _{\mathbf{q},j}} + \omega _{\mathbf{q},j}\frac{c_{\mathbf{q},j'}}{\omega _{\mathbf{q},j'}}\right)\times \\ &\times\frac{\Gamma _{\mathbf{q},j} + \Gamma _{\mathbf{q},j'}}{(\omega _{\mathbf{q},j} - \omega _{\mathbf{q},j'})^2 + (\Gamma _{\mathbf{q},j} + \Gamma _{\mathbf{q},j'})^2}   \\
	\kappa _{A}^{xy} = \frac{1}{2NV}&\sum  _{\mathbf{q},j,j'}v^{x}_{\mathbf{q},j,j'}v^{y*}_{\mathbf{q},j,j'}\left(\omega _{\mathbf{q},j'}\frac{c_{\mathbf{q},j}}{\omega _{\mathbf{q},j}} + \omega _{\mathbf{q},j}\frac{c_{\mathbf{q},j'}}{\omega _{\mathbf{q},j'}}\right)\times \\ &\times\frac{\Gamma _{\mathbf{q},j} + \Gamma _{\mathbf{q},j'}}{(\omega _{\mathbf{q},j} + \omega _{\mathbf{q},j'})^2 + (\Gamma _{\mathbf{q},j} + \Gamma _{\mathbf{q},j'})^2}.\numberthis
	\label{pert_limit}
\end{align*}
Here $\kappa _{R}^{xy}$ is the resonant term and  $\kappa _{A}^{xy}$ the anti-resonant one. A similar splitting of the off-diagonal terms has been described in  Refs.~\cite{Isaeva2019, Caldarelli}.
The resonant term has a square of the difference of phonon frequencies in the denominator and, because of this, it is the dominant one. Additionally, for $j=j'$ this term reduces to Eq.~\eqref{kappa_bte}, the solution of the Boltzmann transport equation in the single relaxation time approximation. This resonant part should have a considerable contribution only in the case when the difference between phonon frequencies is much smaller than the phonon linewidths. Current results for resonant and anti-resonant terms differ slightly from the ones obtained in Ref.~\cite{Caldarelli}. The difference can be traced to the approximation scheme employed for the evaluation of the overlap integrals since the definitions of the lattice thermal conductivity in the Lorentzian approximation are identical. The different approximation schemes are shown in the Supplementary Material. Using the same approximation scheme yields identical results. It is important to note that the differences between Eq.~\eqref{pert_limit} and the one in Ref.~\cite{Caldarelli} get smaller as $\omega _{\mathbf{q},j} - \omega _{\mathbf{q},j'} \to 0$, which is also the case when this transport mode is most effective. We have compared explicitly different approximations of the off-diagonal contributions to the lattice thermal conductivity in the Supplementary Material~\cite{supp_mat}. 

\subsection{Technical details of the implementation}

The calculation of the generalized phonon group velocities is done by employing the analytical formula~\cite{Hardy}:
\begin{equation}
v^{x}_{\mathbf{q},j,j'} = \frac{i}{2\sqrt{\omega _{\mathbf{q},j}\omega _{\mathbf{q},j'}}}\sum_{a,b}e^{a*}_{\mathbf{q},j}e^{b}_{\mathbf{q},j'}\sum_{\mathbf{r}_{ab}}\frac{\Phi_{ab}(\mathbf{r}_{ab})}{\sqrt{M_aM_b}}r_{ab}^{x}e^{i\mathbf{q}\mathbf{r}_{ab}}.
\end{equation}
Here $\omega _{\mathbf{q},j}^2$ and $\mathbf{e}_{\mathbf{q},j}$ are eigenvalues and eigenvectors of the dynamical matrices in reciprocal space constructed from the SSCHA auxiliary force constants $\Phi_{ab}(\mathbf{r}_{ab})$, and $a$ and $b$ denote Cartesian and atom indexes restricted to the unit cell. For the Fourier interpolation of phonon frequencies and eigenvectors, we are using the smooth convention for the phase factor~\cite{Simoncelli2022}. Thus $\mathbf{r}_{ab}$ is the vector connecting atoms $a$ and $b$. We calculate the group velocities on the entire grid of $\mathbf{q}$ points. Then, we symmetrize the group velocities with respect to the little group of the wave vector $\mathbf{q}$ by applying the point group part of the spacegroup symmetry to the group velocity vector. Finally, in Eq.~\eqref{kappa_eq} and Eq.~\eqref{kappa_bte} we perform the sum over the whole set of $\mathbf{q}$ points and not only over the irreducible zone. 

The SSCHA code implements two ways of calculating the lattice thermal conductivity. It first considers the use of Eq.~\eqref{kappa_bte}, which is practically the solution of the Boltzmann transport equation. The phonon lifetimes can be calculated in the perturbative limit, one-shot approximation, or self-consistently as described above. In the case of solving the equation self-consistently, the phonon self-energy is sampled on a fixed grid of frequencies and linearly interpolated when needed. In this case, the off-diagonal contribution to the lattice thermal conductivity can be calculated using Eq.~\eqref{pert_limit}.

The second way of calculating the lattice thermal conductivity is the implementation of the full Green-Kubo method, see Eq.~\eqref{kappa_eq}. Similarly to the calculation of self-consistent lifetimes, in this case the phonon self-energy is sampled on a fixed grid of frequencies too, and then the integral in the equation is evaluated numerically. The spacing between frequencies in this grid should be of the order of the smallest phonon linewidth, but stable results should be possible even with larger values. In principle, the number of frequency steps is an additional convergence parameter in addition to the $\mathbf{q}$-point grid.

In order to obtain the correlation functions from the phonon Green function we need to take the imaginary part of the limit $\lim _{\epsilon\rightarrow 0}G(\Omega)$ (see Eq.~\eqref{eq:lambda}). Here we can think of $\epsilon$ as the smearing parameter. This procedure we will call Lorentzian smearing. Alternatively, since $\epsilon$ is effectively 0, we can use the following identity $\frac{1}{x\pm i\epsilon} = \mathcal{P}\frac{1}{x}\mp i\pi\delta(x)$. Now the imaginary part of the self-energy contains Dirac delta functions that can be approximated with a Gaussian function of width $\epsilon$. We refer to this procedure as Gaussian smearing and, in this case, the real part of the self-energy is obtained using the Kramers-Kronig relation.

Next, we would like to discuss the technical issues with regard to using the Green-Kubo method. From Eq.~\eqref{kappa_eq} we can see that we have a possible divergence in $\frac{1}{\Omega ^2}$ as $\Omega \rightarrow 0$. This implies that phonon spectral functions have to decay at least linearly with $\Omega \rightarrow 0$ to have stable results. If the decay of the spectral function is slower, or if the spectral function is not decaying at all (as it happens in CsPbBr$_3$ at low temperatures in the $P4/mbm$ and $Pm\bar{3}m$ phases, and at high temperatures in the $Pnma$ phase, see below), we can have unphysical results. This is a good thing, because it signals us that the structure is not stable at the conditions (pressure, temperature) we are simulating it on. The BTE does not have this technical issue, but its results are not any more meaningful. 

Finally, let us discuss the similarities and differences between the current implementation and other Green-Kubo approaches for calculating the lattice thermal conductivity. The most common way to calculate the lattice thermal conductivity using the Green-Kubo method is with molecular dynamics (MD) simulations~\cite{IvanaMD,BaroniMD,CarbognoMD}. In this method heat current autocorrelation functions are directly sampled on an MD trajectory. The main advantages of the MD method are that it fully includes anharmonicity and that it goes beyond the harmonic approximation in the definition of the heat current, which is our case (see Eq.~\eqref{heatcurrent}). Additionally, with MD simulations it is straightforward to include additional scattering mechanisms such as impurities, grain boundaries, etc. On the other hand, these calculations are notoriously hard to converge and usually not possible with density functional theory (DFT), although recently there has been some progress in this area~\cite{BaroniMD}. Moreover, in MD methods it is difficult to distinguish structures that are related to temperature-induced displacive phase transitions. Also, determining which are the phonon modes that contribute the most to the thermal conductivity is hard in these approaches. In the method presented here, we do not need to run any MD simulations to consider anharmonicity at a non-perturbative level, as the basic ingredients that are needed are provided by the SSCHA, which further provides a clear way of determining the crystal structure at any temperature and gives access to phonon spectra.

There are derivations of the Green-Kubo approach in the literature in the same spirit as our implementation~\cite{DepBehera,srivastava,Semwal}. Our approach uses a different definition of the heat current for the diagonal part of $\kappa$ compared to these previous reports. For the non-diagonal part of the lattice thermal conductivity, previous studies are using the same heat current definition as we are~\cite{Semwal}. However, although we are using the same definition for the non-diagonal part of the lattice thermal conductivity, we get slightly different results compared to Ref.~\cite{Semwal}. This is due to different expressions for the momentum correlation function ($J_{BB}$). Our definition of the momentum correlation function is consistent with the ones obtained in Refs.~\cite{PhilAllen, TDSSCHA}. Finally, we extend the theory to derive the expression for the complex dynamical part of the lattice thermal conductivity (see Supplementary Material~\cite{supp_mat}). 

The described methods are implemented in the SSCHA code within \emph{ThermalConductivity} method~\cite{sschacode}. 

\section{Lattice thermal conductivity of CsPbBr$_3$}

\begin{figure}
	\centering
	\includegraphics[width=\linewidth]{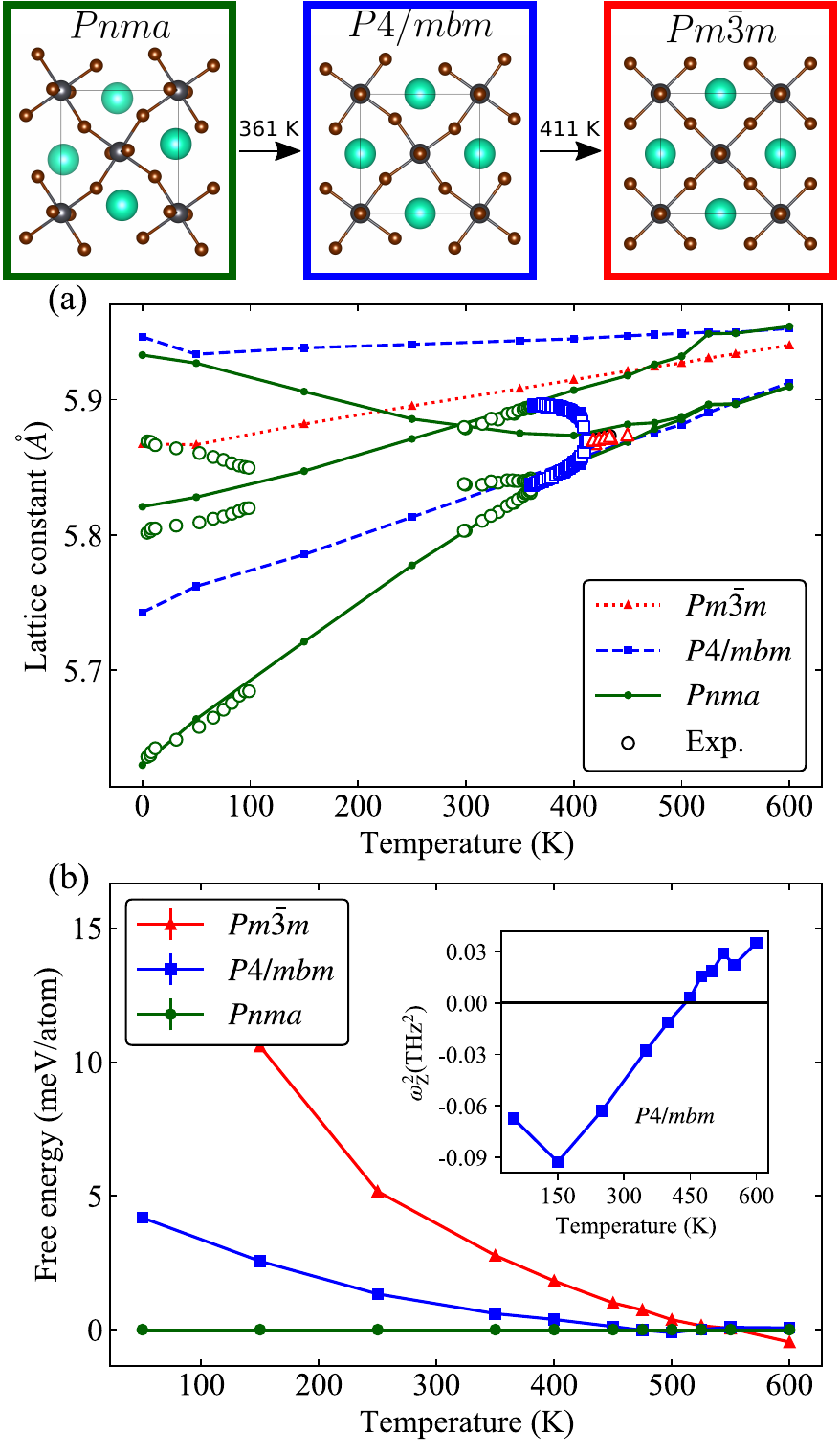}
	\caption{Structure of CsPbBr$_3$ along $z$ direction in three different phases. (a) Thermal expansion of CsPbBr$_3$ calculated with SSCHA and compared with experiments~\cite{CsPbBr3_TE1,CsPbBr3_TE2,CsPbBr3_TE3}. The colors of the experimental points match the colors for different phases. (b) Helmholtz free energy of CsPbBr$_3$ calculated with SSCHA with reference to the orthorhombic phase. The inset shows the lowest eigenvalue of the free energy Hessian matrix at $Z$ high-symmetry point of the tetragonal phase.}
	\label{fig:thermal_expansion}
\end{figure}

\begin{figure*}[t!]
	\centering
	\includegraphics[width=\textwidth]{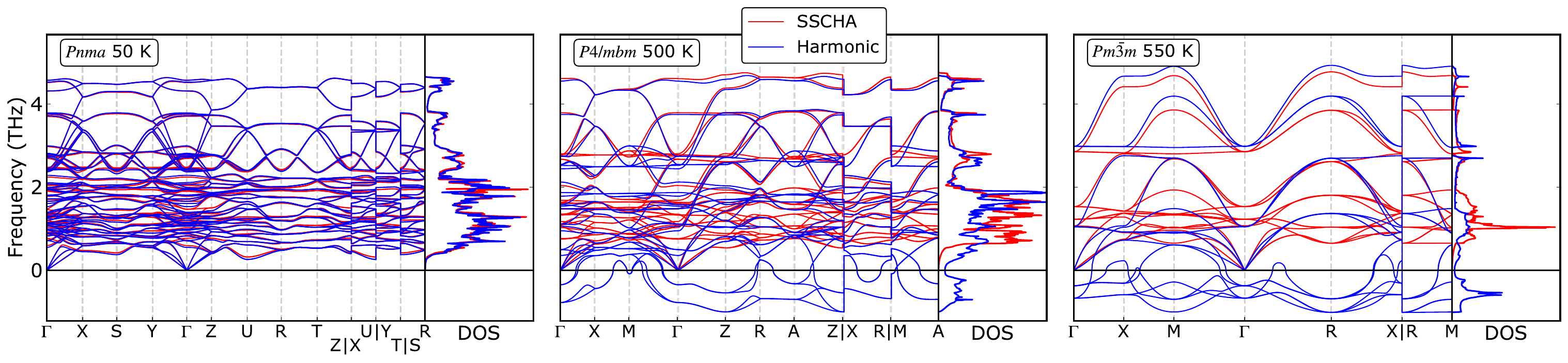}
	\caption{Phonon band structures calculated in the harmonic approximation and with the SSCHA auxiliary force constants for the three different phases of CsPbBr$_3$ at different temperatures. Structures for which we calculated these phonon dispersions are minima of SSCHA total free energy. Sideplots show the corresponding phonon density of states.}
	\label{fig:phonon_bands}
\end{figure*}

In this section we present results on the lattice thermal conductivity of CsPbBr$_3$, using both the Green-Kubo method and its perturbative limit. We chose CsPbBr$_3$ as our test subject for two main reasons. First, CsPbBr$_3$ exhibits a remarkably rich phase diagram, with at least three different phases up to 420 K~\cite{CsPbBr3_TE1, CsPbBr3_TE2, CsPbBr3_TE3}. Below 361 K, CsPbBr$_3$ crystallizes in an orthorhombic $Pmna$ structure, undergoing a second-order phase transition to a tetragonal $P4/mbm$ structure at this temperature. Subsequently, around 411 K, there is a first-order phase transition to the perovskite $Pm\bar{3}m$ cubic structure. As shown in Fig.~\eqref{fig:thermal_expansion}, the phase transition from perovskite $Pm\bar{3}m$ to $P4/mbm$ phase is marked by the rotation of Pb-Br octahedra inside the $xy$ plane, while the phase transition between tetragonal $P4/mbm$ and orthorhombic $Pnma$ phases is manifested by the rotation of the same octahedra from the $z$ axis. The high-temperature phases, $P4/mbm$ and $Pm\bar{3}m$, are dynamically unstable in the harmonic approximation (see Fig.~\eqref{fig:phonon_bands}) and to correctly describe them the SSCHA approach is a necessity. Additionally, one expects that anharmonicity is very strong at these temperatures and that this will lead to a strong renormalization of phonon spectral functions, making it a prime subject for the Green-Kubo method.

Another compelling reason for selecting CsPbBr$_3$ as the subject of our study is its manifestation of unusual thermal transport behavior. The conventional solution of the Boltzmann transport equation significantly underestimates the lattice thermal conductivity of this compound~\cite{Simoncelli2019, CsPbBr3_kappa1}. It has been demonstrated that to accurately model the thermal conductivity ($\kappa$) of this material, one must incorporate the contributions of so-called coherent transport~\cite{Simoncelli2019, Simoncelli2022, Caldarelli}. This phenomenon is not unique to CsPbBr$_3$; it is also expected to have a substantial impact on amorphous materials~\cite{Isaeva2019, Simoncelli_glasses}. Both CsPbBr$_3$ and amorphous materials share the characteristic of having a frequency spacing between different phonon modes comparable to the phonon scattering rate, leading to the emergence of this specific transport regime. The Green-Kubo method naturally includes such contributions, making it intriguing to compare its results with those obtained from coherent transport calculations using the perturbative limit (which is the same as the Boltzmann transport equation).

We model the phase transition in CsPbBr$_3$ using the stochastic self-consistent harmonic approximation (SSCHA). Employing SSCHA enables us to accurately capture the temperature-dependent evolution of the crystal structure, including the temperature-dependent second and third-order force constants as defined in Eq. \eqref{eq:def_phi}. However, the orthorhombic phase of CsPbBr$_3$ consists of 20 atoms per primitive unit cell, making it impractical to use density functional theory (DFT) to obtain the forces, energies, and stresses required for SSCHA's minimization of the total free energy. To overcome this limitation, we opted to create a machine learning potential based on Gaussian Approximation Potentials (GAP)~\cite{GAP1, GAP2, GAP3}. The Supplementary Material contains further details about the fitting procedure and the thorough testing of the interatomic potential~\cite{supp_mat}.

To calculate the second and third-order force constants, we relaxed CsPbBr$_3$ using SSCHA, enabling structural changes and atomic position adjustments during the minimization of the Helmholtz free energy. We considered all three possible phases ($Pnma$, $P4/mbm$, and $Pm\bar{3}m$) at various temperatures. The results are compared with experimental data in Fig.~\eqref{fig:thermal_expansion} (a). The machine learning potential provides good agreement with experimental lattice constants in the orthorhombic phase, except for overestimating the largest lattice constant. This overestimation is also present in the tetragonal and cubic phases. The overestimation originates mainly from the underlying density functional theory (DFT) data, which overestimates the largest orthorhombic lattice constant. Above 450 K, the anisotropy between in-plane lattice constants in the orthorhombic calculation disappears, suggesting a transition. Above this temperature, both orthorhombic and tetragonal structures display practically identical lattice constants, pointing to a second-order phase transition. This is in agreement with the experiments. In contrast, the cubic lattice constants do not converge to any of the lattice constants of the orthorhombic and tetragonal phases, suggesting a first-order phase transition between the tetragonal and cubic phases.

To investigate the nature of the phase transition more carefully, we calculated the Helmholtz free energies for each phase at various temperatures, as depicted in Fig.~\eqref{fig:thermal_expansion} (b). Above 450 K, the free energies of the tetragonal and orthorhombic phases coincide within the error margin allowed by the stochastic sampling of the SSCHA, confirming a second-order phase transition. This conclusion is corroborated by the behavior of the soft mode inducing the phase transition, shown in the inset of Fig.~\eqref{fig:thermal_expansion} (b), where the frequency of the soft mode of the free energy Hessian becomes real exactly at 450 K. The free energy Hessians were calculated with the fourth-order anharmonic contribution for both $P4/mbm$ and $Pm\bar{3}m$ phases, i.e. including $\stackrel{(4)}{\boldsymbol{\mathcal{D}}}_{\mathcal{R}}$ in Eq. \eqref{eq:self-energy}. Further, at 550 K, the free energy of the $Pm\bar{3}m$ phase becomes the lowest, which together with the results of the thermal expansion (the existence of discontinuity of the values of the lattice parameters at this temperature) indicates a first-order phase transition. Concurrently with this crossing of the free energies, the free energy Hessian associated with the soft mode $M$ of the cubic structure becomes positive also at 550 K. Similarly, the Hessian of the tetragonal phase is as well positive at this temperature. This suggests that even if the cubic phase becomes dynamically stable exactly at the temperature at which it becomes the ground state structure, the tetragonal remains still a local minimum of the free energy at this temperature, which is a hallmark of the first-order phase transition. It is important to note that we are overestimating the phase transition temperatures by approximately 100 K in both cases. This overestimation can be attributed to the DFT parametrization used in the training of the GAP potential due to its impact on one of the in-plane lattice constants in the $Pnma$ phase. With a smaller overestimation of the lattice constant, we would expect a $P4/mbm\rightarrow Pnma$ phase transition at a lower temperature, leading to a better agreement with experimental results.

Next, we present the phonon band structures of CsPbBr$_3$ in the three phases at different temperatures within their stability ranges ($Pnma$ at 50 K, $P4/mbm$ at 500 K, and $Pm\bar{3}m$ at 550 K) in Fig.~\eqref{fig:phonon_bands}. We compare the results of the harmonic phonons with those obtained from the SSCHA auxiliary phonon frequencies. This is important to show since these are the numbers that are going into evaluating Eq.~\ref{kappa_bte} and Eq.~\ref{pert_limit}.  At 50 K, the phonons of the orthorhombic phase are relatively unaffected by anharmonicity and quantum effects, as expected. 
In contrast, at 500 K (the tetragonal structure) and 550 K (the cubic structure), the harmonic approximation indicates imaginary frequencies, revealing the dynamical instability of these structures in the classical perspective based on the BOES minima. To investigate further, we calculated the second-order force constants of these structures at the appropriate temperatures using the SSCHA, leading to a significant renormalization of phonon quasiparticles. The SSCHA-calculated auxiliary phonon frequencies demonstrate only modest softening for relevant $\mathbf{q}$ points in both phases (Z in the tetragonal phase, and M and R in the cubic phase). It is worth noting that in this and subsequent analyses, we have not accounted for the effects of the long-range interaction of phonon modes (LO-TO splitting). 
These effects primarily influence optical modes close to the $\Gamma$ point, which generally do not significantly contribute to thermal transport. 

\begin{figure}
	\centering
	\includegraphics[width=\linewidth]{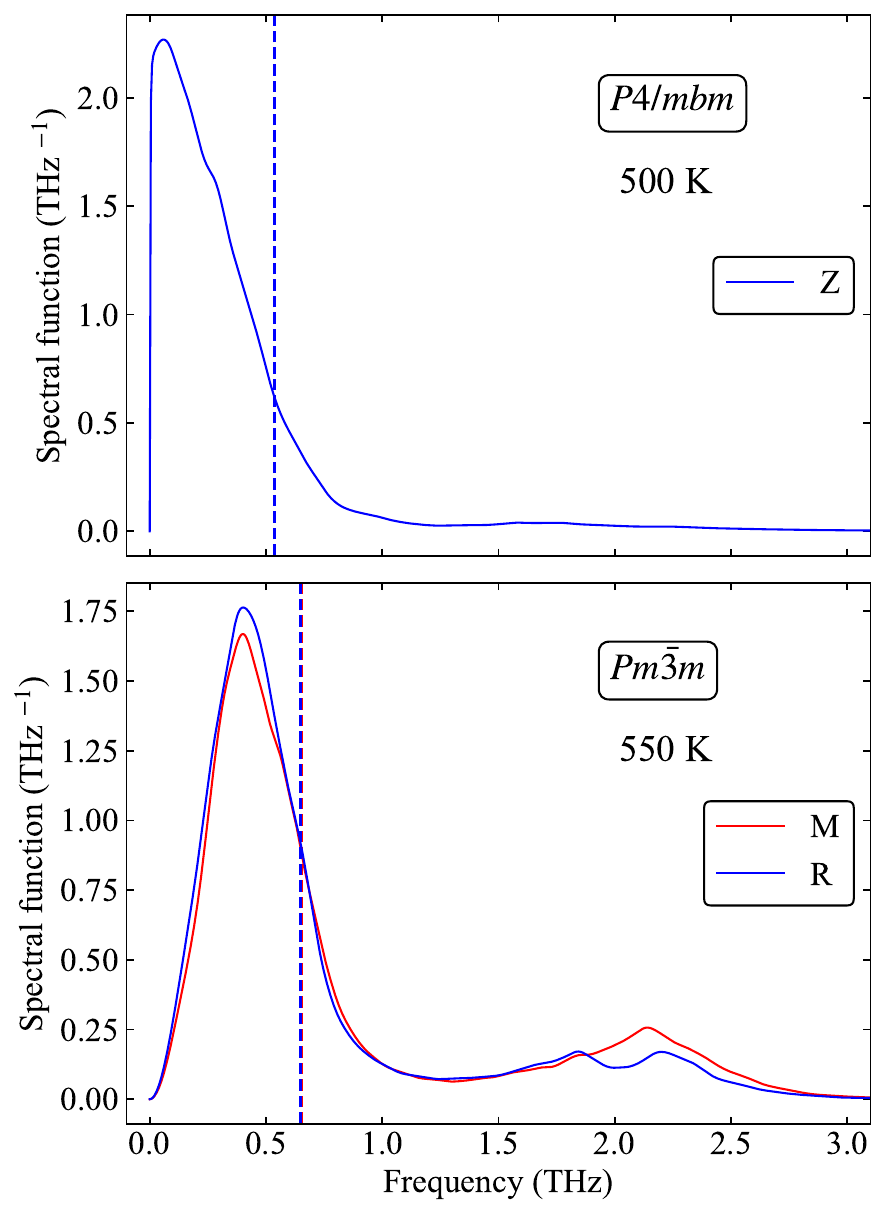}
	\caption{Phonon spectral functions of soft modes in the tetragonal $P4/mbm$ phase (Z \textbf{q}-point) and the cubic $Pm\bar{3}m$ phase (M and R \textbf{q}-points). Vertical dashed lines indicate SSCHA auxiliary frequencies for these modes.}
	\label{fig:spec_func}
\end{figure}

However, SSCHA auxiliary phonons are not real experimentally observable quantities, but just variational parameters of the SSCHA minimization. However, they are a better representation of the vibrational properties of a material compared to a harmonic approximation due to the inclusion of anharmonicity in a non-perturbative way as shown in Fig.~\ref{fig:self_energy}. Even a better representation of the vibrational properties of a material is provided by the SSCHA dynamical Green function (double line in Fig.~\ref{fig:self_energy}). For that reason, we show in Fig.~\ref{fig:spec_func} spectral functions of soft phonon modes in the tetragonal (Z $\mathbf{q}$ point) and cubic (M and R $\mathbf{q}$ points) phases at 500 K and 550 K, respectively,  neglecting the $\stackrel{(4)}{\boldsymbol{\mathcal{D}}}_{\mathcal{R}}$ term in the phonon self-energy (see Fig.~\eqref{fig:spec_func}). We have fixed the value of the spectral function to be 0 at $\Omega = 0$ to avoid numerical instabilities. In the tetragonal phase at 500 K, the Z mode is highly softened. However, it is important to note that at this temperature, the free energy Hessian (calculated including $\stackrel{(4)}{\boldsymbol{\mathcal{D}}}_{\mathcal{R}}$) does not exhibit imaginary frequencies, indicating that the structure is dynamically stable with highly damped vibrations. This is consistent with the first-order phase transition described above between the tetragonal and cubic phases. The peak of the spectral function occurs around 0.06 THz, and the value of the spectral function appears to approach 0 as $\Omega \rightarrow 0$. The overdamped dynamics in the cubic phase have been experimentally observed in $M$ and $R$ points~\cite{dampedCsPbBr3}, and, at $\Gamma$~\cite{CsPbBr_QE_peak} as well. The $R$ point of the cubic phase folds back to the $Z$ point of the tetragonal phase, where we observe this overdamped behavior. 

Similarly, in the cubic phase, the soft modes are significantly softened compared to their auxiliary frequency values, although not to the same extent as in the tetragonal $P4/mbm$ structure. All spectral functions exhibit a non-Lorentzian lineshape, with the spectral function of the cubic phase showing satellite peaks at significantly higher energies. These results clearly indicate that vibrations in CsPbBr$_3$ within this temperature range are overdamped.

We wish to emphasize another technical challenge inherent in modeling highly anharmonic materials. Specifically, the orthorhombic phase of CsPbBr$_3$ should, in theory, exhibit dynamic stability up to the second-order phase transition to the tetragonal phase (above 450 K). However, our calculations reveal an intriguing discrepancy: at elevated temperatures ($>$300 K), this phase manifests negative total free energy curvature—evidenced by imaginary Hessian frequencies. 
The primary factor underpinning this limitation is our omission of $\stackrel{(4)}{\boldsymbol{\mathcal{D}}}_{\mathcal{R}}$ in the calculation of the free energy Hessian in this phase due to the substantial computational demands associated with its inclusion (refer to the Supplementary Material for details). Including this term would shift the emergence of negative eigenvalues in the Hessian at 450 K, where the orthorhombic phase undergoes a second-order phase transition to the tetragonal phase. Indeed, this is an example that in a second-order phase transition also the phonon driving the transition softens in the low-symmetry phase when increasing the temperature~\cite{Holt2001X-Ray}. The pronounced softening of a particular phonon mode, as discerned through Hessian calculations, will cause numerical instabilities in the computation of the lattice thermal conductivity via the Green-Kubo method, as we have explained in the previous section. Similarly, this exact thing happens for the other two phases ($P4/mbm$ and $Pm\bar{3}m$) at lower temperatures. These instabilities are however limited to a small part of the Brillouin zone and, hence, do not have a substantial effect on the overall lattice thermal conductivity. 

Before presenting our own findings on the lattice thermal conductivity of CsPbBr$_3$, it is pertinent to analyze the existing experimental results~\cite{CsPbBr3_kappa1, CsPbBr_kappa2, CsPbBr_kappa3, CsPbBr_kappa4, CsPbBr_kappa5}. Most of these experiments only measured the lattice thermal conductivity in the orthorhombic phase. These experimental studies can be categorized into two distinct groups, each yielding markedly different results. The first category involves investigations conducted on CsPbBr$_3$ nanowires, as detailed in Refs.~\cite{CsPbBr3_kappa1, CsPbBr_kappa2}. These results exhibit a significant level of agreement both within the same study across different samples and also between different studies. Furthermore, it is worth noting that the lattice thermal conductivity from these experiments displays limited dependence on the nanowire cross-sectional area (all nanowires' cross-section dimensions are $>$ 100 nm), implying that $\kappa$ should attain the bulk limit. This is further confirmed by our calculations of the phonon mean free path at 250 K which rarely surpasses 10 nm (see Supp. Material). These nanowire-based investigations were subjected to comparison with preceding theoretical findings~\cite{Simoncelli2019, CsPbBr3_newkappa}, revealing a good agreement between theoretical predictions and empirical observations for the low-temperature orthorhombic phase. Conversely, the second category of experiments involves examinations of single crystal samples of CsPbBr$_3$~\cite{CsPbBr_kappa3, CsPbBr_kappa4, CsPbBr_kappa5}. These results consistently report higher thermal conductivity values in comparison to those observed in nanowires. However, these single-crystal results do not exhibit strong agreement amongst themselves. It is noteworthy that, aside from one instance~\cite{CsPbBr_kappa5}, the majority of these studies omit reporting the experimental error, thereby introducing challenges in the comparison of results across different experiments. One of the references pertaining to single crystal results notes that their material is virtually insulating, thereby negating the possibility that the electronic contribution to the thermal conductivity can explain the difference in results between nanowires and single crystals. Additionally, while one of the experiments demonstrates a correlation between higher symmetry structures and elevated thermal conductivities~\cite{CsPbBr_kappa4}, another study fails to identify any distinct changes between various phases~\cite{CsPbBr_kappa3}. The disparity between the thermal conductivity results from single-crystal and nanowire examinations, as well as the inconsistencies among the single-crystal results, underscores the need for further comprehensive experimental investigations aimed at elucidating these discrepancies.

\begin{figure}
	\centering
	\includegraphics[width=\linewidth]{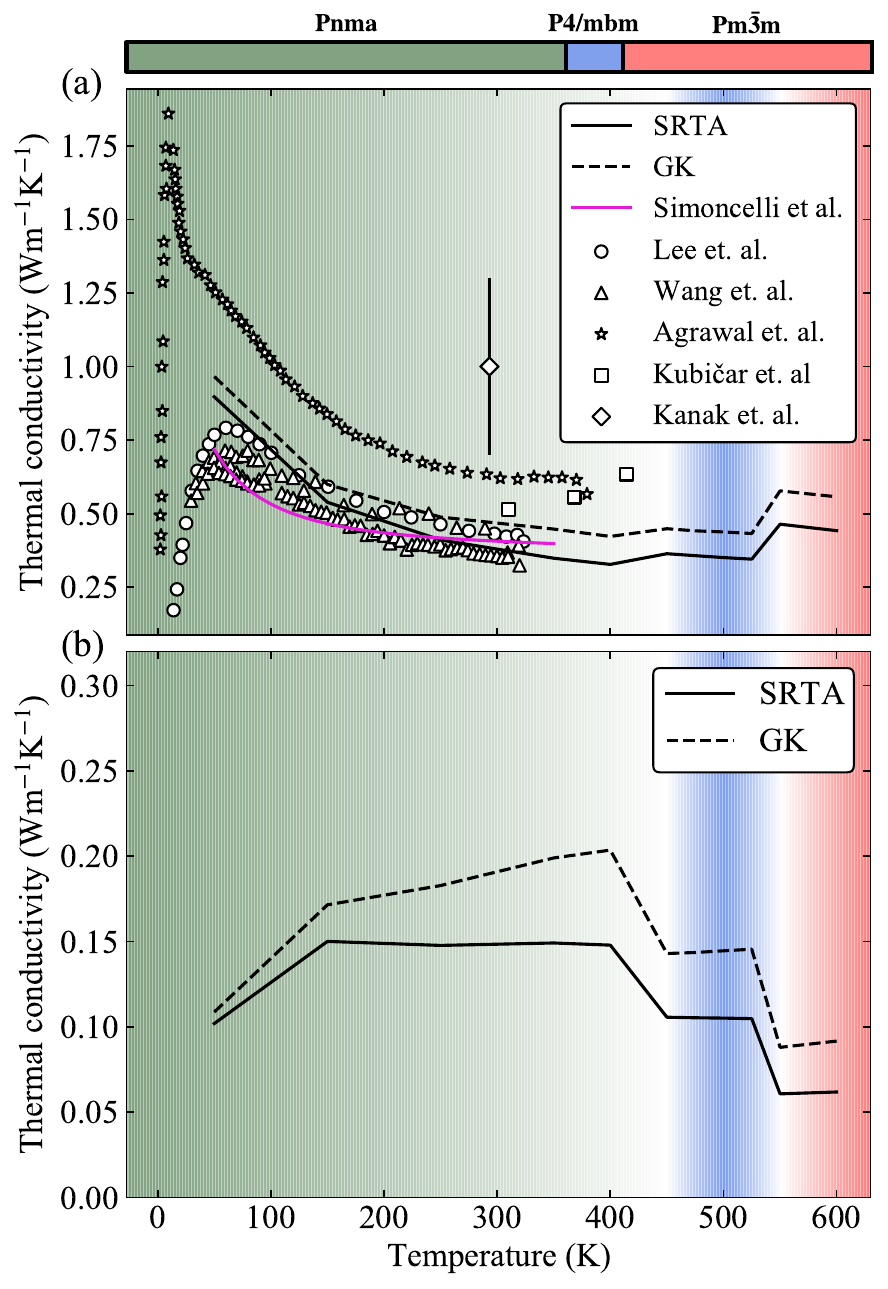}
	\caption{The bar above the graph represents the experimental phase diagram of CsPbBr$_3$ from Ref.~\cite{CsPbBr3_TE1}. The background color shows the phase diagram as obtained by the SSCHA and corresponds to the colors for the experimental bar above the graph. (a) Total lattice thermal conductivity of CsPbBr$_3$ calculated for different phases in perturbative limit (full lines) and Green-Kubo method (dashed lines) in $x$ Cartesian direction. The presented values contain both diagonal and non-diagonal contributions. The magenta line is the theoretical result from Ref.~\cite{Simoncelli2019}. Experiments are given with empty symbols from Refs.~\cite{CsPbBr3_kappa1, CsPbBr_kappa2, CsPbBr_kappa3, CsPbBr_kappa4, CsPbBr_kappa5}. (b) The coherent part of the lattice thermal conductivity was calculated for different phases in the perturbative limit using Eq.~\eqref{pert_limit} (full lines) and with the Green-Kubo method (dashed lines).}
	\label{fig:kappa}
\end{figure}

In our study, we have computed the lattice thermal conductivity of CsPbBr$_3$ across the entire temperature range for all three structural phases, regardless of their stability at each temperature. However, in Fig.~\eqref{fig:kappa} we only represent the results of the total lattice thermal conductivity (particle plus coherent part) in $x$ Cartesian direction for the structures that are the minimum of the total free energy at that temperature. The computed values of the lattice thermal conductivity exhibit a consistent trend: for all temperatures, the cubic $Pm\bar{3}m$ phase shows the highest thermal conductivity, followed by the tetragonal $P4/mbm$ phase, and finally the orthorhombic $Pnma$ phase, in agreement with measurements~\cite{CsPbBr_kappa4} and recent theoretical results~\cite{CsPbBr3_newkappa}. The greater phonon lifetimes for low-frequency modes in both the cubic and tetragonal structures contribute to their higher thermal conductivities ($\kappa$). Additionally, the cubic phase displays larger phonon group velocities compared to the tetragonal phase, further enhancing its $\kappa$. Notably, our results fall within the spectrum between nanowire and single-crystal samples. While overestimating the experimental results could potentially be attributed to factors such as isotope scattering, grain boundaries, and crystalline impurities, explaining the underestimation relative to single crystal samples poses a significant challenge, particularly considering the assertion by the authors of Ref.~\cite{CsPbBr_kappa3} that they possess an insulating crystal. One possible solution to this discrepancy could be the contribution of the anharmonic heat flux which we do not include here~\cite{Hardy}. This contribution might be considerable in strongly anharmonic materials and could lead to the observed discrepancy between our results and single crystal measurements. Finally, we also compare our results with previously reported ones in Ref.~\cite{Simoncelli2019}. Our results differ slightly from these ones because the force constants that are being used in these two studies are different, and because in Ref.~\cite{Simoncelli2019} isotope scattering is included.

Green-Kubo results consistently give larger values for the lattice thermal conductivity compared to the perturbative limit given by Eqs.~\ref{kappa_bte} and ~\ref{pert_limit} (10~\%-25~\%), similar to the case of GeTe~\cite{GeTe}. The main reason for that is the larger effective heat capacities in the Green-Kubo method due to the shift of the quasiparticle peak to lower frequencies. The inclusion of the fourth-order scattering would probably shift the peaks of the phonon spectral functions to higher frequencies and lead to a lower lattice thermal conductivity. This would increase the agreement with nanowire samples and further decrease the agreement with single-crystal results unless the fourth-order scattering beyond the SSCHA self-energy significantly increases the coherent transport contribution. These extra terms in the self-energy would increase phonon linewidths~\cite{CsPbBr3_newkappa}, which would lead to a larger overlap between spectral functions of different phonon branches, thus increasing the coherent contribution. In the orthorhombic phase, we have small anisotropy between $\kappa$ in-plane and out-of-plane values, while in the tetragonal phase, this anisotropy is larger (see Supp. Material for more information). 

As we previously highlighted, the influence of coherent transport on the calculated lattice thermal conductivity in CsPbBr$_3$ can be substantial~\cite{Simoncelli2019, CsPbBr3_newkappa}. Fig.~\eqref{fig:kappa} (b) presents our results for this phonon transport mechanism, calculated for all phases using both the Green-Kubo method and the perturbative limit. Notably, we observe a non-monotonic variation of the coherent term with temperature in the perturbative limit. This finding contradicts earlier calculations and is likely due to the structure and temperature dependence of the SSCHA interatomic force constants. Intriguingly, the coherent transport contribution reaches its maximum at 150 K within the perturbative limit for all phases. The Green-Kubo method consistently yields larger coherent transport values across all phases and temperatures. In the perturbative limit, a large contribution to the coherent transport is expected only when two harmonic frequencies are similar in magnitude. The Green-Kubo method does not have that limitation and instead calculates the overlap between spectral functions of different phonon modes with the same wavevector. This leads to a larger contribution to the coherent transport. 
It is noteworthy that the magnitude of the coherent transport contribution is most pronounced in the $Pnma$ phase and weakest in the $Pm\bar{3}m$ phase, thus displaying an inverse correlation with the overall $\kappa$. This can be explained by the fact that the total lattice conductivity is inversely proportional to the phonon linewidth, while the coherent part is directly proportional to it.

\begin{figure}
	\centering
	\includegraphics[width=0.9\linewidth]{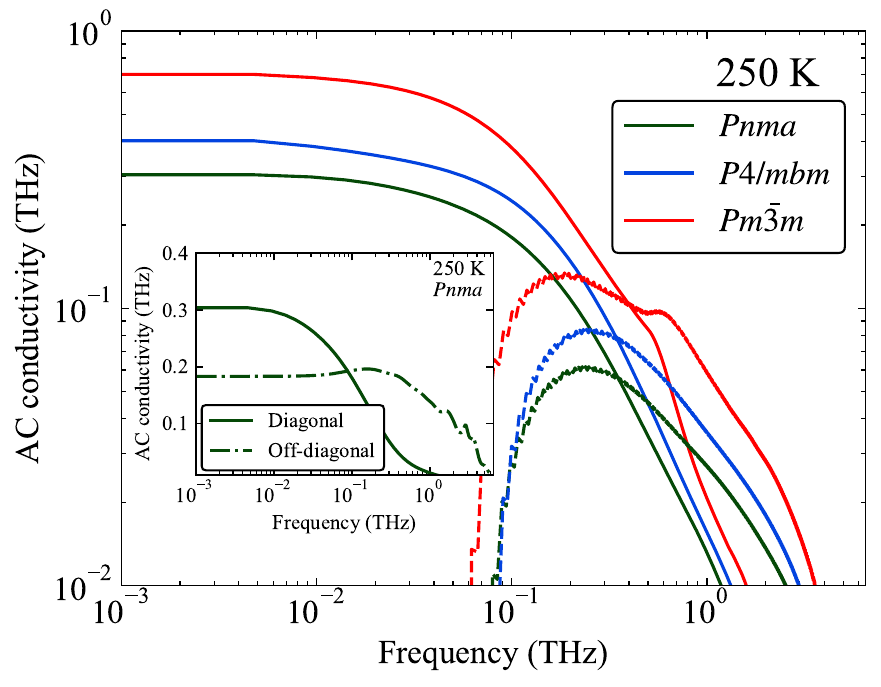}
	\caption{AC conductivity of CsPbBr$_3$ at 250 K for all phases (only diagonal contribution). Full lines are the real part of the conductivity, while dashed lines represent the imaginary part of the lattice thermal conductivity. The inset shows diagonal and off-diagonal contributions for the AC lattice thermal conductivity of $Pnma$ phase of CsPbBr$_3$ at 250 K.}
	\label{fig:ac_cond}
\end{figure}

In Fig.~\eqref{fig:ac_cond} we present our results on the dynamical lattice thermal conductivity. We see a characteristic decay of the lattice thermal conductivity with increasing frequency, as it is observed for Si and Ge~\cite{Thermoreflectance3}. The imaginary part of the AC conductivity was obtained by the Kramers-Kronig relation. If we take an analogy with the Drude model for electrical conductivity, the peak of the imaginary part can be associated with the inverse of the characteristic transport lifetime. As expected, the $Pm\bar{3}m$ phase has the highest characteristic lifetime (1.05 ps), while $P4/mbm$ and $Pnma$ phases have similar effective phonon lifetimes (0.89 and 0.92 ps respectively). The real part of the lattice thermal conductivity, however, does not fit well with the Drude model with constant relaxation time. If we assume that the relaxation time is inversely proportional to the frequency the fit improves significantly (see Supp. Material~\cite{supp_mat}). Fitted values of characteristic lifetimes are slightly larger than those extracted from the imaginary part of $\kappa$, but overall agree qualitatively. 

Now we analyze the results for the diagonal and off-diagonal contributions to the AC lattice thermal conductivity, see the inset of Fig.~\eqref{fig:ac_cond}. The diagonal part does not show any structure in the high-frequency range and has a shape expected from the Drude model. On the other hand, the off-diagonal lattice thermal conductivity contribution shows a much more interesting behavior (see the inset of Fig.~\eqref{fig:ac_cond}). It does not fall to zero with increasing frequency but rather has a structure that does not resemble any other signatures of the lattice thermal conductivity (phonon density of states or spectral lattice thermal conductivity). From the analysis of the perturbative limit of dynamical lattice thermal conductivity (see Supp. Mat.~\cite{supp_mat}), we can see that the structure comes from the spectral density of the frequency spacing between phonon modes with the same wave vector. Since it significantly deviates from the diagonal part at high frequencies, this fact could be used to experimentally study this mode of heat transport. It is important to point out that the current experiments can only probe dynamical lattice thermal conductivity up to 100 MHz, while the appreciable changes in dynamical lattice thermal conductivity happen around 1 THz. An interesting avenue of development might be the experimental investigation of the lattice thermal conductivity in the GHz and maybe THz range, where the interesting behavior happens in these highly anharmonic materials.  

\begin{figure}
	\centering
	\includegraphics[width=\linewidth]{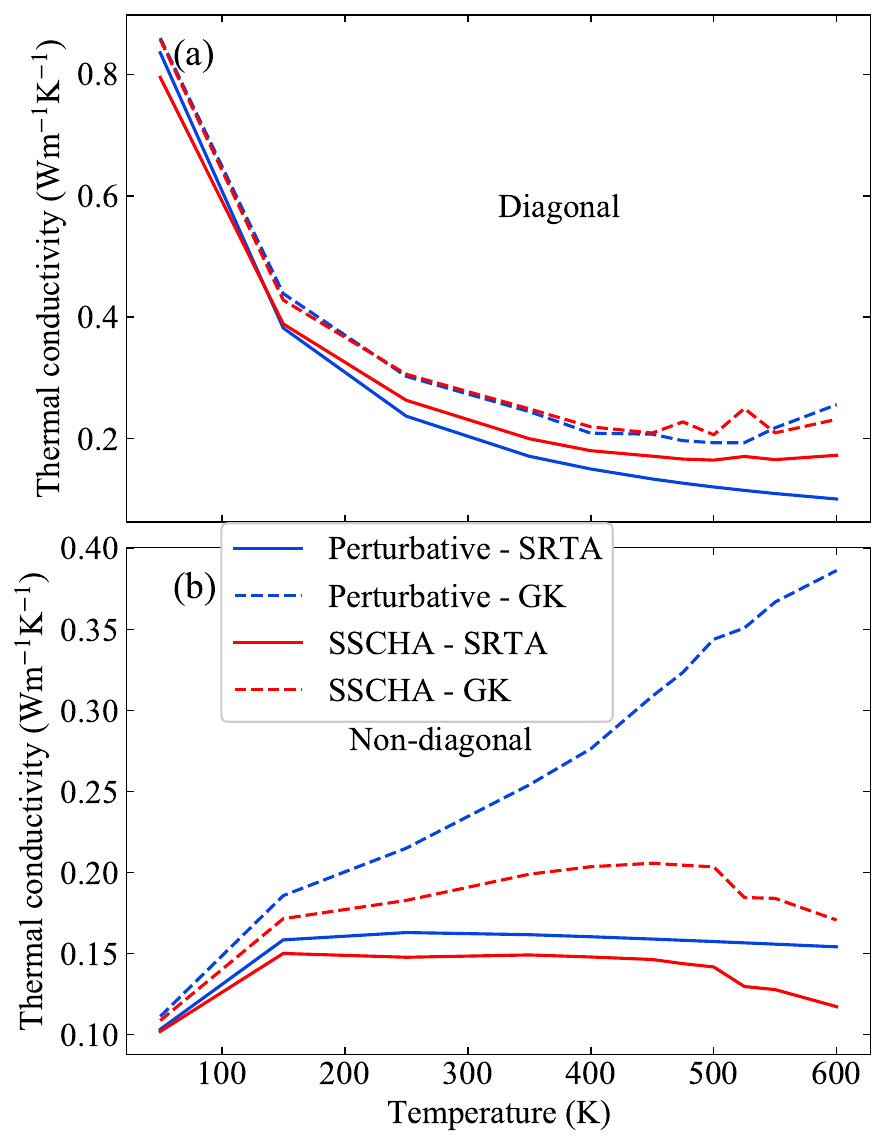}
	\caption{(a) Diagonal part and (b) the coherent part of the lattice thermal conductivity of CsPbBr$_3$ calculated for the orthorhombic $Pnma$ phase with the perturbative limit of the Green-Kubo approach (Eqs.~\ref{kappa_bte} and~\ref{pert_limit} full lines) and the Green-Kubo method (dashed lines) in $x$ Cartesian direction with temperature and structure-dependent SSCHA force constants and perturbative force constants obtained by finite difference.}
	\label{fig:comp_w_harm}
\end{figure}

Finally, it is worth exploring what is the consequence of accounting for the temperature and structure dependence of the interatomic force constants in the phonon self-energy, as in the SSCHA self-energy, compared to those obtained directly as derivatives of the BOES as in perturbation theory and not as expectation values. We calculated perturbative second and third-order force constants for the orthorhombic phase for the structure that minimizes the Born-Oppenheimer energy surface by using finite difference approaches as implemented in the PHONOPY/PHONO3PY code~\cite{phonopy-phono3py-JPSJ}. With these force constants, we calculated the lattice thermal conductivity for the same temperature range as in the SSCHA study, see Fig.~\eqref{fig:comp_w_harm}. In the perturbative limit, we see that the lattice thermal conductivity using force constants obtained from numerical derivatives is lower at higher temperatures. This is due to the fact that temperature tends to reduce SSCHA non-perturbative force constants, as it has been already observed in GeTe~\cite{GeTe} (see Supp. Material for more details~\cite{supp_mat}). Remarkably, the Green-Kubo results for the diagonal contribution to $\kappa$ agree remarkably well between finite difference and SSCHA force constants below 400 K. This shows that the quasiparticle picture holds up very well up to this temperature, roughly until the orthorhombic structure is the minimum of the total free energy. For the off-diagonal contribution, there is a significant difference between finite difference and SSCHA force constants in the Green-Kubo approach, with the finite difference approach giving much larger values. This can be viewed as a result of the stronger dampening of the phonon modes in the case of finite difference force constants, which increases the overlap between the spectral functions of different phonon modes for the same wave vector \textbf{q}. 

\section{Conclusions}

In this paper, we presented the Green-Kubo method for calculating the lattice thermal conductivity of highly anharmonic materials, which is capable of handling situations with the presence of displacive phase transitions. This Green-Kubo method uses the same ingredients as the standard Boltzmann transport equation approach, i.e. second and third-order force constants. However, in contrast to the perturbative limit, the Green-Kubo method uses phonon spectral functions, rather than phonon lifetimes, to describe the lattice thermal conductivity. This means that the Green-Kubo method is applicable even in the case of overdamped phonon modes, where a clear definition of phonon lifetimes and energies is impossible. The Green-Kubo method naturally includes the off-diagonal in phonon branches contribution to the lattice thermal conductivity, meaning it is able to model coherent transport, which is the dominant contribution in complex crystals and amorphous materials. We extended the theory to calculate the complex dynamical lattice thermal conductivity. The results of the dynamical lattice thermal conductivity can be used to analyze time-dependent thermoreflectance measurements.

We apply the Green-Kubo method to calculate the lattice thermal conductivity of CsPbBr$_3$ across different crystal structures. We model the phase transitions in this material with the stochastic self-consistent harmonic approximation. Our results overestimate the transition temperatures by around 100 K, which is a consequence of the parametrization of the exchange-correlation functional used in the training of the machine learning potential we employ. Strong softening of the phonon modes and departure from Lorentzian lineshape of phonon spectral functions is observed in $P4/mbm$ and $Pm\bar{3}m$ phases in their stability range of temperatures. 

We calculate the lattice thermal conductivity of all three phases of interest ($Pnma$, $P4/mbm$, and $Pm\bar{3}m$) in the entire temperature range (50 K - 600 K). We find that $Pm\bar{3}$m phase has the highest lattice thermal conductivity in the entire temperature range, followed by the $P4/mbm$ phase, with $Pnma$ having the lowest $\kappa$. The reason for this hierarchy is the longer phonon lifetimes and higher group velocities the higher the symmetry. We compared our SSCHA results for the orthorhombic phase with results obtained with perturbative force constants. In the perturbative limit $\kappa$ is smaller at high temperatures for perturbative force constants since the SSCHA reduces third-order force constants considerably. The Green-Kubo method on the other hand shows remarkable agreement between the two approaches for the diagonal part of the conductivity. 

Differences between perturbative limit and GK results are of the order of 20\%, larger than the differences coming from the temperature renormalization of the force constants. However, in the case of CsPbBr$_3$ it is hard to discern whether the Green-Kubo approach leads to a practical increase in the predictive power for the lattice thermal conductivity. More studies focusing on the application of the Green-Kubo method are needed in order to definitely gauge the impact of the non-Lorentzian lineshapes on the lattice thermal conductivity. 
 
\section{Acknowledgments}

The authors thank Lorenzo Monacelli and Francesco Mauri for useful discussions. This work was supported by the European Research Council (ERC) under the European Union's Horizon 2020 research and innovation program (grant agreement No. 802533), the Spanish Ministry of Science and Innovation (Grant No. PID2022-142861NA-I00), and the Department of Education, Universities and Research of the Eusko Jaurlaritza and the University of the Basque Country UPV/EHU (Grant No. IT1527-22). 

\bibliography{main}

\providecommand{\noopsort}[1]{}\providecommand{\singleletter}[1]{#1}%
\begin{thebibliography}{90}%
\makeatletter
\providecommand \@ifxundefined [1]{%
 \@ifx{#1\undefined}
}%
\providecommand \@ifnum [1]{%
 \ifnum #1\expandafter \@firstoftwo
 \else \expandafter \@secondoftwo
 \fi
}%
\providecommand \@ifx [1]{%
 \ifx #1\expandafter \@firstoftwo
 \else \expandafter \@secondoftwo
 \fi
}%
\providecommand \natexlab [1]{#1}%
\providecommand \enquote  [1]{``#1''}%
\providecommand \bibnamefont  [1]{#1}%
\providecommand \bibfnamefont [1]{#1}%
\providecommand \citenamefont [1]{#1}%
\providecommand \href@noop [0]{\@secondoftwo}%
\providecommand \href [0]{\begingroup \@sanitize@url \@href}%
\providecommand \@href[1]{\@@startlink{#1}\@@href}%
\providecommand \@@href[1]{\endgroup#1\@@endlink}%
\providecommand \@sanitize@url [0]{\catcode `\\12\catcode `\$12\catcode
  `\&12\catcode `\#12\catcode `\^12\catcode `\_12\catcode `\%12\relax}%
\providecommand \@@startlink[1]{}%
\providecommand \@@endlink[0]{}%
\providecommand \url  [0]{\begingroup\@sanitize@url \@url }%
\providecommand \@url [1]{\endgroup\@href {#1}{\urlprefix }}%
\providecommand \urlprefix  [0]{URL }%
\providecommand \Eprint [0]{\href }%
\providecommand \doibase [0]{https://doi.org/}%
\providecommand \selectlanguage [0]{\@gobble}%
\providecommand \bibinfo  [0]{\@secondoftwo}%
\providecommand \bibfield  [0]{\@secondoftwo}%
\providecommand \translation [1]{[#1]}%
\providecommand \BibitemOpen [0]{}%
\providecommand \bibitemStop [0]{}%
\providecommand \bibitemNoStop [0]{.\EOS\space}%
\providecommand \EOS [0]{\spacefactor3000\relax}%
\providecommand \BibitemShut  [1]{\csname bibitem#1\endcsname}%
\let\auto@bib@innerbib\@empty
\bibitem [{\citenamefont {Zhao}\ \emph {et~al.}(2017)\citenamefont {Zhao},
  \citenamefont {Pan}, \citenamefont {Wan}, \citenamefont {Qu}, \citenamefont
  {Li},\ and\ \citenamefont {Yang}}]{Thermalinsulation1}%
  \BibitemOpen
  \bibfield  {author} {\bibinfo {author} {\bibfnamefont {M.}~\bibnamefont
  {Zhao}}, \bibinfo {author} {\bibfnamefont {W.}~\bibnamefont {Pan}}, \bibinfo
  {author} {\bibfnamefont {C.}~\bibnamefont {Wan}}, \bibinfo {author}
  {\bibfnamefont {Z.}~\bibnamefont {Qu}}, \bibinfo {author} {\bibfnamefont
  {Z.}~\bibnamefont {Li}},\ and\ \bibinfo {author} {\bibfnamefont
  {J.}~\bibnamefont {Yang}},\ }\bibfield  {title} {\bibinfo {title} {{Defect
  engineering in development of low thermal conductivity materials: A
  review}},\ }\href
  {https://doi.org/https://doi.org/10.1016/j.jeurceramsoc.2016.07.036}
  {\bibfield  {journal} {\bibinfo  {journal} {Journal of the European Ceramic
  Society}\ }\textbf {\bibinfo {volume} {37}},\ \bibinfo {pages} {1} (\bibinfo
  {year} {2017})}\BibitemShut {NoStop}%
\bibitem [{\citenamefont {Li}\ \emph {et~al.}(2019)\citenamefont {Li},
  \citenamefont {Zhou}, \citenamefont {Liu}, \citenamefont {Liang},\ and\
  \citenamefont {Zhang}}]{Thermalinsulation2}%
  \BibitemOpen
  \bibfield  {author} {\bibinfo {author} {\bibfnamefont {F.}~\bibnamefont
  {Li}}, \bibinfo {author} {\bibfnamefont {L.}~\bibnamefont {Zhou}}, \bibinfo
  {author} {\bibfnamefont {J.-X.}\ \bibnamefont {Liu}}, \bibinfo {author}
  {\bibfnamefont {Y.}~\bibnamefont {Liang}},\ and\ \bibinfo {author}
  {\bibfnamefont {G.-J.}\ \bibnamefont {Zhang}},\ }\bibfield  {title} {\bibinfo
  {title} {High-entropy pyrochlores with low thermal conductivity for thermal
  barrier coating materials},\ }\href
  {https://doi.org/10.1007/s40145-019-0342-4} {\bibfield  {journal} {\bibinfo
  {journal} {Journal of Advanced Ceramics}\ }\textbf {\bibinfo {volume} {8}},\
  \bibinfo {pages} {576} (\bibinfo {year} {2019})}\BibitemShut {NoStop}%
\bibitem [{\citenamefont {Dangi{\'{c}}}\ \emph {et~al.}(2021)\citenamefont
  {Dangi{\'{c}}}, \citenamefont {Hellman}, \citenamefont {Fahy},\ and\
  \citenamefont {Savi{\'{c}}}}]{GeTe}%
  \BibitemOpen
  \bibfield  {author} {\bibinfo {author} {\bibfnamefont {{\DJ}.}~\bibnamefont
  {Dangi{\'{c}}}}, \bibinfo {author} {\bibfnamefont {O.}~\bibnamefont
  {Hellman}}, \bibinfo {author} {\bibfnamefont {S.}~\bibnamefont {Fahy}},\ and\
  \bibinfo {author} {\bibfnamefont {I.}~\bibnamefont {Savi{\'{c}}}},\
  }\bibfield  {title} {\bibinfo {title} {{The origin of the lattice thermal
  conductivity enhancement at the ferroelectric phase transition in GeTe}},\
  }\href {https://doi.org/10.1038/s41524-021-00523-7} {\bibfield  {journal}
  {\bibinfo  {journal} {npj Computational Materials}\ }\textbf {\bibinfo
  {volume} {7}},\ \bibinfo {pages} {57} (\bibinfo {year} {2021})}\BibitemShut
  {NoStop}%
\bibitem [{\citenamefont {Aseginolaza}\ \emph {et~al.}(2019)\citenamefont
  {Aseginolaza}, \citenamefont {Bianco}, \citenamefont {Monacelli},
  \citenamefont {Paulatto}, \citenamefont {Calandra}, \citenamefont {Mauri},
  \citenamefont {Bergara},\ and\ \citenamefont {Errea}}]{IonSnSe}%
  \BibitemOpen
  \bibfield  {author} {\bibinfo {author} {\bibfnamefont {U.}~\bibnamefont
  {Aseginolaza}}, \bibinfo {author} {\bibfnamefont {R.}~\bibnamefont {Bianco}},
  \bibinfo {author} {\bibfnamefont {L.}~\bibnamefont {Monacelli}}, \bibinfo
  {author} {\bibfnamefont {L.}~\bibnamefont {Paulatto}}, \bibinfo {author}
  {\bibfnamefont {M.}~\bibnamefont {Calandra}}, \bibinfo {author}
  {\bibfnamefont {F.}~\bibnamefont {Mauri}}, \bibinfo {author} {\bibfnamefont
  {A.}~\bibnamefont {Bergara}},\ and\ \bibinfo {author} {\bibfnamefont
  {I.}~\bibnamefont {Errea}},\ }\bibfield  {title} {\bibinfo {title} {{Phonon
  Collapse and Second-Order Phase Transition in Thermoelectric SnSe}},\ }\href
  {https://doi.org/10.1103/PhysRevLett.122.075901} {\bibfield  {journal}
  {\bibinfo  {journal} {Phys. Rev. Lett.}\ }\textbf {\bibinfo {volume} {122}},\
  \bibinfo {pages} {075901} (\bibinfo {year} {2019})}\BibitemShut {NoStop}%
\bibitem [{\citenamefont {Murphy}\ \emph {et~al.}(2016)\citenamefont {Murphy},
  \citenamefont {Murray}, \citenamefont {Fahy},\ and\ \citenamefont
  {Savi\ifmmode~\acute{c}\else \'{c}\fi{}}}]{PbTe}%
  \BibitemOpen
  \bibfield  {author} {\bibinfo {author} {\bibfnamefont {R.~M.}\ \bibnamefont
  {Murphy}}, \bibinfo {author} {\bibfnamefont {E.~D.}\ \bibnamefont {Murray}},
  \bibinfo {author} {\bibfnamefont {S.}~\bibnamefont {Fahy}},\ and\ \bibinfo
  {author} {\bibfnamefont {I.}~\bibnamefont {Savi\ifmmode~\acute{c}\else
  \'{c}\fi{}}},\ }\bibfield  {title} {\bibinfo {title} {{Broadband phonon
  scattering in PbTe-based materials driven near ferroelectric phase transition
  by strain or alloying}},\ }\href {https://doi.org/10.1103/PhysRevB.93.104304}
  {\bibfield  {journal} {\bibinfo  {journal} {Phys. Rev. B}\ }\textbf {\bibinfo
  {volume} {93}},\ \bibinfo {pages} {104304} (\bibinfo {year}
  {2016})}\BibitemShut {NoStop}%
\bibitem [{\citenamefont {Murphy}\ \emph {et~al.}(2017)\citenamefont {Murphy},
  \citenamefont {Murray}, \citenamefont {Fahy},\ and\ \citenamefont
  {Savi\ifmmode~\acute{c}\else \'{c}\fi{}}}]{PbGeTe}%
  \BibitemOpen
  \bibfield  {author} {\bibinfo {author} {\bibfnamefont {R.~M.}\ \bibnamefont
  {Murphy}}, \bibinfo {author} {\bibfnamefont {E.~D.}\ \bibnamefont {Murray}},
  \bibinfo {author} {\bibfnamefont {S.}~\bibnamefont {Fahy}},\ and\ \bibinfo
  {author} {\bibfnamefont {I.}~\bibnamefont {Savi\ifmmode~\acute{c}\else
  \'{c}\fi{}}},\ }\bibfield  {title} {\bibinfo {title} {{Ferroelectric phase
  transition and the lattice thermal conductivity of
  ${\mathrm{Pb}}_{1\ensuremath{-}x}{\mathrm{Ge}}_{x}\mathrm{Te}$ alloys}},\
  }\href {https://doi.org/10.1103/PhysRevB.95.144302} {\bibfield  {journal}
  {\bibinfo  {journal} {Phys. Rev. B}\ }\textbf {\bibinfo {volume} {95}},\
  \bibinfo {pages} {144302} (\bibinfo {year} {2017})}\BibitemShut {NoStop}%
\bibitem [{\citenamefont {Monacelli}\ \emph
  {et~al.}(2021{\natexlab{a}})\citenamefont {Monacelli}, \citenamefont
  {Bianco}, \citenamefont {Cherubini}, \citenamefont {Calandra}, \citenamefont
  {Errea},\ and\ \citenamefont {Mauri}}]{SSCHA1}%
  \BibitemOpen
  \bibfield  {author} {\bibinfo {author} {\bibfnamefont {L.}~\bibnamefont
  {Monacelli}}, \bibinfo {author} {\bibfnamefont {R.}~\bibnamefont {Bianco}},
  \bibinfo {author} {\bibfnamefont {M.}~\bibnamefont {Cherubini}}, \bibinfo
  {author} {\bibfnamefont {M.}~\bibnamefont {Calandra}}, \bibinfo {author}
  {\bibfnamefont {I.}~\bibnamefont {Errea}},\ and\ \bibinfo {author}
  {\bibfnamefont {F.}~\bibnamefont {Mauri}},\ }\bibfield  {title} {\bibinfo
  {title} {The stochastic self-consistent harmonic approximation: calculating
  vibrational properties of materials with full quantum and anharmonic
  effects},\ }\href {https://doi.org/10.1088/1361-648X/ac066b} {\bibfield
  {journal} {\bibinfo  {journal} {Journal of Physics: Condensed Matter}\
  }\textbf {\bibinfo {volume} {33}},\ \bibinfo {pages} {363001} (\bibinfo
  {year} {2021}{\natexlab{a}})}\BibitemShut {NoStop}%
\bibitem [{\citenamefont {Errea}\ \emph {et~al.}(2013)\citenamefont {Errea},
  \citenamefont {Calandra},\ and\ \citenamefont {Mauri}}]{SSCHA2}%
  \BibitemOpen
  \bibfield  {author} {\bibinfo {author} {\bibfnamefont {I.}~\bibnamefont
  {Errea}}, \bibinfo {author} {\bibfnamefont {M.}~\bibnamefont {Calandra}},\
  and\ \bibinfo {author} {\bibfnamefont {F.}~\bibnamefont {Mauri}},\ }\bibfield
   {title} {\bibinfo {title} {{First-Principles Theory of Anharmonicity and the
  Inverse Isotope Effect in Superconducting Palladium-Hydride Compounds}},\
  }\href {https://doi.org/10.1103/PhysRevLett.111.177002} {\bibfield  {journal}
  {\bibinfo  {journal} {Phys. Rev. Lett.}\ }\textbf {\bibinfo {volume} {111}},\
  \bibinfo {pages} {177002} (\bibinfo {year} {2013})}\BibitemShut {NoStop}%
\bibitem [{\citenamefont {Errea}\ \emph {et~al.}(2014)\citenamefont {Errea},
  \citenamefont {Calandra},\ and\ \citenamefont {Mauri}}]{SSCHA3}%
  \BibitemOpen
  \bibfield  {author} {\bibinfo {author} {\bibfnamefont {I.}~\bibnamefont
  {Errea}}, \bibinfo {author} {\bibfnamefont {M.}~\bibnamefont {Calandra}},\
  and\ \bibinfo {author} {\bibfnamefont {F.}~\bibnamefont {Mauri}},\ }\bibfield
   {title} {\bibinfo {title} {{Anharmonic free energies and phonon dispersions
  from the stochastic self-consistent harmonic approximation: Application to
  platinum and palladium hydrides}},\ }\href
  {https://doi.org/10.1103/PhysRevB.89.064302} {\bibfield  {journal} {\bibinfo
  {journal} {Phys. Rev. B}\ }\textbf {\bibinfo {volume} {89}},\ \bibinfo
  {pages} {064302} (\bibinfo {year} {2014})}\BibitemShut {NoStop}%
\bibitem [{\citenamefont {Hellman}\ and\ \citenamefont
  {Abrikosov}(2013)}]{TDEP}%
  \BibitemOpen
  \bibfield  {author} {\bibinfo {author} {\bibfnamefont {O.}~\bibnamefont
  {Hellman}}\ and\ \bibinfo {author} {\bibfnamefont {I.~A.}\ \bibnamefont
  {Abrikosov}},\ }\bibfield  {title} {\bibinfo {title} {Temperature-dependent
  effective third-order interatomic force constants from first principles},\
  }\href {https://doi.org/10.1103/PhysRevB.88.144301} {\bibfield  {journal}
  {\bibinfo  {journal} {Phys. Rev. B}\ }\textbf {\bibinfo {volume} {88}},\
  \bibinfo {pages} {144301} (\bibinfo {year} {2013})}\BibitemShut {NoStop}%
\bibitem [{\citenamefont
  {Monacelli}(2024)}]{monacelli2024simulatinganharmoniccrystalslights}%
  \BibitemOpen
  \bibfield  {author} {\bibinfo {author} {\bibfnamefont {L.}~\bibnamefont
  {Monacelli}},\ }\href {https://arxiv.org/abs/2407.03090} {\bibinfo {title}
  {Simulating anharmonic crystals: Lights and shadows of first-principles
  approaches}} (\bibinfo {year} {2024}),\ \Eprint
  {https://arxiv.org/abs/2407.03090} {arXiv:2407.03090 [cond-mat.mtrl-sci]}
  \BibitemShut {NoStop}%
\bibitem [{\citenamefont {Ghim}\ \emph {et~al.}(2023)\citenamefont {Ghim},
  \citenamefont {Choi},\ and\ \citenamefont {Jhi}}]{vacancies}%
  \BibitemOpen
  \bibfield  {author} {\bibinfo {author} {\bibfnamefont {M.}~\bibnamefont
  {Ghim}}, \bibinfo {author} {\bibfnamefont {Y.-J.}\ \bibnamefont {Choi}},\
  and\ \bibinfo {author} {\bibfnamefont {S.-H.}\ \bibnamefont {Jhi}},\
  }\bibfield  {title} {\bibinfo {title} {{Lattice thermal conductivity of cubic
  GeTe with vacancy defects}},\ }\href
  {https://doi.org/10.1103/PhysRevB.107.184301} {\bibfield  {journal} {\bibinfo
   {journal} {Phys. Rev. B}\ }\textbf {\bibinfo {volume} {107}},\ \bibinfo
  {pages} {184301} (\bibinfo {year} {2023})}\BibitemShut {NoStop}%
\bibitem [{\citenamefont {Lindsay}\ \emph {et~al.}(2013)\citenamefont
  {Lindsay}, \citenamefont {Broido},\ and\ \citenamefont {Reinecke}}]{isotope}%
  \BibitemOpen
  \bibfield  {author} {\bibinfo {author} {\bibfnamefont {L.}~\bibnamefont
  {Lindsay}}, \bibinfo {author} {\bibfnamefont {D.~A.}\ \bibnamefont
  {Broido}},\ and\ \bibinfo {author} {\bibfnamefont {T.~L.}\ \bibnamefont
  {Reinecke}},\ }\bibfield  {title} {\bibinfo {title} {{Phonon-isotope
  scattering and thermal conductivity in materials with a large isotope effect:
  A first-principles study}},\ }\href
  {https://doi.org/10.1103/PhysRevB.88.144306} {\bibfield  {journal} {\bibinfo
  {journal} {Phys. Rev. B}\ }\textbf {\bibinfo {volume} {88}},\ \bibinfo
  {pages} {144306} (\bibinfo {year} {2013})}\BibitemShut {NoStop}%
\bibitem [{\citenamefont {Dangi\ifmmode~\acute{c}\else \'{c}\fi{}}\ \emph
  {et~al.}(2020)\citenamefont {Dangi\ifmmode~\acute{c}\else \'{c}\fi{}},
  \citenamefont {Murray}, \citenamefont {Fahy},\ and\ \citenamefont
  {Savi\ifmmode~\acute{c}\else \'{c}\fi{}}}]{DWs}%
  \BibitemOpen
  \bibfield  {author} {\bibinfo {author} {\bibfnamefont {{\DJ}.}~\bibnamefont
  {Dangi\ifmmode~\acute{c}\else \'{c}\fi{}}}, \bibinfo {author} {\bibfnamefont
  {E.~D.}\ \bibnamefont {Murray}}, \bibinfo {author} {\bibfnamefont
  {S.}~\bibnamefont {Fahy}},\ and\ \bibinfo {author} {\bibfnamefont
  {I.}~\bibnamefont {Savi\ifmmode~\acute{c}\else \'{c}\fi{}}},\ }\bibfield
  {title} {\bibinfo {title} {{Structural and thermal transport properties of
  ferroelectric domain walls in GeTe from first principles}},\ }\href
  {https://doi.org/10.1103/PhysRevB.101.184110} {\bibfield  {journal} {\bibinfo
   {journal} {Phys. Rev. B}\ }\textbf {\bibinfo {volume} {101}},\ \bibinfo
  {pages} {184110} (\bibinfo {year} {2020})}\BibitemShut {NoStop}%
\bibitem [{\citenamefont {Ravichandran}\ and\ \citenamefont
  {Broido}(2019)}]{BoronArsenide}%
  \BibitemOpen
  \bibfield  {author} {\bibinfo {author} {\bibfnamefont {N.~K.}\ \bibnamefont
  {Ravichandran}}\ and\ \bibinfo {author} {\bibfnamefont {D.}~\bibnamefont
  {Broido}},\ }\bibfield  {title} {\bibinfo {title} {Non-monotonic pressure
  dependence of the thermal conductivity of boron arsenide},\ }\href
  {https://doi.org/10.1038/s41467-019-08713-0} {\bibfield  {journal} {\bibinfo
  {journal} {Nature Communications}\ }\textbf {\bibinfo {volume} {10}},\
  \bibinfo {pages} {827} (\bibinfo {year} {2019})}\BibitemShut {NoStop}%
\bibitem [{\citenamefont {Broido}\ \emph {et~al.}(2007)\citenamefont {Broido},
  \citenamefont {Malorny}, \citenamefont {Birner}, \citenamefont {Mingo},\ and\
  \citenamefont {Stewart}}]{Broido_review}%
  \BibitemOpen
  \bibfield  {author} {\bibinfo {author} {\bibfnamefont {D.~A.}\ \bibnamefont
  {Broido}}, \bibinfo {author} {\bibfnamefont {M.}~\bibnamefont {Malorny}},
  \bibinfo {author} {\bibfnamefont {G.}~\bibnamefont {Birner}}, \bibinfo
  {author} {\bibfnamefont {N.}~\bibnamefont {Mingo}},\ and\ \bibinfo {author}
  {\bibfnamefont {D.~A.}\ \bibnamefont {Stewart}},\ }\bibfield  {title}
  {\bibinfo {title} {{Intrinsic lattice thermal conductivity of semiconductors
  from first principles}},\ }\href {https://doi.org/10.1063/1.2822891}
  {\bibfield  {journal} {\bibinfo  {journal} {Applied Physics Letters}\
  }\textbf {\bibinfo {volume} {91}},\ \bibinfo {pages} {231922} (\bibinfo
  {year} {2007})}\BibitemShut {NoStop}%
\bibitem [{\citenamefont {Fugallo}\ and\ \citenamefont
  {Colombo}(2018)}]{Fugallo_2018}%
  \BibitemOpen
  \bibfield  {author} {\bibinfo {author} {\bibfnamefont {G.}~\bibnamefont
  {Fugallo}}\ and\ \bibinfo {author} {\bibfnamefont {L.}~\bibnamefont
  {Colombo}},\ }\bibfield  {title} {\bibinfo {title} {Calculating lattice
  thermal conductivity: a synopsis},\ }\href
  {https://doi.org/10.1088/1402-4896/aaa6f3} {\bibfield  {journal} {\bibinfo
  {journal} {Physica Scripta}\ }\textbf {\bibinfo {volume} {93}},\ \bibinfo
  {pages} {043002} (\bibinfo {year} {2018})}\BibitemShut {NoStop}%
\bibitem [{\citenamefont {Knoop}\ \emph {et~al.}(2023)\citenamefont {Knoop},
  \citenamefont {Purcell}, \citenamefont {Scheffler},\ and\ \citenamefont
  {Carbogno}}]{Anharmonicity_Knoop}%
  \BibitemOpen
  \bibfield  {author} {\bibinfo {author} {\bibfnamefont {F.}~\bibnamefont
  {Knoop}}, \bibinfo {author} {\bibfnamefont {T.~A.~R.}\ \bibnamefont
  {Purcell}}, \bibinfo {author} {\bibfnamefont {M.}~\bibnamefont {Scheffler}},\
  and\ \bibinfo {author} {\bibfnamefont {C.}~\bibnamefont {Carbogno}},\
  }\bibfield  {title} {\bibinfo {title} {{Anharmonicity in Thermal Insulators:
  An Analysis from First Principles}},\ }\href
  {https://doi.org/10.1103/PhysRevLett.130.236301} {\bibfield  {journal}
  {\bibinfo  {journal} {Phys. Rev. Lett.}\ }\textbf {\bibinfo {volume} {130}},\
  \bibinfo {pages} {236301} (\bibinfo {year} {2023})}\BibitemShut {NoStop}%
\bibitem [{\citenamefont {Kim}\ \emph {et~al.}(2020)\citenamefont {Kim},
  \citenamefont {Hellman}, \citenamefont {Shulumba}, \citenamefont {Saunders},
  \citenamefont {Lin}, \citenamefont {Smith}, \citenamefont {Herriman},
  \citenamefont {Niedziela}, \citenamefont {Abernathy}, \citenamefont {Li},\
  and\ \citenamefont {Fultz}}]{Si_lifetimes}%
  \BibitemOpen
  \bibfield  {author} {\bibinfo {author} {\bibfnamefont {D.~S.}\ \bibnamefont
  {Kim}}, \bibinfo {author} {\bibfnamefont {O.}~\bibnamefont {Hellman}},
  \bibinfo {author} {\bibfnamefont {N.}~\bibnamefont {Shulumba}}, \bibinfo
  {author} {\bibfnamefont {C.~N.}\ \bibnamefont {Saunders}}, \bibinfo {author}
  {\bibfnamefont {J.~Y.~Y.}\ \bibnamefont {Lin}}, \bibinfo {author}
  {\bibfnamefont {H.~L.}\ \bibnamefont {Smith}}, \bibinfo {author}
  {\bibfnamefont {J.~E.}\ \bibnamefont {Herriman}}, \bibinfo {author}
  {\bibfnamefont {J.~L.}\ \bibnamefont {Niedziela}}, \bibinfo {author}
  {\bibfnamefont {D.~L.}\ \bibnamefont {Abernathy}}, \bibinfo {author}
  {\bibfnamefont {C.~W.}\ \bibnamefont {Li}},\ and\ \bibinfo {author}
  {\bibfnamefont {B.}~\bibnamefont {Fultz}},\ }\bibfield  {title} {\bibinfo
  {title} {Temperature-dependent phonon lifetimes and thermal conductivity of
  silicon by inelastic neutron scattering and ab initio calculations},\ }\href
  {https://doi.org/10.1103/PhysRevB.102.174311} {\bibfield  {journal} {\bibinfo
   {journal} {Phys. Rev. B}\ }\textbf {\bibinfo {volume} {102}},\ \bibinfo
  {pages} {174311} (\bibinfo {year} {2020})}\BibitemShut {NoStop}%
\bibitem [{\citenamefont {Togo}(2023)}]{phonopy-phono3py-JPSJ}%
  \BibitemOpen
  \bibfield  {author} {\bibinfo {author} {\bibfnamefont {A.}~\bibnamefont
  {Togo}},\ }\bibfield  {title} {\bibinfo {title} {{First-principles Phonon
  Calculations with Phonopy and Phono3py}},\ }\href
  {https://doi.org/10.7566/JPSJ.92.012001} {\bibfield  {journal} {\bibinfo
  {journal} {J. Phys. Soc. Jpn.}\ }\textbf {\bibinfo {volume} {92}},\ \bibinfo
  {pages} {012001} (\bibinfo {year} {2023})}\BibitemShut {NoStop}%
\bibitem [{\citenamefont {Tadano}\ \emph {et~al.}(2014)\citenamefont {Tadano},
  \citenamefont {Gohda},\ and\ \citenamefont {Tsuneyuki}}]{Alamode}%
  \BibitemOpen
  \bibfield  {author} {\bibinfo {author} {\bibfnamefont {T.}~\bibnamefont
  {Tadano}}, \bibinfo {author} {\bibfnamefont {Y.}~\bibnamefont {Gohda}},\ and\
  \bibinfo {author} {\bibfnamefont {S.}~\bibnamefont {Tsuneyuki}},\ }\bibfield
  {title} {\bibinfo {title} {Anharmonic force constants extracted from
  first-principles molecular dynamics: applications to heat transfer
  simulations},\ }\href {https://doi.org/10.1088/0953-8984/26/22/225402}
  {\bibfield  {journal} {\bibinfo  {journal} {Journal of Physics: Condensed
  Matter}\ }\textbf {\bibinfo {volume} {26}},\ \bibinfo {pages} {225402}
  (\bibinfo {year} {2014})}\BibitemShut {NoStop}%
\bibitem [{\citenamefont {Li}\ \emph {et~al.}(2014{\natexlab{a}})\citenamefont
  {Li}, \citenamefont {Carrete}, \citenamefont {{A. Katcho}},\ and\
  \citenamefont {Mingo}}]{ShengBTE}%
  \BibitemOpen
  \bibfield  {author} {\bibinfo {author} {\bibfnamefont {W.}~\bibnamefont
  {Li}}, \bibinfo {author} {\bibfnamefont {J.}~\bibnamefont {Carrete}},
  \bibinfo {author} {\bibfnamefont {N.}~\bibnamefont {{A. Katcho}}},\ and\
  \bibinfo {author} {\bibfnamefont {N.}~\bibnamefont {Mingo}},\ }\bibfield
  {title} {\bibinfo {title} {{ShengBTE: A solver of the Boltzmann transport
  equation for phonons}},\ }\href
  {https://doi.org/https://doi.org/10.1016/j.cpc.2014.02.015} {\bibfield
  {journal} {\bibinfo  {journal} {Computer Physics Communications}\ }\textbf
  {\bibinfo {volume} {185}},\ \bibinfo {pages} {1747} (\bibinfo {year}
  {2014}{\natexlab{a}})}\BibitemShut {NoStop}%
\bibitem [{\citenamefont {Dangi\ifmmode~\acute{c}\else \'{c}\fi{}}\ \emph
  {et~al.}(2018)\citenamefont {Dangi\ifmmode~\acute{c}\else \'{c}\fi{}},
  \citenamefont {Murphy}, \citenamefont {Murray}, \citenamefont {Fahy},\ and\
  \citenamefont {Savi\ifmmode~\acute{c}\else \'{c}\fi{}}}]{GeTeTexp}%
  \BibitemOpen
  \bibfield  {author} {\bibinfo {author} {\bibfnamefont {{\DJ}.}~\bibnamefont
  {Dangi\ifmmode~\acute{c}\else \'{c}\fi{}}}, \bibinfo {author} {\bibfnamefont
  {A.~R.}\ \bibnamefont {Murphy}}, \bibinfo {author} {\bibfnamefont {E.~D.}\
  \bibnamefont {Murray}}, \bibinfo {author} {\bibfnamefont {S.}~\bibnamefont
  {Fahy}},\ and\ \bibinfo {author} {\bibfnamefont {I.}~\bibnamefont
  {Savi\ifmmode~\acute{c}\else \'{c}\fi{}}},\ }\bibfield  {title} {\bibinfo
  {title} {{Coupling between acoustic and soft transverse optical phonons leads
  to negative thermal expansion of GeTe near the ferroelectric phase
  transition}},\ }\href {https://doi.org/10.1103/PhysRevB.97.224106} {\bibfield
   {journal} {\bibinfo  {journal} {Phys. Rev. B}\ }\textbf {\bibinfo {volume}
  {97}},\ \bibinfo {pages} {224106} (\bibinfo {year} {2018})}\BibitemShut
  {NoStop}%
\bibitem [{\citenamefont {Schelling}\ and\ \citenamefont
  {Keblinski}(2003)}]{QHarm}%
  \BibitemOpen
  \bibfield  {author} {\bibinfo {author} {\bibfnamefont {P.~K.}\ \bibnamefont
  {Schelling}}\ and\ \bibinfo {author} {\bibfnamefont {P.}~\bibnamefont
  {Keblinski}},\ }\bibfield  {title} {\bibinfo {title} {Thermal expansion of
  carbon structures},\ }\href {https://doi.org/10.1103/PhysRevB.68.035425}
  {\bibfield  {journal} {\bibinfo  {journal} {Phys. Rev. B}\ }\textbf {\bibinfo
  {volume} {68}},\ \bibinfo {pages} {035425} (\bibinfo {year}
  {2003})}\BibitemShut {NoStop}%
\bibitem [{\citenamefont {Diego}\ \emph {et~al.}(2021)\citenamefont {Diego},
  \citenamefont {Said}, \citenamefont {Mahatha}, \citenamefont {Bianco},
  \citenamefont {Monacelli}, \citenamefont {Calandra}, \citenamefont {Mauri},
  \citenamefont {Rossnagel}, \citenamefont {Errea},\ and\ \citenamefont
  {Blanco-Canosa}}]{CDW}%
  \BibitemOpen
  \bibfield  {author} {\bibinfo {author} {\bibfnamefont {J.}~\bibnamefont
  {Diego}}, \bibinfo {author} {\bibfnamefont {A.~H.}\ \bibnamefont {Said}},
  \bibinfo {author} {\bibfnamefont {S.~K.}\ \bibnamefont {Mahatha}}, \bibinfo
  {author} {\bibfnamefont {R.}~\bibnamefont {Bianco}}, \bibinfo {author}
  {\bibfnamefont {L.}~\bibnamefont {Monacelli}}, \bibinfo {author}
  {\bibfnamefont {M.}~\bibnamefont {Calandra}}, \bibinfo {author}
  {\bibfnamefont {F.}~\bibnamefont {Mauri}}, \bibinfo {author} {\bibfnamefont
  {K.}~\bibnamefont {Rossnagel}}, \bibinfo {author} {\bibfnamefont
  {I.}~\bibnamefont {Errea}},\ and\ \bibinfo {author} {\bibfnamefont
  {S.}~\bibnamefont {Blanco-Canosa}},\ }\bibfield  {title} {\bibinfo {title}
  {van der waals driven anharmonic melting of the 3d charge density wave in
  vse2},\ }\href {https://doi.org/10.1038/s41467-020-20829-2} {\bibfield
  {journal} {\bibinfo  {journal} {Nature Communications}\ }\textbf {\bibinfo
  {volume} {12}},\ \bibinfo {pages} {598} (\bibinfo {year} {2021})}\BibitemShut
  {NoStop}%
\bibitem [{\citenamefont {{van Roekeghem}}\ \emph {et~al.}(2021)\citenamefont
  {{van Roekeghem}}, \citenamefont {Carrete},\ and\ \citenamefont
  {Mingo}}]{QSCAILD}%
  \BibitemOpen
  \bibfield  {author} {\bibinfo {author} {\bibfnamefont {A.}~\bibnamefont {{van
  Roekeghem}}}, \bibinfo {author} {\bibfnamefont {J.}~\bibnamefont {Carrete}},\
  and\ \bibinfo {author} {\bibfnamefont {N.}~\bibnamefont {Mingo}},\ }\bibfield
   {title} {\bibinfo {title} {Quantum self-consistent ab-initio lattice
  dynamics},\ }\href
  {https://doi.org/https://doi.org/10.1016/j.cpc.2021.107945} {\bibfield
  {journal} {\bibinfo  {journal} {Computer Physics Communications}\ }\textbf
  {\bibinfo {volume} {263}},\ \bibinfo {pages} {107945} (\bibinfo {year}
  {2021})}\BibitemShut {NoStop}%
\bibitem [{\citenamefont {Zacharias}\ \emph {et~al.}(2023)\citenamefont
  {Zacharias}, \citenamefont {Volonakis}, \citenamefont {Giustino},\ and\
  \citenamefont {Even}}]{SDM}%
  \BibitemOpen
  \bibfield  {author} {\bibinfo {author} {\bibfnamefont {M.}~\bibnamefont
  {Zacharias}}, \bibinfo {author} {\bibfnamefont {G.}~\bibnamefont
  {Volonakis}}, \bibinfo {author} {\bibfnamefont {F.}~\bibnamefont
  {Giustino}},\ and\ \bibinfo {author} {\bibfnamefont {J.}~\bibnamefont
  {Even}},\ }\bibfield  {title} {\bibinfo {title} {Anharmonic lattice dynamics
  via the special displacement method},\ }\href
  {https://doi.org/10.1103/PhysRevB.108.035155} {\bibfield  {journal} {\bibinfo
   {journal} {Phys. Rev. B}\ }\textbf {\bibinfo {volume} {108}},\ \bibinfo
  {pages} {035155} (\bibinfo {year} {2023})}\BibitemShut {NoStop}%
\bibitem [{\citenamefont {Bianco}\ \emph {et~al.}(2017)\citenamefont {Bianco},
  \citenamefont {Errea}, \citenamefont {Paulatto}, \citenamefont {Calandra},\
  and\ \citenamefont {Mauri}}]{SSCHA4}%
  \BibitemOpen
  \bibfield  {author} {\bibinfo {author} {\bibfnamefont {R.}~\bibnamefont
  {Bianco}}, \bibinfo {author} {\bibfnamefont {I.}~\bibnamefont {Errea}},
  \bibinfo {author} {\bibfnamefont {L.}~\bibnamefont {Paulatto}}, \bibinfo
  {author} {\bibfnamefont {M.}~\bibnamefont {Calandra}},\ and\ \bibinfo
  {author} {\bibfnamefont {F.}~\bibnamefont {Mauri}},\ }\bibfield  {title}
  {\bibinfo {title} {{Second-order structural phase transitions, free energy
  curvature, and temperature-dependent anharmonic phonons in the
  self-consistent harmonic approximation: Theory and stochastic
  implementation}},\ }\href {https://doi.org/10.1103/PhysRevB.96.014111}
  {\bibfield  {journal} {\bibinfo  {journal} {Phys. Rev. B}\ }\textbf {\bibinfo
  {volume} {96}},\ \bibinfo {pages} {014111} (\bibinfo {year}
  {2017})}\BibitemShut {NoStop}%
\bibitem [{\citenamefont {Monacelli}\ \emph {et~al.}(2018)\citenamefont
  {Monacelli}, \citenamefont {Errea}, \citenamefont {Calandra},\ and\
  \citenamefont {Mauri}}]{SSCHA5}%
  \BibitemOpen
  \bibfield  {author} {\bibinfo {author} {\bibfnamefont {L.}~\bibnamefont
  {Monacelli}}, \bibinfo {author} {\bibfnamefont {I.}~\bibnamefont {Errea}},
  \bibinfo {author} {\bibfnamefont {M.}~\bibnamefont {Calandra}},\ and\
  \bibinfo {author} {\bibfnamefont {F.}~\bibnamefont {Mauri}},\ }\bibfield
  {title} {\bibinfo {title} {Pressure and stress tensor of complex anharmonic
  crystals within the stochastic self-consistent harmonic approximation},\
  }\href {https://doi.org/10.1103/PhysRevB.98.024106} {\bibfield  {journal}
  {\bibinfo  {journal} {Phys. Rev. B}\ }\textbf {\bibinfo {volume} {98}},\
  \bibinfo {pages} {024106} (\bibinfo {year} {2018})}\BibitemShut {NoStop}%
\bibitem [{\citenamefont {Monacelli}\ \emph
  {et~al.}(2021{\natexlab{b}})\citenamefont {Monacelli}, \citenamefont {Errea},
  \citenamefont {Calandra},\ and\ \citenamefont {Mauri}}]{Monacelli2021}%
  \BibitemOpen
  \bibfield  {author} {\bibinfo {author} {\bibfnamefont {L.}~\bibnamefont
  {Monacelli}}, \bibinfo {author} {\bibfnamefont {I.}~\bibnamefont {Errea}},
  \bibinfo {author} {\bibfnamefont {M.}~\bibnamefont {Calandra}},\ and\
  \bibinfo {author} {\bibfnamefont {F.}~\bibnamefont {Mauri}},\ }\bibfield
  {title} {\bibinfo {title} {{Black metal hydrogen above 360{\thinspace}GPa
  driven by proton quantum fluctuations}},\ }\href
  {https://doi.org/10.1038/s41567-020-1009-3} {\bibfield  {journal} {\bibinfo
  {journal} {Nature Physics}\ }\textbf {\bibinfo {volume} {17}},\ \bibinfo
  {pages} {63} (\bibinfo {year} {2021}{\natexlab{b}})}\BibitemShut {NoStop}%
\bibitem [{\citenamefont {Ranalli}\ \emph {et~al.}(2023)\citenamefont
  {Ranalli}, \citenamefont {Verdi}, \citenamefont {Monacelli}, \citenamefont
  {Kresse}, \citenamefont {Calandra},\ and\ \citenamefont {Franchini}}]{KTa03}%
  \BibitemOpen
  \bibfield  {author} {\bibinfo {author} {\bibfnamefont {L.}~\bibnamefont
  {Ranalli}}, \bibinfo {author} {\bibfnamefont {C.}~\bibnamefont {Verdi}},
  \bibinfo {author} {\bibfnamefont {L.}~\bibnamefont {Monacelli}}, \bibinfo
  {author} {\bibfnamefont {G.}~\bibnamefont {Kresse}}, \bibinfo {author}
  {\bibfnamefont {M.}~\bibnamefont {Calandra}},\ and\ \bibinfo {author}
  {\bibfnamefont {C.}~\bibnamefont {Franchini}},\ }\bibfield  {title} {\bibinfo
  {title} {{Temperature-Dependent Anharmonic Phonons in Quantum Paraelectric
  KTaO3 by First Principles and Machine-Learned Force Fields}},\ }\href
  {https://doi.org/https://doi.org/10.1002/qute.202200131} {\bibfield
  {journal} {\bibinfo  {journal} {Advanced Quantum Technologies}\ }\textbf
  {\bibinfo {volume} {6}},\ \bibinfo {pages} {2200131} (\bibinfo {year}
  {2023})}\BibitemShut {NoStop}%
\bibitem [{\citenamefont {Fransson}\ \emph {et~al.}(2023)\citenamefont
  {Fransson}, \citenamefont {Rosander}, \citenamefont {Eriksson}, \citenamefont
  {Rahm}, \citenamefont {Tadano},\ and\ \citenamefont
  {Erhart}}]{overdamped_theory}%
  \BibitemOpen
  \bibfield  {author} {\bibinfo {author} {\bibfnamefont {E.}~\bibnamefont
  {Fransson}}, \bibinfo {author} {\bibfnamefont {P.}~\bibnamefont {Rosander}},
  \bibinfo {author} {\bibfnamefont {F.}~\bibnamefont {Eriksson}}, \bibinfo
  {author} {\bibfnamefont {J.~M.}\ \bibnamefont {Rahm}}, \bibinfo {author}
  {\bibfnamefont {T.}~\bibnamefont {Tadano}},\ and\ \bibinfo {author}
  {\bibfnamefont {P.}~\bibnamefont {Erhart}},\ }\bibfield  {title} {\bibinfo
  {title} {Limits of the phonon quasi-particle picture at the
  cubic-to-tetragonal phase transition in halide perovskites},\ }\href
  {https://doi.org/10.1038/s42005-023-01297-8} {\bibfield  {journal} {\bibinfo
  {journal} {Communications Physics}\ }\textbf {\bibinfo {volume} {6}},\
  \bibinfo {pages} {173} (\bibinfo {year} {2023})}\BibitemShut {NoStop}%
\bibitem [{\citenamefont {Li}\ \emph {et~al.}(2014{\natexlab{b}})\citenamefont
  {Li}, \citenamefont {Hellman}, \citenamefont {Ma}, \citenamefont {May},
  \citenamefont {Cao}, \citenamefont {Chen}, \citenamefont {Christianson},
  \citenamefont {Ehlers}, \citenamefont {Singh}, \citenamefont {Sales},\ and\
  \citenamefont {Delaire}}]{SnTe_PbTe}%
  \BibitemOpen
  \bibfield  {author} {\bibinfo {author} {\bibfnamefont {C.~W.}\ \bibnamefont
  {Li}}, \bibinfo {author} {\bibfnamefont {O.}~\bibnamefont {Hellman}},
  \bibinfo {author} {\bibfnamefont {J.}~\bibnamefont {Ma}}, \bibinfo {author}
  {\bibfnamefont {A.~F.}\ \bibnamefont {May}}, \bibinfo {author} {\bibfnamefont
  {H.~B.}\ \bibnamefont {Cao}}, \bibinfo {author} {\bibfnamefont
  {X.}~\bibnamefont {Chen}}, \bibinfo {author} {\bibfnamefont {A.~D.}\
  \bibnamefont {Christianson}}, \bibinfo {author} {\bibfnamefont
  {G.}~\bibnamefont {Ehlers}}, \bibinfo {author} {\bibfnamefont {D.~J.}\
  \bibnamefont {Singh}}, \bibinfo {author} {\bibfnamefont {B.~C.}\ \bibnamefont
  {Sales}},\ and\ \bibinfo {author} {\bibfnamefont {O.}~\bibnamefont
  {Delaire}},\ }\bibfield  {title} {\bibinfo {title} {{Phonon Self-Energy and
  Origin of Anomalous Neutron Scattering Spectra in SnTe and PbTe
  Thermoelectrics}},\ }\href {https://doi.org/10.1103/PhysRevLett.112.175501}
  {\bibfield  {journal} {\bibinfo  {journal} {Phys. Rev. Lett.}\ }\textbf
  {\bibinfo {volume} {112}},\ \bibinfo {pages} {175501} (\bibinfo {year}
  {2014}{\natexlab{b}})}\BibitemShut {NoStop}%
\bibitem [{\citenamefont {Lanigan-Atkins}\ \emph {et~al.}(2021)\citenamefont
  {Lanigan-Atkins}, \citenamefont {He}, \citenamefont {Krogstad}, \citenamefont
  {Pajerowski}, \citenamefont {Abernathy}, \citenamefont {Xu}, \citenamefont
  {Xu}, \citenamefont {Chung}, \citenamefont {Kanatzidis}, \citenamefont
  {Rosenkranz}, \citenamefont {Osborn},\ and\ \citenamefont
  {Delaire}}]{dampedCsPbBr3}%
  \BibitemOpen
  \bibfield  {author} {\bibinfo {author} {\bibfnamefont {T.}~\bibnamefont
  {Lanigan-Atkins}}, \bibinfo {author} {\bibfnamefont {X.}~\bibnamefont {He}},
  \bibinfo {author} {\bibfnamefont {M.~J.}\ \bibnamefont {Krogstad}}, \bibinfo
  {author} {\bibfnamefont {D.~M.}\ \bibnamefont {Pajerowski}}, \bibinfo
  {author} {\bibfnamefont {D.~L.}\ \bibnamefont {Abernathy}}, \bibinfo {author}
  {\bibfnamefont {G.~N. M.~N.}\ \bibnamefont {Xu}}, \bibinfo {author}
  {\bibfnamefont {Z.}~\bibnamefont {Xu}}, \bibinfo {author} {\bibfnamefont
  {D.-Y.}\ \bibnamefont {Chung}}, \bibinfo {author} {\bibfnamefont {M.~G.}\
  \bibnamefont {Kanatzidis}}, \bibinfo {author} {\bibfnamefont
  {S.}~\bibnamefont {Rosenkranz}}, \bibinfo {author} {\bibfnamefont
  {R.}~\bibnamefont {Osborn}},\ and\ \bibinfo {author} {\bibfnamefont
  {O.}~\bibnamefont {Delaire}},\ }\bibfield  {title} {\bibinfo {title}
  {{Two-dimensional overdamped fluctuations of the soft perovskite lattice in
  CsPbBr3}},\ }\href {https://doi.org/10.1038/s41563-021-00947-y} {\bibfield
  {journal} {\bibinfo  {journal} {Nature Materials}\ }\textbf {\bibinfo
  {volume} {20}},\ \bibinfo {pages} {977} (\bibinfo {year} {2021})}\BibitemShut
  {NoStop}%
\bibitem [{\citenamefont {Niedziela}\ \emph {et~al.}(2019)\citenamefont
  {Niedziela}, \citenamefont {Bansal}, \citenamefont {May}, \citenamefont
  {Ding}, \citenamefont {Lanigan-Atkins}, \citenamefont {Ehlers}, \citenamefont
  {Abernathy}, \citenamefont {Said},\ and\ \citenamefont
  {Delaire}}]{Niedziela2019}%
  \BibitemOpen
  \bibfield  {author} {\bibinfo {author} {\bibfnamefont {J.~L.}\ \bibnamefont
  {Niedziela}}, \bibinfo {author} {\bibfnamefont {D.}~\bibnamefont {Bansal}},
  \bibinfo {author} {\bibfnamefont {A.~F.}\ \bibnamefont {May}}, \bibinfo
  {author} {\bibfnamefont {J.}~\bibnamefont {Ding}}, \bibinfo {author}
  {\bibfnamefont {T.}~\bibnamefont {Lanigan-Atkins}}, \bibinfo {author}
  {\bibfnamefont {G.}~\bibnamefont {Ehlers}}, \bibinfo {author} {\bibfnamefont
  {D.~L.}\ \bibnamefont {Abernathy}}, \bibinfo {author} {\bibfnamefont
  {A.}~\bibnamefont {Said}},\ and\ \bibinfo {author} {\bibfnamefont
  {O.}~\bibnamefont {Delaire}},\ }\bibfield  {title} {\bibinfo {title}
  {{Selective breakdown of phonon quasiparticles across superionic transition
  in CuCrSe2}},\ }\href {https://doi.org/10.1038/s41567-018-0298-2} {\bibfield
  {journal} {\bibinfo  {journal} {Nature Physics}\ }\textbf {\bibinfo {volume}
  {15}},\ \bibinfo {pages} {73} (\bibinfo {year} {2019})}\BibitemShut {NoStop}%
\bibitem [{\citenamefont {Simoncelli}\ \emph {et~al.}(2019)\citenamefont
  {Simoncelli}, \citenamefont {Marzari},\ and\ \citenamefont
  {Mauri}}]{Simoncelli2019}%
  \BibitemOpen
  \bibfield  {author} {\bibinfo {author} {\bibfnamefont {M.}~\bibnamefont
  {Simoncelli}}, \bibinfo {author} {\bibfnamefont {N.}~\bibnamefont
  {Marzari}},\ and\ \bibinfo {author} {\bibfnamefont {F.}~\bibnamefont
  {Mauri}},\ }\bibfield  {title} {\bibinfo {title} {{Unified theory of thermal
  transport in crystals and glasses}},\ }\href
  {https://doi.org/10.1038/s41567-019-0520-x} {\bibfield  {journal} {\bibinfo
  {journal} {Nature Physics}\ }\textbf {\bibinfo {volume} {15}},\ \bibinfo
  {pages} {809} (\bibinfo {year} {2019})}\BibitemShut {NoStop}%
\bibitem [{\citenamefont {Simoncelli}\ \emph {et~al.}(2022)\citenamefont
  {Simoncelli}, \citenamefont {Marzari},\ and\ \citenamefont
  {Mauri}}]{Simoncelli2022}%
  \BibitemOpen
  \bibfield  {author} {\bibinfo {author} {\bibfnamefont {M.}~\bibnamefont
  {Simoncelli}}, \bibinfo {author} {\bibfnamefont {N.}~\bibnamefont
  {Marzari}},\ and\ \bibinfo {author} {\bibfnamefont {F.}~\bibnamefont
  {Mauri}},\ }\bibfield  {title} {\bibinfo {title} {{Wigner Formulation of
  Thermal Transport in Solids}},\ }\href
  {https://doi.org/10.1103/PhysRevX.12.041011} {\bibfield  {journal} {\bibinfo
  {journal} {Phys. Rev. X}\ }\textbf {\bibinfo {volume} {12}},\ \bibinfo
  {pages} {041011} (\bibinfo {year} {2022})}\BibitemShut {NoStop}%
\bibitem [{\citenamefont {Simoncelli}\ \emph {et~al.}(2023)\citenamefont
  {Simoncelli}, \citenamefont {Mauri},\ and\ \citenamefont
  {Marzari}}]{Simoncelli_glasses}%
  \BibitemOpen
  \bibfield  {author} {\bibinfo {author} {\bibfnamefont {M.}~\bibnamefont
  {Simoncelli}}, \bibinfo {author} {\bibfnamefont {F.}~\bibnamefont {Mauri}},\
  and\ \bibinfo {author} {\bibfnamefont {N.}~\bibnamefont {Marzari}},\
  }\bibfield  {title} {\bibinfo {title} {Thermal conductivity of glasses:
  first-principles theory and applications},\ }\href
  {https://doi.org/10.1038/s41524-023-01033-4} {\bibfield  {journal} {\bibinfo
  {journal} {npj Computational Materials}\ }\textbf {\bibinfo {volume} {9}},\
  \bibinfo {pages} {106} (\bibinfo {year} {2023})}\BibitemShut {NoStop}%
\bibitem [{\citenamefont {Isaeva}\ \emph {et~al.}(2019)\citenamefont {Isaeva},
  \citenamefont {Barbalinardo}, \citenamefont {Donadio},\ and\ \citenamefont
  {Baroni}}]{Isaeva2019}%
  \BibitemOpen
  \bibfield  {author} {\bibinfo {author} {\bibfnamefont {L.}~\bibnamefont
  {Isaeva}}, \bibinfo {author} {\bibfnamefont {G.}~\bibnamefont
  {Barbalinardo}}, \bibinfo {author} {\bibfnamefont {D.}~\bibnamefont
  {Donadio}},\ and\ \bibinfo {author} {\bibfnamefont {S.}~\bibnamefont
  {Baroni}},\ }\bibfield  {title} {\bibinfo {title} {Modeling heat transport in
  crystals and glasses from a unified lattice-dynamical approach},\ }\href
  {https://doi.org/10.1038/s41467-019-11572-4} {\bibfield  {journal} {\bibinfo
  {journal} {Nature Communications}\ }\textbf {\bibinfo {volume} {10}},\
  \bibinfo {pages} {3853} (\bibinfo {year} {2019})}\BibitemShut {NoStop}%
\bibitem [{\citenamefont {Fiorentino}\ and\ \citenamefont
  {Baroni}(2023)}]{Fiorentino}%
  \BibitemOpen
  \bibfield  {author} {\bibinfo {author} {\bibfnamefont {A.}~\bibnamefont
  {Fiorentino}}\ and\ \bibinfo {author} {\bibfnamefont {S.}~\bibnamefont
  {Baroni}},\ }\bibfield  {title} {\bibinfo {title} {{From Green-Kubo to the
  full Boltzmann kinetic approach to heat transport in crystals and glasses}},\
  }\href {https://doi.org/10.1103/PhysRevB.107.054311} {\bibfield  {journal}
  {\bibinfo  {journal} {Phys. Rev. B}\ }\textbf {\bibinfo {volume} {107}},\
  \bibinfo {pages} {054311} (\bibinfo {year} {2023})}\BibitemShut {NoStop}%
\bibitem [{\citenamefont {Fiorentino}\ \emph {et~al.}(2023)\citenamefont
  {Fiorentino}, \citenamefont {Pegolo},\ and\ \citenamefont
  {Baroni}}]{fiorentino2023hydrodynamic}%
  \BibitemOpen
  \bibfield  {author} {\bibinfo {author} {\bibfnamefont {A.}~\bibnamefont
  {Fiorentino}}, \bibinfo {author} {\bibfnamefont {P.}~\bibnamefont {Pegolo}},\
  and\ \bibinfo {author} {\bibfnamefont {S.}~\bibnamefont {Baroni}},\
  }\href@noop {} {\bibinfo {title} {Hydrodynamic finite-size scaling of the
  thermal conductivity in glasses}} (\bibinfo {year} {2023})\BibitemShut
  {NoStop}%
\bibitem [{\citenamefont {Monacelli}\ and\ \citenamefont
  {Mauri}(2021)}]{TDSSCHA}%
  \BibitemOpen
  \bibfield  {author} {\bibinfo {author} {\bibfnamefont {L.}~\bibnamefont
  {Monacelli}}\ and\ \bibinfo {author} {\bibfnamefont {F.}~\bibnamefont
  {Mauri}},\ }\bibfield  {title} {\bibinfo {title} {{Time-dependent
  self-consistent harmonic approximation: Anharmonic nuclear quantum dynamics
  and time correlation functions}},\ }\href
  {https://doi.org/10.1103/PhysRevB.103.104305} {\bibfield  {journal} {\bibinfo
   {journal} {Phys. Rev. B}\ }\textbf {\bibinfo {volume} {103}},\ \bibinfo
  {pages} {104305} (\bibinfo {year} {2021})}\BibitemShut {NoStop}%
\bibitem [{\citenamefont {Lihm}\ and\ \citenamefont
  {Park}(2021)}]{PhysRevResearch.3.L032017}%
  \BibitemOpen
  \bibfield  {author} {\bibinfo {author} {\bibfnamefont {J.-M.}\ \bibnamefont
  {Lihm}}\ and\ \bibinfo {author} {\bibfnamefont {C.-H.}\ \bibnamefont
  {Park}},\ }\bibfield  {title} {\bibinfo {title} {Gaussian time-dependent
  variational principle for the finite-temperature anharmonic lattice
  dynamics},\ }\href {https://doi.org/10.1103/PhysRevResearch.3.L032017}
  {\bibfield  {journal} {\bibinfo  {journal} {Phys. Rev. Res.}\ }\textbf
  {\bibinfo {volume} {3}},\ \bibinfo {pages} {L032017} (\bibinfo {year}
  {2021})}\BibitemShut {NoStop}%
\bibitem [{\citenamefont {Han}\ \emph {et~al.}(2022)\citenamefont {Han},
  \citenamefont {Yang}, \citenamefont {Li}, \citenamefont {Feng},\ and\
  \citenamefont {Ruan}}]{fourth1}%
  \BibitemOpen
  \bibfield  {author} {\bibinfo {author} {\bibfnamefont {Z.}~\bibnamefont
  {Han}}, \bibinfo {author} {\bibfnamefont {X.}~\bibnamefont {Yang}}, \bibinfo
  {author} {\bibfnamefont {W.}~\bibnamefont {Li}}, \bibinfo {author}
  {\bibfnamefont {T.}~\bibnamefont {Feng}},\ and\ \bibinfo {author}
  {\bibfnamefont {X.}~\bibnamefont {Ruan}},\ }\bibfield  {title} {\bibinfo
  {title} {{FourPhonon: An extension module to ShengBTE for computing
  four-phonon scattering rates and thermal conductivity}},\ }\href
  {https://doi.org/https://doi.org/10.1016/j.cpc.2021.108179} {\bibfield
  {journal} {\bibinfo  {journal} {Computer Physics Communications}\ }\textbf
  {\bibinfo {volume} {270}},\ \bibinfo {pages} {108179} (\bibinfo {year}
  {2022})}\BibitemShut {NoStop}%
\bibitem [{\citenamefont {Feng}\ \emph {et~al.}(2017)\citenamefont {Feng},
  \citenamefont {Lindsay},\ and\ \citenamefont {Ruan}}]{fourth2}%
  \BibitemOpen
  \bibfield  {author} {\bibinfo {author} {\bibfnamefont {T.}~\bibnamefont
  {Feng}}, \bibinfo {author} {\bibfnamefont {L.}~\bibnamefont {Lindsay}},\ and\
  \bibinfo {author} {\bibfnamefont {X.}~\bibnamefont {Ruan}},\ }\bibfield
  {title} {\bibinfo {title} {Four-phonon scattering significantly reduces
  intrinsic thermal conductivity of solids},\ }\href
  {https://doi.org/10.1103/PhysRevB.96.161201} {\bibfield  {journal} {\bibinfo
  {journal} {Phys. Rev. B}\ }\textbf {\bibinfo {volume} {96}},\ \bibinfo
  {pages} {161201} (\bibinfo {year} {2017})}\BibitemShut {NoStop}%
\bibitem [{\citenamefont {Siciliano}\ \emph {et~al.}(2023)\citenamefont
  {Siciliano}, \citenamefont {Monacelli}, \citenamefont {Caldarelli},\ and\
  \citenamefont {Mauri}}]{Siciliano}%
  \BibitemOpen
  \bibfield  {author} {\bibinfo {author} {\bibfnamefont {A.}~\bibnamefont
  {Siciliano}}, \bibinfo {author} {\bibfnamefont {L.}~\bibnamefont
  {Monacelli}}, \bibinfo {author} {\bibfnamefont {G.}~\bibnamefont
  {Caldarelli}},\ and\ \bibinfo {author} {\bibfnamefont {F.}~\bibnamefont
  {Mauri}},\ }\bibfield  {title} {\bibinfo {title} {{Wigner Gaussian dynamics:
  Simulating the anharmonic and quantum ionic motion}},\ }\href
  {https://doi.org/10.1103/PhysRevB.107.174307} {\bibfield  {journal} {\bibinfo
   {journal} {Phys. Rev. B}\ }\textbf {\bibinfo {volume} {107}},\ \bibinfo
  {pages} {174307} (\bibinfo {year} {2023})}\BibitemShut {NoStop}%
\bibitem [{\citenamefont {Tripathi}\ and\ \citenamefont
  {Pathak}(1974)}]{Tripathi1974}%
  \BibitemOpen
  \bibfield  {author} {\bibinfo {author} {\bibfnamefont {R.~S.}\ \bibnamefont
  {Tripathi}}\ and\ \bibinfo {author} {\bibfnamefont {K.~N.}\ \bibnamefont
  {Pathak}},\ }\bibfield  {title} {\bibinfo {title} {{Self-energy of phonons in
  an anharmonic crystal to O($\delta$4)}},\ }\href
  {https://doi.org/10.1007/BF02737485} {\bibfield  {journal} {\bibinfo
  {journal} {Il Nuovo Cimento B (1971-1996)}\ }\textbf {\bibinfo {volume}
  {21}},\ \bibinfo {pages} {289} (\bibinfo {year} {1974})}\BibitemShut
  {NoStop}%
\bibitem [{\citenamefont {Monga}\ and\ \citenamefont {Pathak}(1978)}]{Monga1}%
  \BibitemOpen
  \bibfield  {author} {\bibinfo {author} {\bibfnamefont {M.~R.}\ \bibnamefont
  {Monga}}\ and\ \bibinfo {author} {\bibfnamefont {K.~N.}\ \bibnamefont
  {Pathak}},\ }\bibfield  {title} {\bibinfo {title} {{Self-energy of phonons in
  an anharmonic crystal of order ${\ensuremath{\lambda}}^{4}$. II. An
  application to a monatomic linear chain}},\ }\href
  {https://doi.org/10.1103/PhysRevB.18.5859} {\bibfield  {journal} {\bibinfo
  {journal} {Phys. Rev. B}\ }\textbf {\bibinfo {volume} {18}},\ \bibinfo
  {pages} {5859} (\bibinfo {year} {1978})}\BibitemShut {NoStop}%
\bibitem [{\citenamefont {Monga}\ \emph {et~al.}(1979)\citenamefont {Monga},
  \citenamefont {Jindal},\ and\ \citenamefont {Pathak}}]{Monga2}%
  \BibitemOpen
  \bibfield  {author} {\bibinfo {author} {\bibfnamefont {M.~R.}\ \bibnamefont
  {Monga}}, \bibinfo {author} {\bibfnamefont {V.~K.}\ \bibnamefont {Jindal}},\
  and\ \bibinfo {author} {\bibfnamefont {K.~N.}\ \bibnamefont {Pathak}},\
  }\bibfield  {title} {\bibinfo {title} {{Self-energy of phonons in an
  anharmonic crystal of order ${\ensuremath{\lambda}}^{4}$. III. Approximate
  numerical results for ionic crystals}},\ }\href
  {https://doi.org/10.1103/PhysRevB.19.1230} {\bibfield  {journal} {\bibinfo
  {journal} {Phys. Rev. B}\ }\textbf {\bibinfo {volume} {19}},\ \bibinfo
  {pages} {1230} (\bibinfo {year} {1979})}\BibitemShut {NoStop}%
\bibitem [{\citenamefont {Green}(2004)}]{Green}%
  \BibitemOpen
  \bibfield  {author} {\bibinfo {author} {\bibfnamefont {M.~S.}\ \bibnamefont
  {Green}},\ }\bibfield  {title} {\bibinfo {title} {{Markoff Random Processes
  and the Statistical Mechanics of Time‐Dependent Phenomena. II. Irreversible
  Processes in Fluids}},\ }\href {https://doi.org/10.1063/1.1740082} {\bibfield
   {journal} {\bibinfo  {journal} {The Journal of Chemical Physics}\ }\textbf
  {\bibinfo {volume} {22}},\ \bibinfo {pages} {398} (\bibinfo {year}
  {2004})}\BibitemShut {NoStop}%
\bibitem [{\citenamefont {Kubo}(1957)}]{Kubo}%
  \BibitemOpen
  \bibfield  {author} {\bibinfo {author} {\bibfnamefont {R.}~\bibnamefont
  {Kubo}},\ }\bibfield  {title} {\bibinfo {title} {{Statistical-Mechanical
  Theory of Irreversible Processes. I. General Theory and Simple Applications
  to Magnetic and Conduction Problems}},\ }\href
  {https://doi.org/10.1143/JPSJ.12.570} {\bibfield  {journal} {\bibinfo
  {journal} {Journal of the Physical Society of Japan}\ }\textbf {\bibinfo
  {volume} {12}},\ \bibinfo {pages} {570} (\bibinfo {year} {1957})}\BibitemShut
  {NoStop}%
\bibitem [{\citenamefont {Yang}\ \emph {et~al.}(2010)\citenamefont {Yang},
  \citenamefont {Cahill}, \citenamefont {Liu}, \citenamefont {Feldman},
  \citenamefont {Crandall}, \citenamefont {Sperling},\ and\ \citenamefont
  {Abelson}}]{Thermoreflectance1}%
  \BibitemOpen
  \bibfield  {author} {\bibinfo {author} {\bibfnamefont {H.-S.}\ \bibnamefont
  {Yang}}, \bibinfo {author} {\bibfnamefont {D.~G.}\ \bibnamefont {Cahill}},
  \bibinfo {author} {\bibfnamefont {X.}~\bibnamefont {Liu}}, \bibinfo {author}
  {\bibfnamefont {J.~L.}\ \bibnamefont {Feldman}}, \bibinfo {author}
  {\bibfnamefont {R.~S.}\ \bibnamefont {Crandall}}, \bibinfo {author}
  {\bibfnamefont {B.~A.}\ \bibnamefont {Sperling}},\ and\ \bibinfo {author}
  {\bibfnamefont {J.~R.}\ \bibnamefont {Abelson}},\ }\bibfield  {title}
  {\bibinfo {title} {Anomalously high thermal conductivity of amorphous si
  deposited by hot-wire chemical vapor deposition},\ }\href
  {https://doi.org/10.1103/PhysRevB.81.104203} {\bibfield  {journal} {\bibinfo
  {journal} {Phys. Rev. B}\ }\textbf {\bibinfo {volume} {81}},\ \bibinfo
  {pages} {104203} (\bibinfo {year} {2010})}\BibitemShut {NoStop}%
\bibitem [{\citenamefont {Koh}\ and\ \citenamefont
  {Cahill}(2007)}]{Thermoreflectance2}%
  \BibitemOpen
  \bibfield  {author} {\bibinfo {author} {\bibfnamefont {Y.~K.}\ \bibnamefont
  {Koh}}\ and\ \bibinfo {author} {\bibfnamefont {D.~G.}\ \bibnamefont
  {Cahill}},\ }\bibfield  {title} {\bibinfo {title} {Frequency dependence of
  the thermal conductivity of semiconductor alloys},\ }\href
  {https://doi.org/10.1103/PhysRevB.76.075207} {\bibfield  {journal} {\bibinfo
  {journal} {Phys. Rev. B}\ }\textbf {\bibinfo {volume} {76}},\ \bibinfo
  {pages} {075207} (\bibinfo {year} {2007})}\BibitemShut {NoStop}%
\bibitem [{\citenamefont {da~Cruz}\ \emph {et~al.}(2012)\citenamefont
  {da~Cruz}, \citenamefont {Li}, \citenamefont {Katcho},\ and\ \citenamefont
  {Mingo}}]{Thermoreflectance3}%
  \BibitemOpen
  \bibfield  {author} {\bibinfo {author} {\bibfnamefont {C.~A.}\ \bibnamefont
  {da~Cruz}}, \bibinfo {author} {\bibfnamefont {W.}~\bibnamefont {Li}},
  \bibinfo {author} {\bibfnamefont {N.~A.}\ \bibnamefont {Katcho}},\ and\
  \bibinfo {author} {\bibfnamefont {N.}~\bibnamefont {Mingo}},\ }\bibfield
  {title} {\bibinfo {title} {{Role of phonon anharmonicity in time-domain
  thermoreflectance measurements}},\ }\href {https://doi.org/10.1063/1.4746275}
  {\bibfield  {journal} {\bibinfo  {journal} {Applied Physics Letters}\
  }\textbf {\bibinfo {volume} {101}},\ \bibinfo {pages} {083108} (\bibinfo
  {year} {2012})}\BibitemShut {NoStop}%
\bibitem [{\citenamefont {Hardy}(1963)}]{Hardy}%
  \BibitemOpen
  \bibfield  {author} {\bibinfo {author} {\bibfnamefont {R.~J.}\ \bibnamefont
  {Hardy}},\ }\bibfield  {title} {\bibinfo {title} {{Energy-Flux Operator for a
  Lattice}},\ }\href {https://doi.org/10.1103/PhysRev.132.168} {\bibfield
  {journal} {\bibinfo  {journal} {Phys. Rev.}\ }\textbf {\bibinfo {volume}
  {132}},\ \bibinfo {pages} {168} (\bibinfo {year} {1963})}\BibitemShut
  {NoStop}%
\bibitem [{\citenamefont {Deo}\ and\ \citenamefont {Behera}(1966)}]{DepBehera}%
  \BibitemOpen
  \bibfield  {author} {\bibinfo {author} {\bibfnamefont {B.}~\bibnamefont
  {Deo}}\ and\ \bibinfo {author} {\bibfnamefont {S.~N.}\ \bibnamefont
  {Behera}},\ }\bibfield  {title} {\bibinfo {title} {{Calculation of Thermal
  Conductivity by the Kubo Formula}},\ }\href
  {https://doi.org/10.1103/PhysRev.141.738} {\bibfield  {journal} {\bibinfo
  {journal} {Phys. Rev.}\ }\textbf {\bibinfo {volume} {141}},\ \bibinfo {pages}
  {738} (\bibinfo {year} {1966})}\BibitemShut {NoStop}%
\bibitem [{\citenamefont {Zwanzig}(1965)}]{Zwanzig}%
  \BibitemOpen
  \bibfield  {author} {\bibinfo {author} {\bibfnamefont {R.}~\bibnamefont
  {Zwanzig}},\ }\bibfield  {title} {\bibinfo {title} {{Time-Correlation
  Functions and Transport Coefficients in Statistical Mechanics}},\ }\href
  {https://doi.org/10.1146/annurev.pc.16.100165.000435} {\bibfield  {journal}
  {\bibinfo  {journal} {Annual Review of Physical Chemistry}\ }\textbf
  {\bibinfo {volume} {16}},\ \bibinfo {pages} {67} (\bibinfo {year}
  {1965})}\BibitemShut {NoStop}%
\bibitem [{\citenamefont {Zubarev}(1960)}]{Zubarev}%
  \BibitemOpen
  \bibfield  {author} {\bibinfo {author} {\bibfnamefont {D.~N.}\ \bibnamefont
  {Zubarev}},\ }\bibfield  {title} {\bibinfo {title} {{DOUBLE-TIME GREEN
  FUNCTIONS IN STATISTICAL PHYSICS}},\ }\href
  {https://doi.org/10.1070/PU1960v003n03ABEH003275} {\bibfield  {journal}
  {\bibinfo  {journal} {Soviet Physics Uspekhi}\ }\textbf {\bibinfo {volume}
  {3}},\ \bibinfo {pages} {320} (\bibinfo {year} {1960})}\BibitemShut {NoStop}%
\bibitem [{\citenamefont {Sun}\ and\ \citenamefont {Allen}(2010)}]{PhilAllen}%
  \BibitemOpen
  \bibfield  {author} {\bibinfo {author} {\bibfnamefont {T.}~\bibnamefont
  {Sun}}\ and\ \bibinfo {author} {\bibfnamefont {P.~B.}\ \bibnamefont
  {Allen}},\ }\bibfield  {title} {\bibinfo {title} {Lattice thermal
  conductivity: Computations and theory of the high-temperature breakdown of
  the phonon-gas model},\ }\href {https://doi.org/10.1103/PhysRevB.82.224305}
  {\bibfield  {journal} {\bibinfo  {journal} {Phys. Rev. B}\ }\textbf {\bibinfo
  {volume} {82}},\ \bibinfo {pages} {224305} (\bibinfo {year}
  {2010})}\BibitemShut {NoStop}%
\bibitem [{sup()}]{supp_mat}%
  \BibitemOpen
  \href@noop {} {}\bibinfo {note} {Please see Supplementary Material at XXX for
  more details on the fitting procedure of {GAP ML} potential, dynamical
  thermal conductivity, etc. The {Supplementary Material} includes citations to
  Refs.~\cite{QE1, QE2, QE3, GARRITY2014446, PBESOL, ASE,
  phonopy-phono3py-JPSJ, adaptive_smearing, myMD}}\BibitemShut {NoStop}%
\bibitem [{\citenamefont {Caldarelli}\ \emph {et~al.}(2022)\citenamefont
  {Caldarelli}, \citenamefont {Simoncelli}, \citenamefont {Marzari},
  \citenamefont {Mauri},\ and\ \citenamefont {Benfatto}}]{Caldarelli}%
  \BibitemOpen
  \bibfield  {author} {\bibinfo {author} {\bibfnamefont {G.}~\bibnamefont
  {Caldarelli}}, \bibinfo {author} {\bibfnamefont {M.}~\bibnamefont
  {Simoncelli}}, \bibinfo {author} {\bibfnamefont {N.}~\bibnamefont {Marzari}},
  \bibinfo {author} {\bibfnamefont {F.}~\bibnamefont {Mauri}},\ and\ \bibinfo
  {author} {\bibfnamefont {L.}~\bibnamefont {Benfatto}},\ }\bibfield  {title}
  {\bibinfo {title} {{Many-body Green's function approach to lattice thermal
  transport}},\ }\href {https://doi.org/10.1103/PhysRevB.106.024312} {\bibfield
   {journal} {\bibinfo  {journal} {Phys. Rev. B}\ }\textbf {\bibinfo {volume}
  {106}},\ \bibinfo {pages} {024312} (\bibinfo {year} {2022})}\BibitemShut
  {NoStop}%
\bibitem [{\citenamefont {He}\ \emph {et~al.}(2012)\citenamefont {He},
  \citenamefont {Savić}, \citenamefont {Donadio},\ and\ \citenamefont
  {Galli}}]{IvanaMD}%
  \BibitemOpen
  \bibfield  {author} {\bibinfo {author} {\bibfnamefont {Y.}~\bibnamefont
  {He}}, \bibinfo {author} {\bibfnamefont {I.}~\bibnamefont {Savić}}, \bibinfo
  {author} {\bibfnamefont {D.}~\bibnamefont {Donadio}},\ and\ \bibinfo {author}
  {\bibfnamefont {G.}~\bibnamefont {Galli}},\ }\bibfield  {title} {\bibinfo
  {title} {{Lattice thermal conductivity of semiconducting bulk materials:
  atomistic simulations}},\ }\href {https://doi.org/10.1039/C2CP42394D}
  {\bibfield  {journal} {\bibinfo  {journal} {Phys. Chem. Chem. Phys.}\
  }\textbf {\bibinfo {volume} {14}},\ \bibinfo {pages} {16209} (\bibinfo {year}
  {2012})}\BibitemShut {NoStop}%
\bibitem [{\citenamefont {Marcolongo}\ \emph {et~al.}(2016)\citenamefont
  {Marcolongo}, \citenamefont {Umari},\ and\ \citenamefont
  {Baroni}}]{BaroniMD}%
  \BibitemOpen
  \bibfield  {author} {\bibinfo {author} {\bibfnamefont {A.}~\bibnamefont
  {Marcolongo}}, \bibinfo {author} {\bibfnamefont {P.}~\bibnamefont {Umari}},\
  and\ \bibinfo {author} {\bibfnamefont {S.}~\bibnamefont {Baroni}},\
  }\bibfield  {title} {\bibinfo {title} {Microscopic theory and quantum
  simulation of atomic heat transport},\ }\href
  {https://doi.org/10.1038/nphys3509} {\bibfield  {journal} {\bibinfo
  {journal} {Nature Physics}\ }\textbf {\bibinfo {volume} {12}},\ \bibinfo
  {pages} {80} (\bibinfo {year} {2016})}\BibitemShut {NoStop}%
\bibitem [{\citenamefont {Carbogno}\ \emph {et~al.}(2017)\citenamefont
  {Carbogno}, \citenamefont {Ramprasad},\ and\ \citenamefont
  {Scheffler}}]{CarbognoMD}%
  \BibitemOpen
  \bibfield  {author} {\bibinfo {author} {\bibfnamefont {C.}~\bibnamefont
  {Carbogno}}, \bibinfo {author} {\bibfnamefont {R.}~\bibnamefont
  {Ramprasad}},\ and\ \bibinfo {author} {\bibfnamefont {M.}~\bibnamefont
  {Scheffler}},\ }\bibfield  {title} {\bibinfo {title} {{Ab Initio Green-Kubo
  Approach for the Thermal Conductivity of Solids}},\ }\href
  {https://doi.org/10.1103/PhysRevLett.118.175901} {\bibfield  {journal}
  {\bibinfo  {journal} {Phys. Rev. Lett.}\ }\textbf {\bibinfo {volume} {118}},\
  \bibinfo {pages} {175901} (\bibinfo {year} {2017})}\BibitemShut {NoStop}%
\bibitem [{\citenamefont {Srivastava}(1990)}]{srivastava}%
  \BibitemOpen
  \bibfield  {author} {\bibinfo {author} {\bibfnamefont {G.}~\bibnamefont
  {Srivastava}},\ }\href {https://books.google.es/books?id=OE-bHd2gzVgC} {\emph
  {\bibinfo {title} {{The Physics of Phonons}}}}\ (\bibinfo  {publisher}
  {Taylor \& Francis},\ \bibinfo {year} {1990})\BibitemShut {NoStop}%
\bibitem [{\citenamefont {Semwal}\ and\ \citenamefont {Sharma}(1972)}]{Semwal}%
  \BibitemOpen
  \bibfield  {author} {\bibinfo {author} {\bibfnamefont {B.~S.}\ \bibnamefont
  {Semwal}}\ and\ \bibinfo {author} {\bibfnamefont {P.~K.}\ \bibnamefont
  {Sharma}},\ }\bibfield  {title} {\bibinfo {title} {{Thermal Conductivity of
  an Anharmonic Crystal}},\ }\href {https://doi.org/10.1103/PhysRevB.5.3909}
  {\bibfield  {journal} {\bibinfo  {journal} {Phys. Rev. B}\ }\textbf {\bibinfo
  {volume} {5}},\ \bibinfo {pages} {3909} (\bibinfo {year} {1972})}\BibitemShut
  {NoStop}%
\bibitem [{ssc()}]{sschacode}%
  \BibitemOpen
  \href@noop {} {\bibinfo {title} {Stochastic self-consistent harmonic
  approximation}},\ \bibinfo {howpublished} {\url{http://sscha.eu}},\ \bibinfo
  {note} {accessed: 2025-02-12}\BibitemShut {NoStop}%
\bibitem [{\citenamefont {Rodov{\'a}}\ \emph {et~al.}(2003)\citenamefont
  {Rodov{\'a}}, \citenamefont {Bro{\v{z}}ek}, \citenamefont
  {Kn{\'i}{\v{z}}ek},\ and\ \citenamefont {Nitsch}}]{CsPbBr3_TE1}%
  \BibitemOpen
  \bibfield  {author} {\bibinfo {author} {\bibfnamefont {M.}~\bibnamefont
  {Rodov{\'a}}}, \bibinfo {author} {\bibfnamefont {J.}~\bibnamefont
  {Bro{\v{z}}ek}}, \bibinfo {author} {\bibfnamefont {K.}~\bibnamefont
  {Kn{\'i}{\v{z}}ek}},\ and\ \bibinfo {author} {\bibfnamefont {K.}~\bibnamefont
  {Nitsch}},\ }\bibfield  {title} {\bibinfo {title} {Phase transitions in
  ternary caesium lead bromide},\ }\href
  {https://doi.org/10.1023/A:1022836800820} {\bibfield  {journal} {\bibinfo
  {journal} {Journal of Thermal Analysis and Calorimetry}\ }\textbf {\bibinfo
  {volume} {71}},\ \bibinfo {pages} {667} (\bibinfo {year} {2003})}\BibitemShut
  {NoStop}%
\bibitem [{\citenamefont {Stoumpos}\ \emph {et~al.}(2013)\citenamefont
  {Stoumpos}, \citenamefont {Malliakas}, \citenamefont {Peters}, \citenamefont
  {Liu}, \citenamefont {Sebastian}, \citenamefont {Im}, \citenamefont
  {Chasapis}, \citenamefont {Wibowo}, \citenamefont {Chung}, \citenamefont
  {Freeman}, \citenamefont {Wessels},\ and\ \citenamefont
  {Kanatzidis}}]{CsPbBr3_TE2}%
  \BibitemOpen
  \bibfield  {author} {\bibinfo {author} {\bibfnamefont {C.~C.}\ \bibnamefont
  {Stoumpos}}, \bibinfo {author} {\bibfnamefont {C.~D.}\ \bibnamefont
  {Malliakas}}, \bibinfo {author} {\bibfnamefont {J.~A.}\ \bibnamefont
  {Peters}}, \bibinfo {author} {\bibfnamefont {Z.}~\bibnamefont {Liu}},
  \bibinfo {author} {\bibfnamefont {M.}~\bibnamefont {Sebastian}}, \bibinfo
  {author} {\bibfnamefont {J.}~\bibnamefont {Im}}, \bibinfo {author}
  {\bibfnamefont {T.~C.}\ \bibnamefont {Chasapis}}, \bibinfo {author}
  {\bibfnamefont {A.~C.}\ \bibnamefont {Wibowo}}, \bibinfo {author}
  {\bibfnamefont {D.~Y.}\ \bibnamefont {Chung}}, \bibinfo {author}
  {\bibfnamefont {A.~J.}\ \bibnamefont {Freeman}}, \bibinfo {author}
  {\bibfnamefont {B.~W.}\ \bibnamefont {Wessels}},\ and\ \bibinfo {author}
  {\bibfnamefont {M.~G.}\ \bibnamefont {Kanatzidis}},\ }\bibfield  {title}
  {\bibinfo {title} {{Crystal Growth of the Perovskite Semiconductor CsPbBr3: A
  New Material for High-Energy Radiation Detection}},\ }\href
  {https://doi.org/10.1021/cg400645t} {\bibfield  {journal} {\bibinfo
  {journal} {Crystal Growth \& Design}\ }\textbf {\bibinfo {volume} {13}},\
  \bibinfo {pages} {2722} (\bibinfo {year} {2013})}\BibitemShut {NoStop}%
\bibitem [{\citenamefont {L{\'o}pez}\ \emph {et~al.}(2020)\citenamefont
  {L{\'o}pez}, \citenamefont {Abia}, \citenamefont {Alvarez-Galv{\'a}n},
  \citenamefont {Hong}, \citenamefont {Mart{\'i}nez-Huerta}, \citenamefont
  {Serrano-S{\'a}nchez}, \citenamefont {Carrascoso}, \citenamefont
  {Castellanos-G{\'o}mez}, \citenamefont {Fernandez-Diaz},\ and\ \citenamefont
  {Alonso}}]{CsPbBr3_TE3}%
  \BibitemOpen
  \bibfield  {author} {\bibinfo {author} {\bibfnamefont {C.~A.}\ \bibnamefont
  {L{\'o}pez}}, \bibinfo {author} {\bibfnamefont {C.}~\bibnamefont {Abia}},
  \bibinfo {author} {\bibfnamefont {M.~C.}\ \bibnamefont {Alvarez-Galv{\'a}n}},
  \bibinfo {author} {\bibfnamefont {B.-K.}\ \bibnamefont {Hong}}, \bibinfo
  {author} {\bibfnamefont {M.~V.}\ \bibnamefont {Mart{\'i}nez-Huerta}},
  \bibinfo {author} {\bibfnamefont {F.}~\bibnamefont {Serrano-S{\'a}nchez}},
  \bibinfo {author} {\bibfnamefont {F.}~\bibnamefont {Carrascoso}}, \bibinfo
  {author} {\bibfnamefont {A.}~\bibnamefont {Castellanos-G{\'o}mez}}, \bibinfo
  {author} {\bibfnamefont {M.~T.}\ \bibnamefont {Fernandez-Diaz}},\ and\
  \bibinfo {author} {\bibfnamefont {J.~A.}\ \bibnamefont {Alonso}},\ }\bibfield
   {title} {\bibinfo {title} {{Crystal Structure Features of CsPbBr3 Perovskite
  Prepared by Mechanochemical Synthesis}},\ }\href
  {https://doi.org/10.1021/acsomega.9b04248} {\bibfield  {journal} {\bibinfo
  {journal} {ACS Omega}\ }\textbf {\bibinfo {volume} {5}},\ \bibinfo {pages}
  {5931} (\bibinfo {year} {2020})},\ \bibinfo {note} {pMID:
  32226873}\BibitemShut {NoStop}%
\bibitem [{\citenamefont {Lee}\ \emph {et~al.}(2017)\citenamefont {Lee},
  \citenamefont {Li}, \citenamefont {Wong}, \citenamefont {Zhang},
  \citenamefont {Lai}, \citenamefont {Yu}, \citenamefont {Kong}, \citenamefont
  {Lin}, \citenamefont {Urban}, \citenamefont {Grossman},\ and\ \citenamefont
  {Yang}}]{CsPbBr3_kappa1}%
  \BibitemOpen
  \bibfield  {author} {\bibinfo {author} {\bibfnamefont {W.}~\bibnamefont
  {Lee}}, \bibinfo {author} {\bibfnamefont {H.}~\bibnamefont {Li}}, \bibinfo
  {author} {\bibfnamefont {A.~B.}\ \bibnamefont {Wong}}, \bibinfo {author}
  {\bibfnamefont {D.}~\bibnamefont {Zhang}}, \bibinfo {author} {\bibfnamefont
  {M.}~\bibnamefont {Lai}}, \bibinfo {author} {\bibfnamefont {Y.}~\bibnamefont
  {Yu}}, \bibinfo {author} {\bibfnamefont {Q.}~\bibnamefont {Kong}}, \bibinfo
  {author} {\bibfnamefont {E.}~\bibnamefont {Lin}}, \bibinfo {author}
  {\bibfnamefont {J.~J.}\ \bibnamefont {Urban}}, \bibinfo {author}
  {\bibfnamefont {J.~C.}\ \bibnamefont {Grossman}},\ and\ \bibinfo {author}
  {\bibfnamefont {P.}~\bibnamefont {Yang}},\ }\bibfield  {title} {\bibinfo
  {title} {Ultralow thermal conductivity in all-inorganic halide perovskites},\
  }\href {https://doi.org/10.1073/pnas.1711744114} {\bibfield  {journal}
  {\bibinfo  {journal} {Proceedings of the National Academy of Sciences}\
  }\textbf {\bibinfo {volume} {114}},\ \bibinfo {pages} {8693} (\bibinfo {year}
  {2017})}\BibitemShut {NoStop}%
\bibitem [{\citenamefont {Bart\'ok}\ \emph {et~al.}(2010)\citenamefont
  {Bart\'ok}, \citenamefont {Payne}, \citenamefont {Kondor},\ and\
  \citenamefont {Cs\'anyi}}]{GAP1}%
  \BibitemOpen
  \bibfield  {author} {\bibinfo {author} {\bibfnamefont {A.~P.}\ \bibnamefont
  {Bart\'ok}}, \bibinfo {author} {\bibfnamefont {M.~C.}\ \bibnamefont {Payne}},
  \bibinfo {author} {\bibfnamefont {R.}~\bibnamefont {Kondor}},\ and\ \bibinfo
  {author} {\bibfnamefont {G.}~\bibnamefont {Cs\'anyi}},\ }\bibfield  {title}
  {\bibinfo {title} {{Gaussian Approximation Potentials: The Accuracy of
  Quantum Mechanics, without the Electrons}},\ }\href
  {https://doi.org/10.1103/PhysRevLett.104.136403} {\bibfield  {journal}
  {\bibinfo  {journal} {Phys. Rev. Lett.}\ }\textbf {\bibinfo {volume} {104}},\
  \bibinfo {pages} {136403} (\bibinfo {year} {2010})}\BibitemShut {NoStop}%
\bibitem [{\citenamefont {Bart\'ok}\ \emph {et~al.}(2013)\citenamefont
  {Bart\'ok}, \citenamefont {Kondor},\ and\ \citenamefont {Cs\'anyi}}]{GAP2}%
  \BibitemOpen
  \bibfield  {author} {\bibinfo {author} {\bibfnamefont {A.~P.}\ \bibnamefont
  {Bart\'ok}}, \bibinfo {author} {\bibfnamefont {R.}~\bibnamefont {Kondor}},\
  and\ \bibinfo {author} {\bibfnamefont {G.}~\bibnamefont {Cs\'anyi}},\
  }\bibfield  {title} {\bibinfo {title} {On representing chemical
  environments},\ }\href {https://doi.org/10.1103/PhysRevB.87.184115}
  {\bibfield  {journal} {\bibinfo  {journal} {Phys. Rev. B}\ }\textbf {\bibinfo
  {volume} {87}},\ \bibinfo {pages} {184115} (\bibinfo {year}
  {2013})}\BibitemShut {NoStop}%
\bibitem [{\citenamefont {Deringer}\ \emph {et~al.}(2021)\citenamefont
  {Deringer}, \citenamefont {Bartók}, \citenamefont {Bernstein}, \citenamefont
  {Wilkins}, \citenamefont {Ceriotti},\ and\ \citenamefont {Csányi}}]{GAP3}%
  \BibitemOpen
  \bibfield  {author} {\bibinfo {author} {\bibfnamefont {V.~L.}\ \bibnamefont
  {Deringer}}, \bibinfo {author} {\bibfnamefont {A.~P.}\ \bibnamefont
  {Bartók}}, \bibinfo {author} {\bibfnamefont {N.}~\bibnamefont {Bernstein}},
  \bibinfo {author} {\bibfnamefont {D.~M.}\ \bibnamefont {Wilkins}}, \bibinfo
  {author} {\bibfnamefont {M.}~\bibnamefont {Ceriotti}},\ and\ \bibinfo
  {author} {\bibfnamefont {G.}~\bibnamefont {Csányi}},\ }\bibfield  {title}
  {\bibinfo {title} {{Gaussian Process Regression for Materials and
  Molecules}},\ }\href {https://doi.org/10.1021/acs.chemrev.1c00022} {\bibfield
   {journal} {\bibinfo  {journal} {Chemical Reviews}\ }\textbf {\bibinfo
  {volume} {121}},\ \bibinfo {pages} {10073} (\bibinfo {year} {2021})},\
  \bibinfo {note} {pMID: 34398616}\BibitemShut {NoStop}%
\bibitem [{\citenamefont {Menahem}\ \emph {et~al.}(2023)\citenamefont
  {Menahem}, \citenamefont {Benshalom}, \citenamefont {Asher}, \citenamefont
  {Aharon}, \citenamefont {Korobko}, \citenamefont {Hellman},\ and\
  \citenamefont {Yaffe}}]{CsPbBr_QE_peak}%
  \BibitemOpen
  \bibfield  {author} {\bibinfo {author} {\bibfnamefont {M.}~\bibnamefont
  {Menahem}}, \bibinfo {author} {\bibfnamefont {N.}~\bibnamefont {Benshalom}},
  \bibinfo {author} {\bibfnamefont {M.}~\bibnamefont {Asher}}, \bibinfo
  {author} {\bibfnamefont {S.}~\bibnamefont {Aharon}}, \bibinfo {author}
  {\bibfnamefont {R.}~\bibnamefont {Korobko}}, \bibinfo {author} {\bibfnamefont
  {O.}~\bibnamefont {Hellman}},\ and\ \bibinfo {author} {\bibfnamefont
  {O.}~\bibnamefont {Yaffe}},\ }\bibfield  {title} {\bibinfo {title} {{Disorder
  origin of Raman scattering in perovskite single crystals}},\ }\href
  {https://doi.org/10.1103/PhysRevMaterials.7.044602} {\bibfield  {journal}
  {\bibinfo  {journal} {Phys. Rev. Mater.}\ }\textbf {\bibinfo {volume} {7}},\
  \bibinfo {pages} {044602} (\bibinfo {year} {2023})}\BibitemShut {NoStop}%
\bibitem [{\citenamefont {Holt}\ \emph {et~al.}(2001)\citenamefont {Holt},
  \citenamefont {Zschack}, \citenamefont {Hong}, \citenamefont {Chou},\ and\
  \citenamefont {Chiang}}]{Holt2001X-Ray}%
  \BibitemOpen
  \bibfield  {author} {\bibinfo {author} {\bibfnamefont {M.}~\bibnamefont
  {Holt}}, \bibinfo {author} {\bibfnamefont {P.}~\bibnamefont {Zschack}},
  \bibinfo {author} {\bibfnamefont {H.}~\bibnamefont {Hong}}, \bibinfo {author}
  {\bibfnamefont {M.~Y.}\ \bibnamefont {Chou}},\ and\ \bibinfo {author}
  {\bibfnamefont {T.-C.}\ \bibnamefont {Chiang}},\ }\bibfield  {title}
  {\bibinfo {title} {X-ray studies of phonon softening in
  ${\mathrm{tise}}_{2}$},\ }\href {https://doi.org/10.1103/PhysRevLett.86.3799}
  {\bibfield  {journal} {\bibinfo  {journal} {Phys. Rev. Lett.}\ }\textbf
  {\bibinfo {volume} {86}},\ \bibinfo {pages} {3799} (\bibinfo {year}
  {2001})}\BibitemShut {NoStop}%
\bibitem [{\citenamefont {Wang}\ \emph {et~al.}(2018)\citenamefont {Wang},
  \citenamefont {Lin}, \citenamefont {Zhu}, \citenamefont {Zheng},
  \citenamefont {Wang}, \citenamefont {Li},\ and\ \citenamefont
  {Zhu}}]{CsPbBr_kappa2}%
  \BibitemOpen
  \bibfield  {author} {\bibinfo {author} {\bibfnamefont {Y.}~\bibnamefont
  {Wang}}, \bibinfo {author} {\bibfnamefont {R.}~\bibnamefont {Lin}}, \bibinfo
  {author} {\bibfnamefont {P.}~\bibnamefont {Zhu}}, \bibinfo {author}
  {\bibfnamefont {Q.}~\bibnamefont {Zheng}}, \bibinfo {author} {\bibfnamefont
  {Q.}~\bibnamefont {Wang}}, \bibinfo {author} {\bibfnamefont {D.}~\bibnamefont
  {Li}},\ and\ \bibinfo {author} {\bibfnamefont {J.}~\bibnamefont {Zhu}},\
  }\bibfield  {title} {\bibinfo {title} {{Cation Dynamics Governed Thermal
  Properties of Lead Halide Perovskite Nanowires}},\ }\href
  {https://doi.org/10.1021/acs.nanolett.7b04437} {\bibfield  {journal}
  {\bibinfo  {journal} {Nano Letters}\ }\textbf {\bibinfo {volume} {18}},\
  \bibinfo {pages} {2772} (\bibinfo {year} {2018})},\ \bibinfo {note} {pMID:
  29618206}\BibitemShut {NoStop}%
\bibitem [{\citenamefont {Agrawal}\ \emph {et~al.}(2022)\citenamefont
  {Agrawal}, \citenamefont {Hasan}, \citenamefont {Blawat}, \citenamefont
  {Mehta}, \citenamefont {Wang}, \citenamefont {Cueto}, \citenamefont
  {Siebenbuerger}, \citenamefont {Kizilkaya}, \citenamefont {Prasad},
  \citenamefont {Dorman}, \citenamefont {Jin},\ and\ \citenamefont
  {Gartia}}]{CsPbBr_kappa3}%
  \BibitemOpen
  \bibfield  {author} {\bibinfo {author} {\bibfnamefont {K.}~\bibnamefont
  {Agrawal}}, \bibinfo {author} {\bibfnamefont {S.~M.~A.}\ \bibnamefont
  {Hasan}}, \bibinfo {author} {\bibfnamefont {J.}~\bibnamefont {Blawat}},
  \bibinfo {author} {\bibfnamefont {N.}~\bibnamefont {Mehta}}, \bibinfo
  {author} {\bibfnamefont {Y.}~\bibnamefont {Wang}}, \bibinfo {author}
  {\bibfnamefont {R.}~\bibnamefont {Cueto}}, \bibinfo {author} {\bibfnamefont
  {M.}~\bibnamefont {Siebenbuerger}}, \bibinfo {author} {\bibfnamefont
  {O.}~\bibnamefont {Kizilkaya}}, \bibinfo {author} {\bibfnamefont {N.~S.}\
  \bibnamefont {Prasad}}, \bibinfo {author} {\bibfnamefont {J.}~\bibnamefont
  {Dorman}}, \bibinfo {author} {\bibfnamefont {R.}~\bibnamefont {Jin}},\ and\
  \bibinfo {author} {\bibfnamefont {M.~R.}\ \bibnamefont {Gartia}},\ }\bibfield
   {title} {\bibinfo {title} {{Thermal, Physical, and Optical Properties of the
  Solution and Melt Synthesized Single Crystal CsPbBr3 Halide Perovskite}},\
  }\href {https://doi.org/10.3390/chemosensors10090369} {\bibfield  {journal}
  {\bibinfo  {journal} {Chemosensors}\ }\textbf {\bibinfo {volume} {10}},\
  \bibinfo {pages} {369} (\bibinfo {year} {2022})}\BibitemShut {NoStop}%
\bibitem [{\citenamefont {Kubi{\v{c}}{\'{a}}r}\ \emph
  {et~al.}(2008)\citenamefont {Kubi{\v{c}}{\'{a}}r}, \citenamefont
  {Vreten{\'{a}}r},\ and\ \citenamefont {Boh{\'{a}}{\v{c}}}}]{CsPbBr_kappa4}%
  \BibitemOpen
  \bibfield  {author} {\bibinfo {author} {\bibfnamefont {L.}~\bibnamefont
  {Kubi{\v{c}}{\'{a}}r}}, \bibinfo {author} {\bibfnamefont {V.}~\bibnamefont
  {Vreten{\'{a}}r}},\ and\ \bibinfo {author} {\bibfnamefont {V.}~\bibnamefont
  {Boh{\'{a}}{\v{c}}}},\ }\bibfield  {title} {\bibinfo {title} {{Study of Phase
  Transitions by Transient Methods}},\ }\href
  {https://doi.org/10.4028/www.scientific.net/SSP.138.3} {\bibfield  {journal}
  {\bibinfo  {journal} {Solid State Phenomena}\ }\textbf {\bibinfo {volume}
  {138}},\ \bibinfo {pages} {3} (\bibinfo {year} {2008})}\BibitemShut {NoStop}%
\bibitem [{\citenamefont {Kanak}\ \emph {et~al.}(2020)\citenamefont {Kanak},
  \citenamefont {Lishchuk}, \citenamefont {Kuryliuk}, \citenamefont {Kuzmich},
  \citenamefont {Lacroix}, \citenamefont {Khalavka},\ and\ \citenamefont
  {Isaiev}}]{CsPbBr_kappa5}%
  \BibitemOpen
  \bibfield  {author} {\bibinfo {author} {\bibfnamefont {A.}~\bibnamefont
  {Kanak}}, \bibinfo {author} {\bibfnamefont {P.}~\bibnamefont {Lishchuk}},
  \bibinfo {author} {\bibfnamefont {V.}~\bibnamefont {Kuryliuk}}, \bibinfo
  {author} {\bibfnamefont {A.}~\bibnamefont {Kuzmich}}, \bibinfo {author}
  {\bibfnamefont {D.}~\bibnamefont {Lacroix}}, \bibinfo {author} {\bibfnamefont
  {Y.}~\bibnamefont {Khalavka}},\ and\ \bibinfo {author} {\bibfnamefont
  {M.}~\bibnamefont {Isaiev}},\ }\bibfield  {title} {\bibinfo {title} {{Thermal
  conductivity of CsPbBr3 halide perovskite: Photoacoustic measurements and
  molecular dynamics analysis}},\ }\href {https://doi.org/10.1063/5.0033821}
  {\bibfield  {journal} {\bibinfo  {journal} {AIP Conference Proceedings}\
  }\textbf {\bibinfo {volume} {2305}},\ \bibinfo {pages} {020006} (\bibinfo
  {year} {2020})}\BibitemShut {NoStop}%
\bibitem [{\citenamefont {Wang}\ \emph {et~al.}(2023)\citenamefont {Wang},
  \citenamefont {Gao}, \citenamefont {Zhu}, \citenamefont {Ren}, \citenamefont
  {Hu}, \citenamefont {Sun}, \citenamefont {Ding}, \citenamefont {Xia},\ and\
  \citenamefont {Li}}]{CsPbBr3_newkappa}%
  \BibitemOpen
  \bibfield  {author} {\bibinfo {author} {\bibfnamefont {X.}~\bibnamefont
  {Wang}}, \bibinfo {author} {\bibfnamefont {Z.}~\bibnamefont {Gao}}, \bibinfo
  {author} {\bibfnamefont {G.}~\bibnamefont {Zhu}}, \bibinfo {author}
  {\bibfnamefont {J.}~\bibnamefont {Ren}}, \bibinfo {author} {\bibfnamefont
  {L.}~\bibnamefont {Hu}}, \bibinfo {author} {\bibfnamefont {J.}~\bibnamefont
  {Sun}}, \bibinfo {author} {\bibfnamefont {X.}~\bibnamefont {Ding}}, \bibinfo
  {author} {\bibfnamefont {Y.}~\bibnamefont {Xia}},\ and\ \bibinfo {author}
  {\bibfnamefont {B.}~\bibnamefont {Li}},\ }\bibfield  {title} {\bibinfo
  {title} {{Role of high-order anharmonicity and off-diagonal terms in thermal
  conductivity: A case study of multiphase ${\mathrm{CsPbBr}}_{3}$}},\ }\href
  {https://doi.org/10.1103/PhysRevB.107.214308} {\bibfield  {journal} {\bibinfo
   {journal} {Phys. Rev. B}\ }\textbf {\bibinfo {volume} {107}},\ \bibinfo
  {pages} {214308} (\bibinfo {year} {2023})}\BibitemShut {NoStop}%
\bibitem [{\citenamefont {Giannozzi}\ \emph {et~al.}(2020)\citenamefont
  {Giannozzi}, \citenamefont {Baseggio}, \citenamefont {Bonfà}, \citenamefont
  {Brunato}, \citenamefont {Car}, \citenamefont {Carnimeo}, \citenamefont
  {Cavazzoni}, \citenamefont {de~Gironcoli}, \citenamefont {Delugas},
  \citenamefont {Ferrari~Ruffino}, \citenamefont {Ferretti}, \citenamefont
  {Marzari}, \citenamefont {Timrov}, \citenamefont {Urru},\ and\ \citenamefont
  {Baroni}}]{QE1}%
  \BibitemOpen
  \bibfield  {author} {\bibinfo {author} {\bibfnamefont {P.}~\bibnamefont
  {Giannozzi}}, \bibinfo {author} {\bibfnamefont {O.}~\bibnamefont {Baseggio}},
  \bibinfo {author} {\bibfnamefont {P.}~\bibnamefont {Bonfà}}, \bibinfo
  {author} {\bibfnamefont {D.}~\bibnamefont {Brunato}}, \bibinfo {author}
  {\bibfnamefont {R.}~\bibnamefont {Car}}, \bibinfo {author} {\bibfnamefont
  {I.}~\bibnamefont {Carnimeo}}, \bibinfo {author} {\bibfnamefont
  {C.}~\bibnamefont {Cavazzoni}}, \bibinfo {author} {\bibfnamefont
  {S.}~\bibnamefont {de~Gironcoli}}, \bibinfo {author} {\bibfnamefont
  {P.}~\bibnamefont {Delugas}}, \bibinfo {author} {\bibfnamefont
  {F.}~\bibnamefont {Ferrari~Ruffino}}, \bibinfo {author} {\bibfnamefont
  {A.}~\bibnamefont {Ferretti}}, \bibinfo {author} {\bibfnamefont
  {N.}~\bibnamefont {Marzari}}, \bibinfo {author} {\bibfnamefont
  {I.}~\bibnamefont {Timrov}}, \bibinfo {author} {\bibfnamefont
  {A.}~\bibnamefont {Urru}},\ and\ \bibinfo {author} {\bibfnamefont
  {S.}~\bibnamefont {Baroni}},\ }\bibfield  {title} {\bibinfo {title} {{Quantum
  ESPRESSO toward the exascale}},\ }\href {https://doi.org/10.1063/5.0005082}
  {\bibfield  {journal} {\bibinfo  {journal} {The Journal of Chemical Physics}\
  }\textbf {\bibinfo {volume} {152}},\ \bibinfo {pages} {154105} (\bibinfo
  {year} {2020})}\BibitemShut {NoStop}%
\bibitem [{\citenamefont {Giannozzi}\ \emph {et~al.}(2017)\citenamefont
  {Giannozzi}, \citenamefont {Andreussi}, \citenamefont {Brumme}, \citenamefont
  {Bunau}, \citenamefont {Nardelli}, \citenamefont {Calandra}, \citenamefont
  {Car}, \citenamefont {Cavazzoni}, \citenamefont {Ceresoli}, \citenamefont
  {Cococcioni}, \citenamefont {Colonna}, \citenamefont {Carnimeo},
  \citenamefont {Corso}, \citenamefont {de~Gironcoli}, \citenamefont {Delugas},
  \citenamefont {DiStasio}, \citenamefont {Ferretti}, \citenamefont {Floris},
  \citenamefont {Fratesi}, \citenamefont {Fugallo}, \citenamefont {Gebauer},
  \citenamefont {Gerstmann}, \citenamefont {Giustino}, \citenamefont {Gorni},
  \citenamefont {Jia}, \citenamefont {Kawamura}, \citenamefont {Ko},
  \citenamefont {Kokalj}, \citenamefont {Küçükbenli}, \citenamefont
  {Lazzeri}, \citenamefont {Marsili}, \citenamefont {Marzari}, \citenamefont
  {Mauri}, \citenamefont {Nguyen}, \citenamefont {Nguyen}, \citenamefont {de-la
  Roza}, \citenamefont {Paulatto}, \citenamefont {Poncé}, \citenamefont
  {Rocca}, \citenamefont {Sabatini}, \citenamefont {Santra}, \citenamefont
  {Schlipf}, \citenamefont {Seitsonen}, \citenamefont {Smogunov}, \citenamefont
  {Timrov}, \citenamefont {Thonhauser}, \citenamefont {Umari}, \citenamefont
  {Vast}, \citenamefont {Wu},\ and\ \citenamefont {Baroni}}]{QE2}%
  \BibitemOpen
  \bibfield  {author} {\bibinfo {author} {\bibfnamefont {P.}~\bibnamefont
  {Giannozzi}}, \bibinfo {author} {\bibfnamefont {O.}~\bibnamefont
  {Andreussi}}, \bibinfo {author} {\bibfnamefont {T.}~\bibnamefont {Brumme}},
  \bibinfo {author} {\bibfnamefont {O.}~\bibnamefont {Bunau}}, \bibinfo
  {author} {\bibfnamefont {M.~B.}\ \bibnamefont {Nardelli}}, \bibinfo {author}
  {\bibfnamefont {M.}~\bibnamefont {Calandra}}, \bibinfo {author}
  {\bibfnamefont {R.}~\bibnamefont {Car}}, \bibinfo {author} {\bibfnamefont
  {C.}~\bibnamefont {Cavazzoni}}, \bibinfo {author} {\bibfnamefont
  {D.}~\bibnamefont {Ceresoli}}, \bibinfo {author} {\bibfnamefont
  {M.}~\bibnamefont {Cococcioni}}, \bibinfo {author} {\bibfnamefont
  {N.}~\bibnamefont {Colonna}}, \bibinfo {author} {\bibfnamefont
  {I.}~\bibnamefont {Carnimeo}}, \bibinfo {author} {\bibfnamefont {A.~D.}\
  \bibnamefont {Corso}}, \bibinfo {author} {\bibfnamefont {S.}~\bibnamefont
  {de~Gironcoli}}, \bibinfo {author} {\bibfnamefont {P.}~\bibnamefont
  {Delugas}}, \bibinfo {author} {\bibfnamefont {R.~A.}\ \bibnamefont
  {DiStasio}}, \bibinfo {author} {\bibfnamefont {A.}~\bibnamefont {Ferretti}},
  \bibinfo {author} {\bibfnamefont {A.}~\bibnamefont {Floris}}, \bibinfo
  {author} {\bibfnamefont {G.}~\bibnamefont {Fratesi}}, \bibinfo {author}
  {\bibfnamefont {G.}~\bibnamefont {Fugallo}}, \bibinfo {author} {\bibfnamefont
  {R.}~\bibnamefont {Gebauer}}, \bibinfo {author} {\bibfnamefont
  {U.}~\bibnamefont {Gerstmann}}, \bibinfo {author} {\bibfnamefont
  {F.}~\bibnamefont {Giustino}}, \bibinfo {author} {\bibfnamefont
  {T.}~\bibnamefont {Gorni}}, \bibinfo {author} {\bibfnamefont
  {J.}~\bibnamefont {Jia}}, \bibinfo {author} {\bibfnamefont {M.}~\bibnamefont
  {Kawamura}}, \bibinfo {author} {\bibfnamefont {H.-Y.}\ \bibnamefont {Ko}},
  \bibinfo {author} {\bibfnamefont {A.}~\bibnamefont {Kokalj}}, \bibinfo
  {author} {\bibfnamefont {E.}~\bibnamefont {Küçükbenli}}, \bibinfo {author}
  {\bibfnamefont {M.}~\bibnamefont {Lazzeri}}, \bibinfo {author} {\bibfnamefont
  {M.}~\bibnamefont {Marsili}}, \bibinfo {author} {\bibfnamefont
  {N.}~\bibnamefont {Marzari}}, \bibinfo {author} {\bibfnamefont
  {F.}~\bibnamefont {Mauri}}, \bibinfo {author} {\bibfnamefont {N.~L.}\
  \bibnamefont {Nguyen}}, \bibinfo {author} {\bibfnamefont {H.-V.}\
  \bibnamefont {Nguyen}}, \bibinfo {author} {\bibfnamefont {A.~O.}\
  \bibnamefont {de-la Roza}}, \bibinfo {author} {\bibfnamefont
  {L.}~\bibnamefont {Paulatto}}, \bibinfo {author} {\bibfnamefont
  {S.}~\bibnamefont {Poncé}}, \bibinfo {author} {\bibfnamefont
  {D.}~\bibnamefont {Rocca}}, \bibinfo {author} {\bibfnamefont
  {R.}~\bibnamefont {Sabatini}}, \bibinfo {author} {\bibfnamefont
  {B.}~\bibnamefont {Santra}}, \bibinfo {author} {\bibfnamefont
  {M.}~\bibnamefont {Schlipf}}, \bibinfo {author} {\bibfnamefont {A.~P.}\
  \bibnamefont {Seitsonen}}, \bibinfo {author} {\bibfnamefont {A.}~\bibnamefont
  {Smogunov}}, \bibinfo {author} {\bibfnamefont {I.}~\bibnamefont {Timrov}},
  \bibinfo {author} {\bibfnamefont {T.}~\bibnamefont {Thonhauser}}, \bibinfo
  {author} {\bibfnamefont {P.}~\bibnamefont {Umari}}, \bibinfo {author}
  {\bibfnamefont {N.}~\bibnamefont {Vast}}, \bibinfo {author} {\bibfnamefont
  {X.}~\bibnamefont {Wu}},\ and\ \bibinfo {author} {\bibfnamefont
  {S.}~\bibnamefont {Baroni}},\ }\bibfield  {title} {\bibinfo {title}
  {{Advanced capabilities for materials modelling with Quantum ESPRESSO}},\
  }\href {https://doi.org/10.1088/1361-648X/aa8f79} {\bibfield  {journal}
  {\bibinfo  {journal} {Journal of Physics: Condensed Matter}\ }\textbf
  {\bibinfo {volume} {29}},\ \bibinfo {pages} {465901} (\bibinfo {year}
  {2017})}\BibitemShut {NoStop}%
\bibitem [{\citenamefont {Giannozzi}\ \emph {et~al.}(2009)\citenamefont
  {Giannozzi}, \citenamefont {Baroni}, \citenamefont {Bonini}, \citenamefont
  {Calandra}, \citenamefont {Car}, \citenamefont {Cavazzoni}, \citenamefont
  {Ceresoli}, \citenamefont {Chiarotti}, \citenamefont {Cococcioni},
  \citenamefont {Dabo}, \citenamefont {Corso}, \citenamefont {de~Gironcoli},
  \citenamefont {Fabris}, \citenamefont {Fratesi}, \citenamefont {Gebauer},
  \citenamefont {Gerstmann}, \citenamefont {Gougoussis}, \citenamefont
  {Kokalj}, \citenamefont {Lazzeri}, \citenamefont {Martin-Samos},
  \citenamefont {Marzari}, \citenamefont {Mauri}, \citenamefont {Mazzarello},
  \citenamefont {Paolini}, \citenamefont {Pasquarello}, \citenamefont
  {Paulatto}, \citenamefont {Sbraccia}, \citenamefont {Scandolo}, \citenamefont
  {Sclauzero}, \citenamefont {Seitsonen}, \citenamefont {Smogunov},
  \citenamefont {Umari},\ and\ \citenamefont {Wentzcovitch}}]{QE3}%
  \BibitemOpen
  \bibfield  {author} {\bibinfo {author} {\bibfnamefont {P.}~\bibnamefont
  {Giannozzi}}, \bibinfo {author} {\bibfnamefont {S.}~\bibnamefont {Baroni}},
  \bibinfo {author} {\bibfnamefont {N.}~\bibnamefont {Bonini}}, \bibinfo
  {author} {\bibfnamefont {M.}~\bibnamefont {Calandra}}, \bibinfo {author}
  {\bibfnamefont {R.}~\bibnamefont {Car}}, \bibinfo {author} {\bibfnamefont
  {C.}~\bibnamefont {Cavazzoni}}, \bibinfo {author} {\bibfnamefont
  {D.}~\bibnamefont {Ceresoli}}, \bibinfo {author} {\bibfnamefont {G.~L.}\
  \bibnamefont {Chiarotti}}, \bibinfo {author} {\bibfnamefont {M.}~\bibnamefont
  {Cococcioni}}, \bibinfo {author} {\bibfnamefont {I.}~\bibnamefont {Dabo}},
  \bibinfo {author} {\bibfnamefont {A.~D.}\ \bibnamefont {Corso}}, \bibinfo
  {author} {\bibfnamefont {S.}~\bibnamefont {de~Gironcoli}}, \bibinfo {author}
  {\bibfnamefont {S.}~\bibnamefont {Fabris}}, \bibinfo {author} {\bibfnamefont
  {G.}~\bibnamefont {Fratesi}}, \bibinfo {author} {\bibfnamefont
  {R.}~\bibnamefont {Gebauer}}, \bibinfo {author} {\bibfnamefont
  {U.}~\bibnamefont {Gerstmann}}, \bibinfo {author} {\bibfnamefont
  {C.}~\bibnamefont {Gougoussis}}, \bibinfo {author} {\bibfnamefont
  {A.}~\bibnamefont {Kokalj}}, \bibinfo {author} {\bibfnamefont
  {M.}~\bibnamefont {Lazzeri}}, \bibinfo {author} {\bibfnamefont
  {L.}~\bibnamefont {Martin-Samos}}, \bibinfo {author} {\bibfnamefont
  {N.}~\bibnamefont {Marzari}}, \bibinfo {author} {\bibfnamefont
  {F.}~\bibnamefont {Mauri}}, \bibinfo {author} {\bibfnamefont
  {R.}~\bibnamefont {Mazzarello}}, \bibinfo {author} {\bibfnamefont
  {S.}~\bibnamefont {Paolini}}, \bibinfo {author} {\bibfnamefont
  {A.}~\bibnamefont {Pasquarello}}, \bibinfo {author} {\bibfnamefont
  {L.}~\bibnamefont {Paulatto}}, \bibinfo {author} {\bibfnamefont
  {C.}~\bibnamefont {Sbraccia}}, \bibinfo {author} {\bibfnamefont
  {S.}~\bibnamefont {Scandolo}}, \bibinfo {author} {\bibfnamefont
  {G.}~\bibnamefont {Sclauzero}}, \bibinfo {author} {\bibfnamefont {A.~P.}\
  \bibnamefont {Seitsonen}}, \bibinfo {author} {\bibfnamefont {A.}~\bibnamefont
  {Smogunov}}, \bibinfo {author} {\bibfnamefont {P.}~\bibnamefont {Umari}},\
  and\ \bibinfo {author} {\bibfnamefont {R.~M.}\ \bibnamefont {Wentzcovitch}},\
  }\bibfield  {title} {\bibinfo {title} {{QUANTUM ESPRESSO: a modular and
  open-source software project for quantum simulations of materials}},\ }\href
  {https://doi.org/10.1088/0953-8984/21/39/395502} {\bibfield  {journal}
  {\bibinfo  {journal} {Journal of Physics: Condensed Matter}\ }\textbf
  {\bibinfo {volume} {21}},\ \bibinfo {pages} {395502} (\bibinfo {year}
  {2009})}\BibitemShut {NoStop}%
\bibitem [{\citenamefont {Garrity}\ \emph {et~al.}(2014)\citenamefont
  {Garrity}, \citenamefont {Bennett}, \citenamefont {Rabe},\ and\ \citenamefont
  {Vanderbilt}}]{GARRITY2014446}%
  \BibitemOpen
  \bibfield  {author} {\bibinfo {author} {\bibfnamefont {K.~F.}\ \bibnamefont
  {Garrity}}, \bibinfo {author} {\bibfnamefont {J.~W.}\ \bibnamefont
  {Bennett}}, \bibinfo {author} {\bibfnamefont {K.~M.}\ \bibnamefont {Rabe}},\
  and\ \bibinfo {author} {\bibfnamefont {D.}~\bibnamefont {Vanderbilt}},\
  }\bibfield  {title} {\bibinfo {title} {{Pseudopotentials for high-throughput
  DFT calculations}},\ }\href
  {https://doi.org/https://doi.org/10.1016/j.commatsci.2013.08.053} {\bibfield
  {journal} {\bibinfo  {journal} {Computational Materials Science}\ }\textbf
  {\bibinfo {volume} {81}},\ \bibinfo {pages} {446} (\bibinfo {year}
  {2014})}\BibitemShut {NoStop}%
\bibitem [{\citenamefont {Perdew}\ \emph {et~al.}(2008)\citenamefont {Perdew},
  \citenamefont {Ruzsinszky}, \citenamefont {Csonka}, \citenamefont {Vydrov},
  \citenamefont {Scuseria}, \citenamefont {Constantin}, \citenamefont {Zhou},\
  and\ \citenamefont {Burke}}]{PBESOL}%
  \BibitemOpen
  \bibfield  {author} {\bibinfo {author} {\bibfnamefont {J.~P.}\ \bibnamefont
  {Perdew}}, \bibinfo {author} {\bibfnamefont {A.}~\bibnamefont {Ruzsinszky}},
  \bibinfo {author} {\bibfnamefont {G.~I.}\ \bibnamefont {Csonka}}, \bibinfo
  {author} {\bibfnamefont {O.~A.}\ \bibnamefont {Vydrov}}, \bibinfo {author}
  {\bibfnamefont {G.~E.}\ \bibnamefont {Scuseria}}, \bibinfo {author}
  {\bibfnamefont {L.~A.}\ \bibnamefont {Constantin}}, \bibinfo {author}
  {\bibfnamefont {X.}~\bibnamefont {Zhou}},\ and\ \bibinfo {author}
  {\bibfnamefont {K.}~\bibnamefont {Burke}},\ }\bibfield  {title} {\bibinfo
  {title} {{Restoring the Density-Gradient Expansion for Exchange in Solids and
  Surfaces}},\ }\href {https://doi.org/10.1103/PhysRevLett.100.136406}
  {\bibfield  {journal} {\bibinfo  {journal} {Phys. Rev. Lett.}\ }\textbf
  {\bibinfo {volume} {100}},\ \bibinfo {pages} {136406} (\bibinfo {year}
  {2008})}\BibitemShut {NoStop}%
\bibitem [{\citenamefont {Larsen}\ \emph {et~al.}(2017)\citenamefont {Larsen},
  \citenamefont {Mortensen}, \citenamefont {Blomqvist}, \citenamefont
  {Castelli}, \citenamefont {Christensen}, \citenamefont {Dułak},
  \citenamefont {Friis}, \citenamefont {Groves}, \citenamefont {Hammer},
  \citenamefont {Hargus}, \citenamefont {Hermes}, \citenamefont {Jennings},
  \citenamefont {Jensen}, \citenamefont {Kermode}, \citenamefont {Kitchin},
  \citenamefont {Kolsbjerg}, \citenamefont {Kubal}, \citenamefont {Kaasbjerg},
  \citenamefont {Lysgaard}, \citenamefont {Maronsson}, \citenamefont {Maxson},
  \citenamefont {Olsen}, \citenamefont {Pastewka}, \citenamefont {Peterson},
  \citenamefont {Rostgaard}, \citenamefont {Schiøtz}, \citenamefont {Schütt},
  \citenamefont {Strange}, \citenamefont {Thygesen}, \citenamefont {Vegge},
  \citenamefont {Vilhelmsen}, \citenamefont {Walter}, \citenamefont {Zeng},\
  and\ \citenamefont {Jacobsen}}]{ASE}%
  \BibitemOpen
  \bibfield  {author} {\bibinfo {author} {\bibfnamefont {A.~H.}\ \bibnamefont
  {Larsen}}, \bibinfo {author} {\bibfnamefont {J.~J.}\ \bibnamefont
  {Mortensen}}, \bibinfo {author} {\bibfnamefont {J.}~\bibnamefont
  {Blomqvist}}, \bibinfo {author} {\bibfnamefont {I.~E.}\ \bibnamefont
  {Castelli}}, \bibinfo {author} {\bibfnamefont {R.}~\bibnamefont
  {Christensen}}, \bibinfo {author} {\bibfnamefont {M.}~\bibnamefont {Dułak}},
  \bibinfo {author} {\bibfnamefont {J.}~\bibnamefont {Friis}}, \bibinfo
  {author} {\bibfnamefont {M.~N.}\ \bibnamefont {Groves}}, \bibinfo {author}
  {\bibfnamefont {B.}~\bibnamefont {Hammer}}, \bibinfo {author} {\bibfnamefont
  {C.}~\bibnamefont {Hargus}}, \bibinfo {author} {\bibfnamefont {E.~D.}\
  \bibnamefont {Hermes}}, \bibinfo {author} {\bibfnamefont {P.~C.}\
  \bibnamefont {Jennings}}, \bibinfo {author} {\bibfnamefont {P.~B.}\
  \bibnamefont {Jensen}}, \bibinfo {author} {\bibfnamefont {J.}~\bibnamefont
  {Kermode}}, \bibinfo {author} {\bibfnamefont {J.~R.}\ \bibnamefont
  {Kitchin}}, \bibinfo {author} {\bibfnamefont {E.~L.}\ \bibnamefont
  {Kolsbjerg}}, \bibinfo {author} {\bibfnamefont {J.}~\bibnamefont {Kubal}},
  \bibinfo {author} {\bibfnamefont {K.}~\bibnamefont {Kaasbjerg}}, \bibinfo
  {author} {\bibfnamefont {S.}~\bibnamefont {Lysgaard}}, \bibinfo {author}
  {\bibfnamefont {J.~B.}\ \bibnamefont {Maronsson}}, \bibinfo {author}
  {\bibfnamefont {T.}~\bibnamefont {Maxson}}, \bibinfo {author} {\bibfnamefont
  {T.}~\bibnamefont {Olsen}}, \bibinfo {author} {\bibfnamefont
  {L.}~\bibnamefont {Pastewka}}, \bibinfo {author} {\bibfnamefont
  {A.}~\bibnamefont {Peterson}}, \bibinfo {author} {\bibfnamefont
  {C.}~\bibnamefont {Rostgaard}}, \bibinfo {author} {\bibfnamefont
  {J.}~\bibnamefont {Schiøtz}}, \bibinfo {author} {\bibfnamefont
  {O.}~\bibnamefont {Schütt}}, \bibinfo {author} {\bibfnamefont
  {M.}~\bibnamefont {Strange}}, \bibinfo {author} {\bibfnamefont {K.~S.}\
  \bibnamefont {Thygesen}}, \bibinfo {author} {\bibfnamefont {T.}~\bibnamefont
  {Vegge}}, \bibinfo {author} {\bibfnamefont {L.}~\bibnamefont {Vilhelmsen}},
  \bibinfo {author} {\bibfnamefont {M.}~\bibnamefont {Walter}}, \bibinfo
  {author} {\bibfnamefont {Z.}~\bibnamefont {Zeng}},\ and\ \bibinfo {author}
  {\bibfnamefont {K.~W.}\ \bibnamefont {Jacobsen}},\ }\bibfield  {title}
  {\bibinfo {title} {{The atomic simulation environment—a Python library for
  working with atoms}},\ }\href {https://doi.org/10.1088/1361-648X/aa680e}
  {\bibfield  {journal} {\bibinfo  {journal} {Journal of Physics: Condensed
  Matter}\ }\textbf {\bibinfo {volume} {29}},\ \bibinfo {pages} {273002}
  (\bibinfo {year} {2017})}\BibitemShut {NoStop}%
\bibitem [{\citenamefont {Yates}\ \emph {et~al.}(2007)\citenamefont {Yates},
  \citenamefont {Wang}, \citenamefont {Vanderbilt},\ and\ \citenamefont
  {Souza}}]{adaptive_smearing}%
  \BibitemOpen
  \bibfield  {author} {\bibinfo {author} {\bibfnamefont {J.~R.}\ \bibnamefont
  {Yates}}, \bibinfo {author} {\bibfnamefont {X.}~\bibnamefont {Wang}},
  \bibinfo {author} {\bibfnamefont {D.}~\bibnamefont {Vanderbilt}},\ and\
  \bibinfo {author} {\bibfnamefont {I.}~\bibnamefont {Souza}},\ }\bibfield
  {title} {\bibinfo {title} {Spectral and fermi surface properties from wannier
  interpolation},\ }\href {https://doi.org/10.1103/PhysRevB.75.195121}
  {\bibfield  {journal} {\bibinfo  {journal} {Phys. Rev. B}\ }\textbf {\bibinfo
  {volume} {75}},\ \bibinfo {pages} {195121} (\bibinfo {year}
  {2007})}\BibitemShut {NoStop}%
\bibitem [{\citenamefont {Dangi\'{c}}\ \emph {et~al.}(2022)\citenamefont
  {Dangi\'{c}}, \citenamefont {Fahy},\ and\ \citenamefont {Savi\'{c}}}]{myMD}%
  \BibitemOpen
  \bibfield  {author} {\bibinfo {author} {\bibfnamefont {D.}~\bibnamefont
  {Dangi\'{c}}}, \bibinfo {author} {\bibfnamefont {S.}~\bibnamefont {Fahy}},\
  and\ \bibinfo {author} {\bibfnamefont {I.}~\bibnamefont {Savi\'{c}}},\
  }\bibfield  {title} {\bibinfo {title} {Molecular dynamics simulation of the
  ferroelectric phase transition in gete: Displacive or order-disorder
  character},\ }\href {https://doi.org/10.1103/PhysRevB.106.134113} {\bibfield
  {journal} {\bibinfo  {journal} {Phys. Rev. B}\ }\textbf {\bibinfo {volume}
  {106}},\ \bibinfo {pages} {134113} (\bibinfo {year} {2022})}\BibitemShut
  {NoStop}%
\bibitem [{Note1()}]{Note1}%
  \BibitemOpen
  \bibinfo {note} {To obtain the expression Eq.~\ref {eq:spec_kappa} we used
  the fact that phonon spectral function is even with respect to $\Omega
  $.}\BibitemShut {Stop}%
\end{thebibliography}%

	\onecolumngrid
    \renewcommand{\figurename}{Supplementary Figure}
    \renewcommand{\tablename}{Supplementary Table}
    \setcounter{figure}{0}
    \setcounter{equation}{0}
	\newpage
	\newpage
	\newpage

\section{Supplementary material for: Lattice thermal conductivity in the  anharmonic overdamped regime}
\subsection{Computation details}

To calculate atomic energies, forces, and stresses we used Gaussian Approximation Potentials trained on SSCHA data iteratively. We started with harmonic dynamical matrices available from Ref.~\cite{Simoncelli2022} (in $Pmna$ structure) and generated ensembles of atomic configurations at several different temperatures for which we calculated density functional theory (DFT) forces, energies, and stresses. With this first training set, we generated the first iteration of the GAP potential. We then relaxed SSCHA structures at different temperatures and different phases using this GAP potential. We then extracted several atomic configurations for each of the relaxed structures, recalculated them in DFT, added them to the training and test sets, and repeated the fitting of the interatomic potential. We continued doing this procedure until we reached the desired accuracy of the interatomic potential. 

The DFT calculations were performed with the PBE-SOL parameterization of the generalized gradient approximation exchange-correlation functional~\cite{PBESOL} using the Quantum Espresso software~\cite{QE1,QE2,QE3}. The valence electrons were represented using Vanderbilt ultrasoft pseudopotentials~\cite{GARRITY2014446}. The electronic wave functions were expanded into a plane wave basis set with a 50 Ry energy cutoff. To maintain constant sampling density of $\mathbf{k}$ points through different phases, we used $0.005\times2\pi$ \AA$^{-1}$ density of $\mathbf{k}$ points for all calculations. To generate and manipulate atomic structures during the fitting procedure we used the Atomic Simulation Environment~\cite{ASE}.

The hyperparameters used for fitting the GAP potentials are given in Supp. Table~\ref{tab:GAP_hyp}.
\begin{table}
\begin{center}
\begin{tabular}{||c | c||} 
 \hline
 Cutoff radius & 6.42 \AA  \\ [0.5ex] 
 \hline
 Cutoff transition width & 0.8 \AA  \\ 
 \hline
 Energy regularization & 0.0009 eV per atom \\
 \hline
 Force regularization & 0.03 eV\AA$^{-1}$ \\
 \hline
 Stress regularization & 0.03 eV\AA$^{-3}$ \\
 \hline
 n$_{sparse}$ & 3200 \\ 
 \hline
 (l$_{max}$, n$_{max}$) & (5,10) \\ [1ex] 
 \hline
\end{tabular}
\caption{The values of hyperparameters that were used for fitting of the GAP potential.}
\label{tab:GAP_hyp}
\end{center}
\end{table}

In order to check the applicability of our GAP potential we show the comparison between DFT and GAP structures that minimize the Born-Oppenheimer energy surface in Supp. Table~\ref{tab:0K_structure}. The agreement we obtained between GAP and DFT is very good. This suggests that the overestimation of the phase transition temperature is due to the underlying DFT approximation.  

\begin{table}
\begin{center}
\begin{tabular}{||c | c | c | c ||} 
 \hline
  & $a$ (\AA) & $b$ (\AA) & $c$ (\AA) \\ [0.5ex] 
 \hline\hline
 $Pnma$ (DFT)       & 7.9636 & 8.3889 & 11.6312  \\ 
 \hline
 $Pnma$ (GAP)       & 7.9619 & 8.3904 & 11.6398 \\
 \hline\hline
 $P4/mbm$ (DFT)     & 8.1021 & 8.1021 & 5.9554 \\
 \hline
 $P4/mbm$ (GAP)     & 8.1215 & 8.1215 & 5.9463 \\
 \hline\hline
 $Pm\bar{3}m$ (DFT) & 5.8675 & 5.8675 & 5.8675 \\ 
 \hline
 $Pm\bar{3}m$ (GAP) & 5.8673 & 5.8673 & 5.8673 \\ [1ex] 
 \hline
\end{tabular}
\caption{The comparison between the lattice parameters of the structures that minimize the Born-Oppenheimer energy surface calculated with GAP and DFT.}
\label{tab:0K_structure}
\end{center}
\end{table}

\begin{figure}
    \centering
    \includegraphics[width=0.95\textwidth]{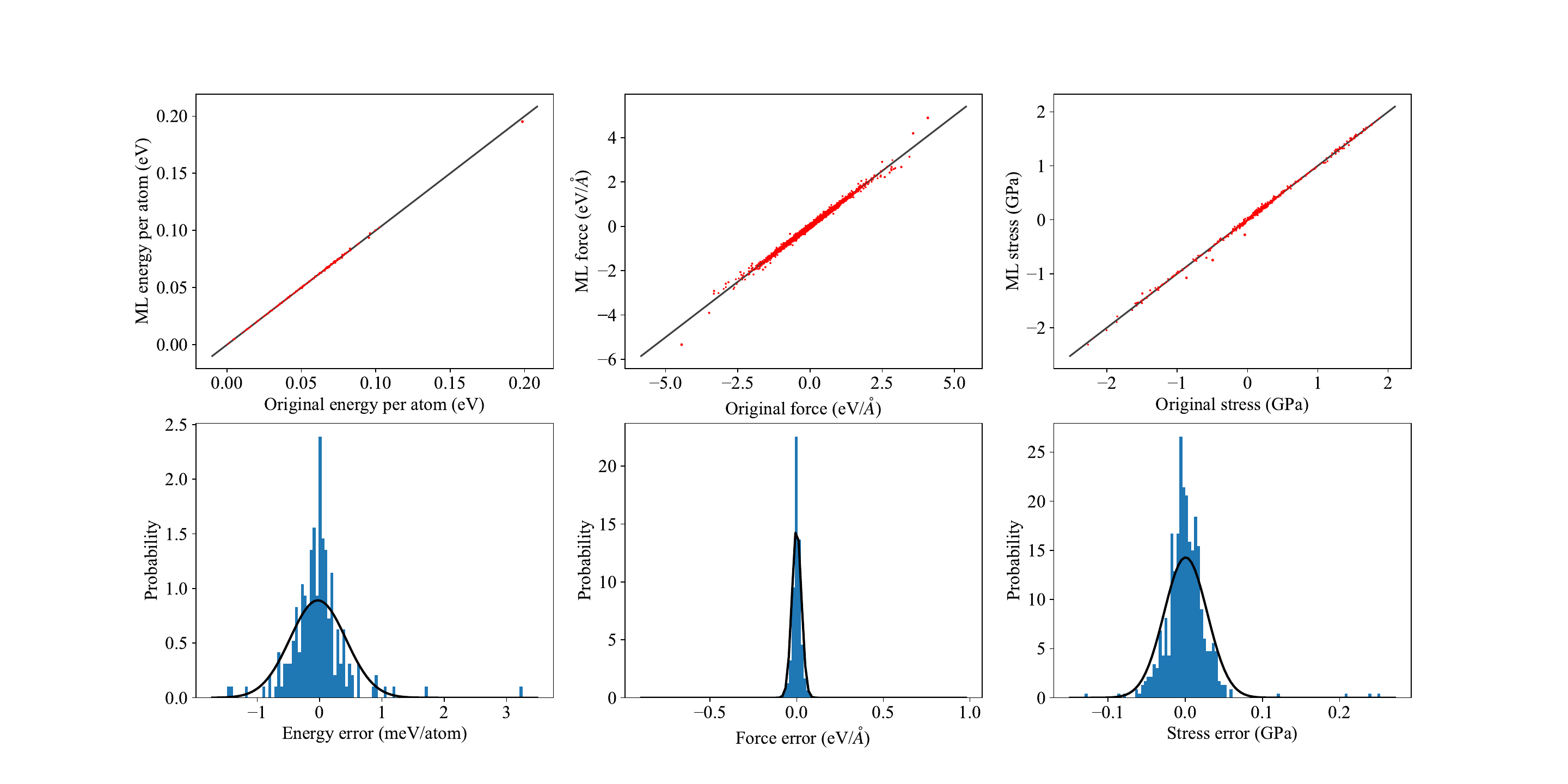}
    \caption{Differences between forces, energies, and stresses calculated using GAP and DFT. The first row shows these quantities calculated in GAP vs calculated in DFT. The perfect agreement would have all points lying on $y=x$ line (black in the figure). The second row shows histograms of errors.}
    \label{fig:errors}
\end{figure}

Finally, we show the errors for the GAP calculated forces, energies, and stresses, see Supp. Fig.~\ref{fig:errors}. Importantly, the structures for which we calculate these errors are not the ones that are included in the training set. All errors have Gaussian-like distribution centered at 0, meaning the potential does not have a bias and is appropriate for the study.

We minimized the total free energy for structures of all phases with the stochastic self-consistent harmonic approximation. The supercell used were 2$\times$2$\times$2 for $P4/mbm$ (80 atoms) and $Pnma$ (160 atoms) phases and 3$\times$3$\times$3 for $Pm\bar{3}m$ (135 atoms) phase. During the minimization process, we used 2000 configurations per population for $Pm\bar{3}m$ phase, and 4000 configurations for $P4/mbm$ and $Pnma$ phases. In some edge cases, we increased these numbers to ease the minimization process. To calculate the free energy of relaxed structure we used 4000 configurations for $Pm\bar{3}m$ and $P4/mbm$ phases and 8000 configurations for $Pnma$ phase and calculated the auxiliary harmonic part of free energy on an interpolated $24\times 24\times 24$ $\mathbf{q}$ point grid. 

To calculate the Hessian of the free energy and third-order force constants we used 25000 configurations for $Pm\bar{3}m$, 30000 configurations for $P4/mbm$ and 40000 configurations for $Pnma$ phase. The Hessian and third-order force constants were calculated on $2\times 2\times 2$ supercells for $Pm\bar{3}m$ to capture the soft phonon modes. The Hessians of the free energy for $Pm\bar{3}m$ and $P4/mbm$ phases were calculated including $\stackrel{(4)}{\boldsymbol{\mathcal{D}}}_{\mathcal{R}}$ (see main text). For the rest of the calculations, including phonon spectral functions and lattice thermal conductivity calculations, we did not include $\stackrel{(4)}{\boldsymbol{\mathcal{D}}}_{\mathcal{R}}$.

To calculate the lattice thermal conductivity we interpolated second and third-order dynamical matrices onto $25\times 25\times 25$ $\mathbf{q}$ grid for the $Pm\bar{3}m$ phase, $15\times 15\times 20$ $\mathbf{q}$ grid for the  $P4/mbm$ phase and $13\times 13\times 9$ $\mathbf{q}$ grid for the $Pnma$ phase. We used an adaptive Gaussian smearing with a smearing scale of 1.0 for the calculation of phonon lifetimes~\cite{adaptive_smearing}. Similarly, we used a 1.0 smearing scale for the Lorentzian smearing in order to calculate phonon spectral functions for the Green-Kubo method. We sampled phonon spectral functions on 2000 frequency steps.

\newpage

\subsection{Expression for the dynamical lattice thermal conductivity}

First, we define the Fourier transform of the $\langle A_{\mathbf{q},j}(0)A_{-\mathbf{q},l}(t)\rangle$:
\begin{align*}
    \langle A_{\mathbf{q},j}(0)A_{-\mathbf{q},l}(t)\rangle = \frac{1}{2\pi}\int ^{\infty}_{-\infty} J^{AA}_{\mathbf{q},j,l}(\Omega)e^{-i\Omega t}\text{d}\Omega.
\end{align*}
Then using the equation of motion for operator $A$, we get the Fourier transform of $\langle A_{\mathbf{q},j}(0)B_{\mathbf{q},l}(t)\rangle$:
\begin{align*}
    -\omega _{\mathbf{q},l}\langle A_{\mathbf{q},j}(0)B_{\mathbf{q},l}(t)\rangle = i \langle A_{\mathbf{q},j}(0)\dot{A}_{-\mathbf{q},l}(t)\rangle = \frac{1}{2\pi}\int ^{\infty}_{-\infty} \Omega J^{AA}_{\mathbf{q},j,l}(\Omega)e^{-i\Omega t}\text{d}\Omega. 
\end{align*}
Next, using the properties of the spectral representation of the correlation function~\cite{Zubarev}, we get $\langle B_{\mathbf{q},j}(0)A_{\mathbf{q},l}(t)\rangle$:
\begin{align*}
    \langle B_{\mathbf{q},j}(0)A_{\mathbf{q},l}(t)\rangle = -\frac{1}{2\pi}\frac{1}{\omega _{\mathbf{q},j}}\int ^{\infty}_{-\infty} e^{\beta\Omega}\Omega J^{AA}_{\mathbf{q},l,j}(\Omega)e^{i\Omega t}\text{d}\Omega.
\end{align*}
Finally, taking the derivative once more of the operator $A$, we obtain the correlation function for $\langle B_{\mathbf{q},j}(0)B_{\mathbf{q},l}(t)\rangle$:
\begin{align*}
    \langle B_{\mathbf{q},j}(0)B_{-\mathbf{q},l}(t)\rangle = -\frac{1}{2\pi}\frac{1}{\omega _{\mathbf{q},j}\omega _{\mathbf{q},l}}\int ^{\infty}_{-\infty} e^{\beta\Omega}\Omega ^2 J^{AA}_{\mathbf{q},l,j}(\Omega)e^{i\Omega t}\text{d}\Omega.
\end{align*}
Substituting these expressions into the definition of the heat current, we end up with two integrals that we can evaluate analytically. The first one:
\begin{align*}
    \frac{1}{\beta}\int ^{\beta}_{0}e^{(\Omega _1 - \Omega _2)s} \text{d}s = \frac{e^{(\Omega _1 - \Omega _2)\beta} - 1}{(\Omega _1 - \Omega _2)\beta},
\end{align*}
and the second one over time (we are keeping the driving frequency ($\nu$)):
\begin{align*}
    \int ^{\infty} _{0} e^{-i(\Omega _1 - (\Omega _2 + \nu))t} = i\lim _{\epsilon \rightarrow 0}\frac{1}{(\Omega _1 - (\Omega _2 + \nu)) + i\epsilon} = \mathcal{P}\frac{i}{(\Omega _1 - (\Omega _2 + \nu))} + \pi\delta(\Omega _1 - (\Omega _2 + \nu)).\numberthis
    \label{eq:identity}
\end{align*}
We are only interested in the real part of the expression. The Dirac delta function will annihilate one of the frequency integrals that come from the Fourier transform of the correlation function and leave us with the final expression for the dynamical lattice thermal conductivity (we cast correlation functions in terms of phonon spectral functions):
\begin{align*}
    \kappa ^{xy} (\nu) = \frac{\pi\beta ^2k_{B}}{NV}\sum _{\mathbf{q},j,j'}v^{x}_{\mathbf{q},j,j'}v^{y*}_{\mathbf{q},j,j'}\omega _{\mathbf{q},j}\omega _{\mathbf{q},j'}\frac{e^{\beta\nu} - 1}{\beta\nu}\int_{-\infty}^{\infty}\textrm{d}\Omega (1+\frac{\Omega}{\Omega + \nu})\frac{e^{\beta\Omega}}{\left(e^{\beta\Omega} - 1\right)\left(e^{\beta(\Omega + \nu)} - 1\right)}\sigma _{\mathbf{q},j'}(\Omega)\sigma_{\mathbf{q},j}(\Omega + \nu). \numberthis
    \label{eq:dyn_kappa1}
\end{align*}
Obviously, in the limit $\nu \rightarrow 0$ we obtain  Eq. (20) in the main part. Strictly, this is the real part of the lattice thermal conductivity because of Eq.~\ref{eq:identity}. The imaginary part of $\kappa$ would be Kramer-Kronig's transformation of the real part.

\newpage

\section{Validity of no mode-mixing approximation}

\begin{figure}
	\centering
	\includegraphics[width=0.5\linewidth]{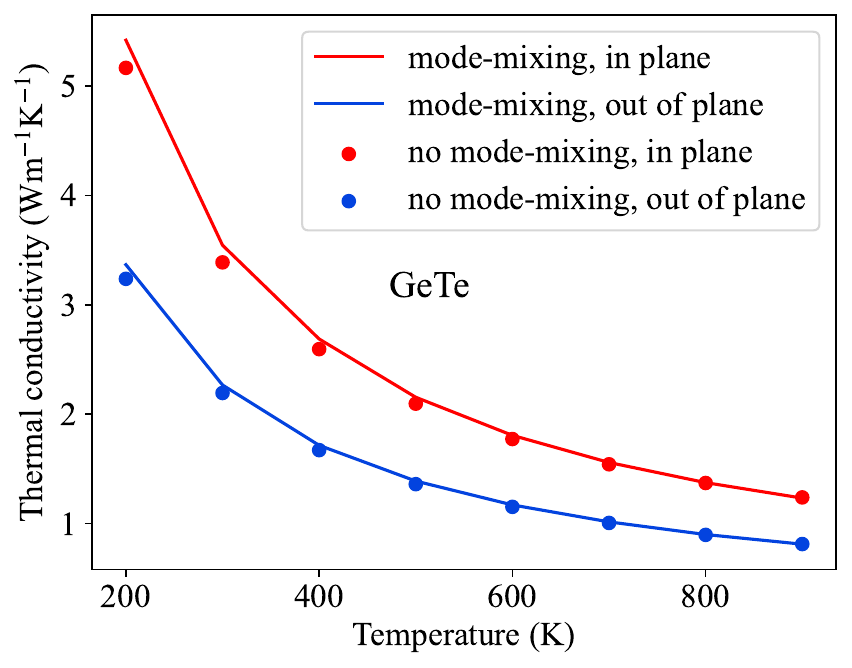}
	\caption{Lattice thermal conductivity of GeTe calculated with Green-Kubo method in mode-mixing (Eq. 4) and no mode-mixing (Eq. 6) approximation.}
	\label{fig:kappa_mm}
\end{figure}

We checked the applicability of the no mode-mixing approximation in the case of Germanium Telluride, a highly anharmonic material. We calculated second and third-order SSCHA force constants at 300 K for a structure that is a minimum of SSCHA free energy at that temperature. To calculate energies, forces, and stresses we used a machine learning potential developed in one of our previous works~\cite{GeTe,myMD}. Results for the lattice thermal conductivity at different temperatures are shown in Figure~\ref{fig:kappa_mm}. We can see that the differences between mode-mixing and no mode-mixing results are minimal.  

\newpage

\subsection{Perturbative limit for the lattice thermal conductivity}

\subsubsection{Diagonal terms}

Let us show the perturbative limit for the diagonal part of the lattice thermal conductivity. In the perturbative limit we assume that the real part of the self-energy does not have a shift, i.e. $\mathrm{Re}\mathcal{Z}_{\mathbf{q},j} (\Omega) = \omega _{\mathbf{q},j}$. Then the imaginary part of the self-energy is constant with frequency $\mathrm{Im}\mathcal{Z}_{\mathbf{q},j} (\Omega) = \Gamma _{\mathbf{q},j}$. In that case the phonon spectral function is simple Lorentzian:
\begin{align*}
    \sigma _{\mathbf{q},j} (\Omega) = \frac{1}{2\pi}\left(\frac{-\Gamma _{\mathbf{q},j}}{(\Omega - \omega _{\mathbf{q},j})^2 + \Gamma ^2 _{\mathbf{q},j}} + \frac{\Gamma _{\mathbf{q},j}}{(\Omega + \omega _{\mathbf{q},j})^2 + \Gamma ^2 _{\mathbf{q},j}}\right). \numberthis
    \label{LA_spec_func} 
\end{align*}

This function strongly peaks in $\pm \omega _{\mathbf{q},j}$ and we can approximate it with Dirac delta function ($\sigma _{\mathbf{q},j} (\Omega) = \frac{1}{2}\delta (\Omega - \omega _{\mathbf{q},j})$). We can substitute this in the definition of the lattice thermal conductivity (Eq. 20. of the main part) to obtain:
\begin{align*}
    \kappa ^{xy} = \frac{\pi\beta ^2k_{B}}{NV}\sum _{\mathbf{q},j}v^{x}_{\mathbf{q},j}v^{y}_{\mathbf{q},j}\omega ^2 _{\mathbf{q},j}\int_{-\infty}^{\infty}\textrm{d}\Omega \frac{\exp(\beta\Omega)}{\left(\exp(\beta\Omega) - 1\right)^2}\delta (\Omega - \omega _{\mathbf{q},j})\sigma_{\mathbf{q},j}(\Omega).\numberthis
\end{align*}
The Dirac delta function will annihilate the integral over $\Omega$ and replace all $\Omega$ with $\omega _{\mathbf{q},j}$. This substitution will simplify the spectral function to:
\begin{align*}
    \sigma _{\mathbf{q},j} (\omega _{\mathbf{q},j}) = \frac{1}{2\pi}\left(\frac{-\Gamma _{\mathbf{q},j}}{\Gamma ^2 _{\mathbf{q},j}} + \frac{\Gamma _{\mathbf{q},j}}{4\omega  ^2 _{\mathbf{q},j} + \Gamma ^2  _{\mathbf{q},j}}\right).
\end{align*}

In case the anharmonicity is small the second term in the brackets will be much smaller than the first which leaves us with the final result for the diagonal term of the lattice thermal conductivity in the perturbative limit (Eq. 22 of the main part):
\begin{align*}
    \kappa ^{xy} = \frac{\beta ^2k_{B}}{NV}\sum _{\mathbf{q},j}v^{x}_{\mathbf{q},j}v^{y}_{\mathbf{q},j}\omega ^2 _{\mathbf{q},j} \frac{\exp(\beta\omega _{\mathbf{q},j})}{\left(\exp(\beta\omega _{\mathbf{q},j}) - 1\right)^2}\frac{-1}{2\Gamma _{\mathbf{q},j}}
    .\numberthis
\end{align*}

\subsubsection{Off-diagonal terms}

Here we will show how to get the perturbative limit for the off-diagonal terms of the lattice thermal conductivity starting with Eq.~(21) in the main part. First we will assume that in the limit of small anharmonicity $\Gamma _{\mathbf{q},j} = \mathrm{Im}\mathcal{Z}_{\mathbf{q},j} (\Omega)$ and $\omega _{\mathbf{q},j} = \mathrm{Re}\mathcal{Z}_{\mathbf{q},j} (\Omega)$. This simplifies the definition of the phonon spectral function to:
\begin{align*}
    \sigma _{\mathbf{q},j} (\Omega) = \frac{1}{2\pi}\left(\frac{-\Gamma _{\mathbf{q},j}}{(\Omega - \omega _{\mathbf{q},j})^2 + \Gamma ^2 _{\mathbf{q},j}} + \frac{\Gamma _{\mathbf{q},j}}{(\Omega + \omega _{\mathbf{q},j})^2 + \Gamma ^2 _{\mathbf{q},j}}\right).
\end{align*}
Substituting this expression for the phonon spectral function in Eq.~(20) we come to four terms inside the $\Omega$ integral. We will look at each of them separately. 

The first two terms involve parts of spectral functions with the pole in the same part of $\Omega$ number line (either both poles are in the positive or negative part). We show the calculation for the poles in the positive part of $\Omega$ and the second case is analogous. The $\Omega$ integral in this case is:
\begin{align*}
    \frac{1}{4\pi^2}\int _{-\infty}^{\infty}\frac{-\Gamma _{\mathbf{q},j}}{(\Omega - \omega _{\mathbf{q},j})^2 + \Gamma ^2 _{\mathbf{q},j}}\frac{-\Gamma _{\mathbf{q},j'}}{(\Omega - \omega _{\mathbf{q},j'})^2 + \Gamma ^2 _{\mathbf{q},j'}}\frac{\exp{(\beta\Omega)}}{(\exp{(\beta\Omega)} - 1)^2}\mathrm{d}\Omega.\numberthis
    \label{integral}
\end{align*}
We now introduce the substitution $\Omega' = \Omega - \omega _{\mathbf{q},j}$, which changes the integral to:
\begin{align*}
    \frac{1}{4\pi^2}\int _{-\infty}^{\infty}\frac{\Gamma _{\mathbf{q},j}}{\Omega'^2 + \Gamma ^2 _{\mathbf{q},j}}\frac{\Gamma _{\mathbf{q},j'}}{(\Omega' + \omega _{\mathbf{q},j} - \omega _{\mathbf{q},j'})^2 + \Gamma ^2 _{\mathbf{q},j'}}\frac{\exp{(\beta(\Omega' + \omega _{\mathbf{q},j}))}}{(\exp{(\beta(\Omega' + \omega _{\mathbf{q},j}))} - 1)^2}\mathrm{d}\Omega'.
\end{align*}
From here we recognize that the above integral is the convolution of two Lorentzians aside from the multiplicative term involving the exponential functions. We approximate this exponential part for the value $\Omega' = 0$, meaning we take it in the limit of auxiliary phonon frequency $\omega _{\mathbf{q},j}$. Since this part is now constant it is taken in front of the integral and we evaluate the convolution analytically to obtain:
\begin{align*}
    \frac{1}{4\pi}\frac{\exp{(\beta\omega _{\mathbf{q},j})}}{(\exp{(\beta\omega _{\mathbf{q},j})} - 1)^2}\frac{\Gamma _{\mathbf{q},j} + \Gamma _{\mathbf{q},j'}}{(\omega _{\mathbf{q},j} - \omega _{\mathbf{q},j'})^2 + (\Gamma _{\mathbf{q},j} + \Gamma _{\mathbf{q},j'})^2}.
\end{align*}
We have arbitiraly decided to perform the substitution $\Omega' = \Omega - \omega _{\mathbf{q},j}$. An equally valid one would be $\Omega' = \Omega - \omega _{\mathbf{q},j'}$. In that case the only difference would be in the exponential part, which would be evaluated at $\omega _{\mathbf{q},j'}$ instead of $\omega _{\mathbf{q},j}$. To account for this we will average between these two options to obtain the final result:
\begin{align*}
    \frac{1}{2}\frac{1}{4\pi}\left(\frac{\exp{(\beta\omega _{\mathbf{q},j})}}{(\exp{(\beta\omega _{\mathbf{q},j})} - 1)^2} + \frac{\exp{(\beta\omega _{\mathbf{q},j'})}}{(\exp{(\beta\omega _{\mathbf{q},j'})} - 1)^2}\right)\frac{\Gamma _{\mathbf{q},j} + \Gamma _{\mathbf{q},j'}}{(\omega _{\mathbf{q},j} - \omega _{\mathbf{q},j'})^2 + (\Gamma _{\mathbf{q},j} + \Gamma _{\mathbf{q},j'})^2}.
\end{align*}
The result follows the same for the product of spectral functions with a positive pole part. This makes the factor of $\frac{1}{2}$ of the above equation to disappear. 

The second two terms involve the product of parts of the phonon spectral function with poles of different signs. We will show the result for one of the terms and the second one will follow the same process. We start with the product of spectral function parts like in the previous case:
\begin{align*}
    \frac{1}{4\pi^2}\int _{-\infty}^{\infty}\frac{-\Gamma _{\mathbf{q},j}}{(\Omega - \omega _{\mathbf{q},j})^2 + \Gamma ^2 _{\mathbf{q},j}}\frac{\Gamma _{\mathbf{q},j'}}{(\Omega + \omega _{\mathbf{q},j'})^2 + \Gamma ^2 _{\mathbf{q},j'}}\frac{\exp{(\beta\Omega)}}{(\exp{(\beta\Omega)} - 1)^2}\mathrm{d}\Omega.
\end{align*}
We apply the same substitution $\Omega' = \Omega - \omega _{\mathbf{q},j}$ and obtain:
\begin{align*}
    -\frac{1}{4\pi^2}\int _{-\infty}^{\infty}\frac{\Gamma _{\mathbf{q},j}}{\Omega'^2 + \Gamma ^2 _{\mathbf{q},j}}\frac{\Gamma _{\mathbf{q},j'}}{(\Omega' - \omega _{\mathbf{q},j} - \omega _{\mathbf{q},j'})^2 + \Gamma ^2 _{\mathbf{q},j'}}\frac{\exp{(\beta(\Omega' + \omega _{\mathbf{q},j}))}}{(\exp{(\beta(\Omega' + \omega _{\mathbf{q},j}))} - 1)^2}\mathrm{d}\Omega'.
\end{align*}
We again pick the exponential part only at $\omega _{\mathbf{q},j}$. This is again the convolution of  Lorentzian functions, which can be evaluated analytically, and gives:
\begin{align*}
    \frac{1}{4\pi}\frac{\exp{(\beta\omega _{\mathbf{q},j})}}{(\exp{(\beta\omega _{\mathbf{q},j})} - 1)^2}\frac{\Gamma _{\mathbf{q},j} + \Gamma _{\mathbf{q},j'}}{(\omega _{\mathbf{q},j} + \omega _{\mathbf{q},j'})^2 + (\Gamma _{\mathbf{q},j} + \Gamma _{\mathbf{q},j'})^2}.
\end{align*}
Again our substitution is arbitrary and we could have chosen $\Omega' = \Omega + \omega _{\mathbf{q},j'}$. As in the previous case, we will average between these two values:
\begin{align*}
    \frac{1}{4\pi}\left(\frac{\exp{(\beta\omega _{\mathbf{q},j})}}{(\exp{(\beta\omega _{\mathbf{q},j})} - 1)^2} + \frac{\exp{(\beta\omega _{\mathbf{q},j'})}}{(\exp{(\beta\omega _{\mathbf{q},j'})} - 1)^2}\right)\frac{\Gamma _{\mathbf{q},j} + \Gamma _{\mathbf{q},j'}}{(\omega _{\mathbf{q},j} + \omega _{\mathbf{q},j'})^2 + (\Gamma _{\mathbf{q},j} + \Gamma _{\mathbf{q},j'})^2}.
\end{align*}
We now substitute these terms inside the expression for the lattice thermal conductivity in Eq.~(20). We obtain two distinct terms:
\begin{align*}
    \kappa _{R}^{xy} &= \frac{1}{2NV}\sum _{\mathbf{q},j,j'}v^{x}_{\mathbf{q},j,j'}v^{y*}_{\mathbf{q},j,j'}\left(\omega _{\mathbf{q},j'}\frac{c_{\mathbf{q},j}}{\omega _{\mathbf{q},j}} + \omega _{\mathbf{q},j}\frac{c_{\mathbf{q},j'}}{\omega _{\mathbf{q},j'}}\right)\frac{\Gamma _{\mathbf{q},j} + \Gamma _{\mathbf{q},j'}}{(\omega _{\mathbf{q},j} - \omega _{\mathbf{q},j'})^2 + (\Gamma _{\mathbf{q},j} + \Gamma _{\mathbf{q},j'})^2},\\
    \kappa _{A}^{xy} &= \frac{1}{2NV}\sum _{\mathbf{q},j,j'}v^{x}_{\mathbf{q},j,j'}v^{y*}_{\mathbf{q},j,j'}\left(\omega _{\mathbf{q},j'}\frac{c_{\mathbf{q},j}}{\omega _{\mathbf{q},j}} + \omega _{\mathbf{q},j}\frac{c_{\mathbf{q},j'}}{\omega _{\mathbf{q},j'}}\right)\frac{\Gamma _{\mathbf{q},j} + \Gamma _{\mathbf{q},j'}}{(\omega _{\mathbf{q},j} + \omega _{\mathbf{q},j'})^2 + (\Gamma _{\mathbf{q},j} + \Gamma _{\mathbf{q},j'})^2}.
\end{align*}
These two terms are analogous to the resonant and antiresonant terms in Ref.~\cite{Isaeva2019,Caldarelli}. They are very similar to the ones obtained in Caldarelli et al.~\cite{Caldarelli}. The main reason for the discrepancy is the approximation scheme used to evaluate integrals of the type featured in Supp. Eq.~\ref{integral}. To show that, we go back to the definition of the lattice thermal conductivity:
\begin{align*}
    \kappa ^{xy} = \frac{2\pi\beta ^2k_{B}}{NV}\sum _{\mathbf{q},j,j'}v^{x}_{\mathbf{q},j,j'}v^{y*}_{\mathbf{q},j,j'}\omega _{\mathbf{q},j}\omega _{\mathbf{q},j'}\int_{-\infty}^{\infty}\textrm{d}\Omega \frac{\exp(\beta\Omega)}{\left(\exp(\beta\Omega) - 1\right)^2}\sigma _{\mathbf{q},j}(\Omega)\sigma_{\mathbf{q},j'}(\Omega)
\end{align*}
Next, we substitute the definition of the Wigner group velocities~\cite{Caldarelli}:
\begin{align*}
    v^{x}_{\mathbf{q},j,j'} = \frac{\omega _{\mathbf{q},j} + \omega _{\mathbf{q},j'}}{2\sqrt{\omega _{\mathbf{q},j}\omega _{\mathbf{q},j'}}}\mathrm{v}^{x}_{\mathbf{q},j,j'}
\end{align*}
and obtain:
\begin{align*}
    \kappa ^{xy} = \frac{2\pi\beta ^2k_{B}}{NV}\sum _{\mathbf{q},j,j'}\frac{(\omega_{\mathbf{q}, j} + \omega_{\mathbf{q}, j'})^2}{4}\mathrm{v}^{x}_{\mathbf{q},j,j'}\mathrm{v}^{y*}_{\mathbf{q},j,j'}\int_{-\infty}^{\infty}\textrm{d}\Omega \frac{\exp(\beta\Omega)}{\left(\exp(\beta\Omega) - 1\right)^2}\sigma _{\mathbf{q},j}(\Omega)\sigma_{\mathbf{q},j'}(\Omega)
\end{align*}
Next, we take one of the sums of frequencies and multiply it with the integral. Then we pick up the value of the exponential part ($\frac{\exp(\beta\Omega)}{\left(\exp(\beta\Omega) - 1\right)^2}$) at the frequency that we multiplied it with, to obtain:
\begin{align*}
    \kappa ^{xy} = \frac{2\pi\beta ^2k_{B}}{NV}\sum _{\mathbf{q},j,j'}\frac{(\omega_{\mathbf{q}, j} + \omega_{\mathbf{q}, j'})}{4}\mathrm{v}^{x}_{\mathbf{q},j,j'}\mathrm{v}^{y*}_{\mathbf{q},j,j'}\left(\frac{\omega _{\mathbf{q}, j}\exp(\beta\omega _{\mathbf{q}, j})}{\left(\exp(\beta\omega _{\mathbf{q}, j}) - 1\right)^2} + \frac{\omega _{\mathbf{q}, j'}\exp(\beta\omega _{\mathbf{q}, j'})}{\left(\exp(\beta\omega _{\mathbf{q}, j'}) - 1\right)^2}\right)\int_{-\infty}^{\infty}\textrm{d}\Omega \sigma _{\mathbf{q},j}(\Omega)\sigma_{\mathbf{q},j'}(\Omega).
\end{align*}
The evaluation of the integral follows the same as in our case and introducing the definition of the phonon mode heat capacity we obtain for the resonant part the same expression as in Refs.~\cite{Caldarelli, Simoncelli2019}:
\begin{align*}
    \kappa _{R}^{xy} &= \frac{1}{NV}\sum _{\mathbf{q},j,j'}\frac{(\omega_{\mathbf{q}, j} + \omega_{\mathbf{q}, j'})}{4}\mathrm{v}^{x}_{\mathbf{q},j,j'}\mathrm{v}^{y*}_{\mathbf{q},j,j'}\left(\frac{c_{\mathbf{q},j}}{\omega _{\mathbf{q},j}} + \frac{c_{\mathbf{q},j'}}{\omega _{\mathbf{q},j'}}\right)\frac{\Gamma _{\mathbf{q},j} + \Gamma _{\mathbf{q},j'}}{(\omega _{\mathbf{q},j} - \omega _{\mathbf{q},j'})^2 + (\Gamma _{\mathbf{q},j} + \Gamma _{\mathbf{q},j'})^2}.
\end{align*}
Here we are labeling the imaginary part of the self-energy as $ \Gamma _{\mathbf{q},j}$, which is different from Ref.~\cite{Caldarelli} (there it was labeled as $\gamma _{\mathbf{q},j}$). Clearly, the different results between the current work and the previous one~\cite{Caldarelli}, come from the different ways of approximating the exponential part inside of the integral in Supp. Eq.~\ref{integral}. However, as we mentioned in the main part, in the limit $\omega _{\mathbf{q}, j} = \omega _{\mathbf{q}, j'}$ results are identical as one would expect.

\begin{figure}[h!]
    \centering
    \includegraphics[width=0.85\textwidth]{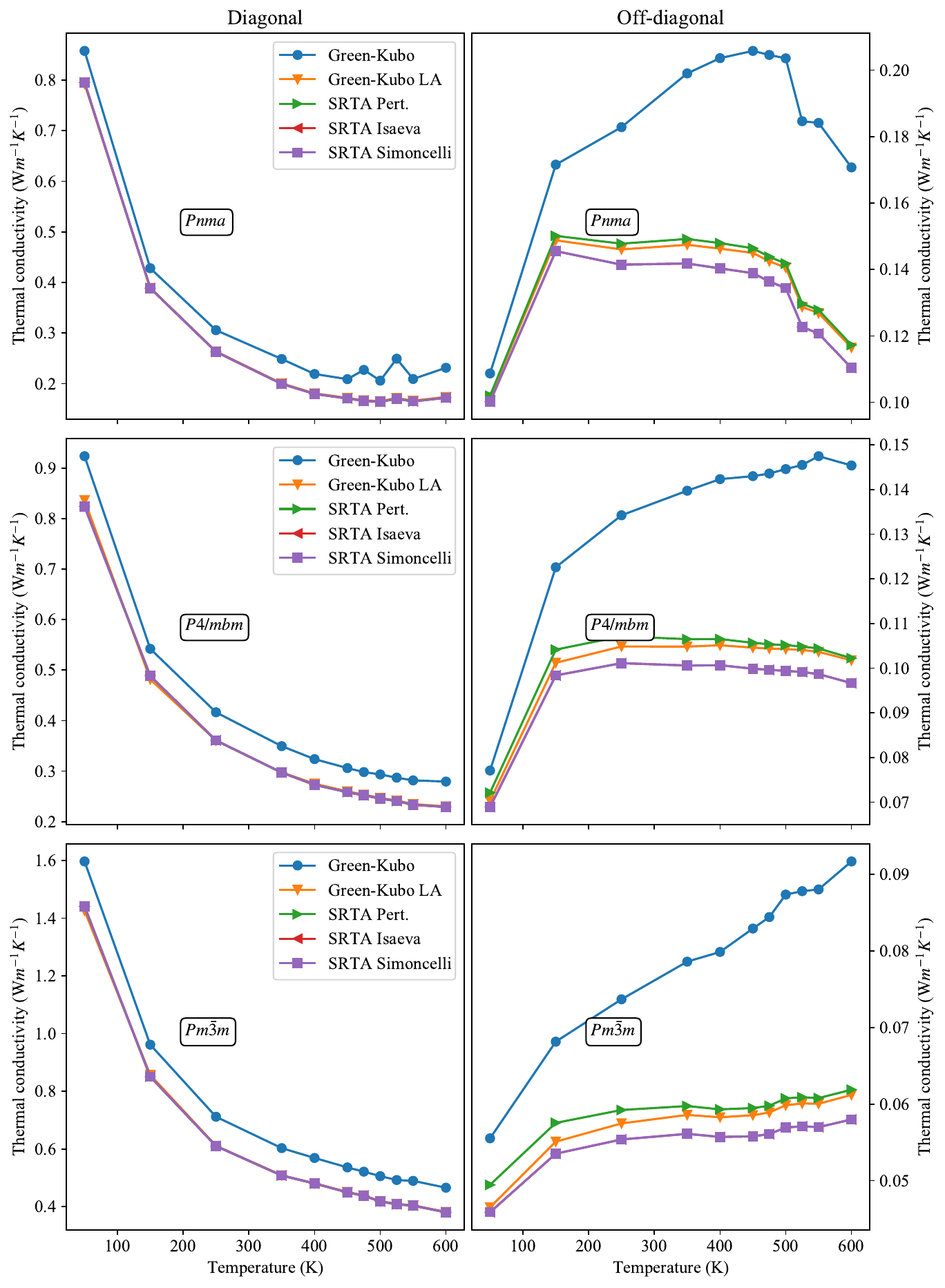}
    \caption{Comparing different approaches for the diagonal and off-diagonal part of the lattice thermal conductivity.}
    \label{fig:compare-offdiag}
\end{figure}

In Supp. Fig.~\ref{fig:compare-offdiag} we compare the numerical results for different definitions of the off-diagonal part of the lattice thermal conductivity. Green-Kubo results stand for calculations done with the Green-Kubo approach using proper anharmonic spectral functions (as defined in Eq. 21 of the main text). Green-Kubo LA refers to the calculations done with the Green-Kubo method, but using phonon spectral functions as defined in Supp. Eq.~\ref{LA_spec_func}. SRTA Isaeva results stand for results obtained using equations from Ref.~\cite{Isaeva2019}, while SRTA Simoncelli results were obtained using equations from Refs.~\cite{Simoncelli2019, Caldarelli}. Finally, SRTA Pert. results were calculated using equations obtained in the perturbative limit of the Green-Kubo result, Eqs. 22 and 24 of the main part.

Results for the diagonal part of the lattice thermal conductivity are identical for all approaches, except for the full Green-Kubo. The discrepancy involving Green-Kubo can be explained by the fact it is the only approach that accounts for the softening of phonon modes due to anharmonicity.

For the off-diagonal contribution, we find that equations from Refs.~\cite{Isaeva2019,Simoncelli2019} give identical results. These results are not far away from our perturbative limit derived in this section. The perturbative limit follows more closely the Green-Kubo LA approximation compared to approaches from mentioned references. Finally, the full Green-Kubo method is the most different, and again it is probably the consequence of the softening of the phonon mode frequencies that are captured with this method.

\newpage

\subsection{Phonon properties at 250 K}

\begin{figure}
    \centering
    \includegraphics[width=0.9\textwidth]{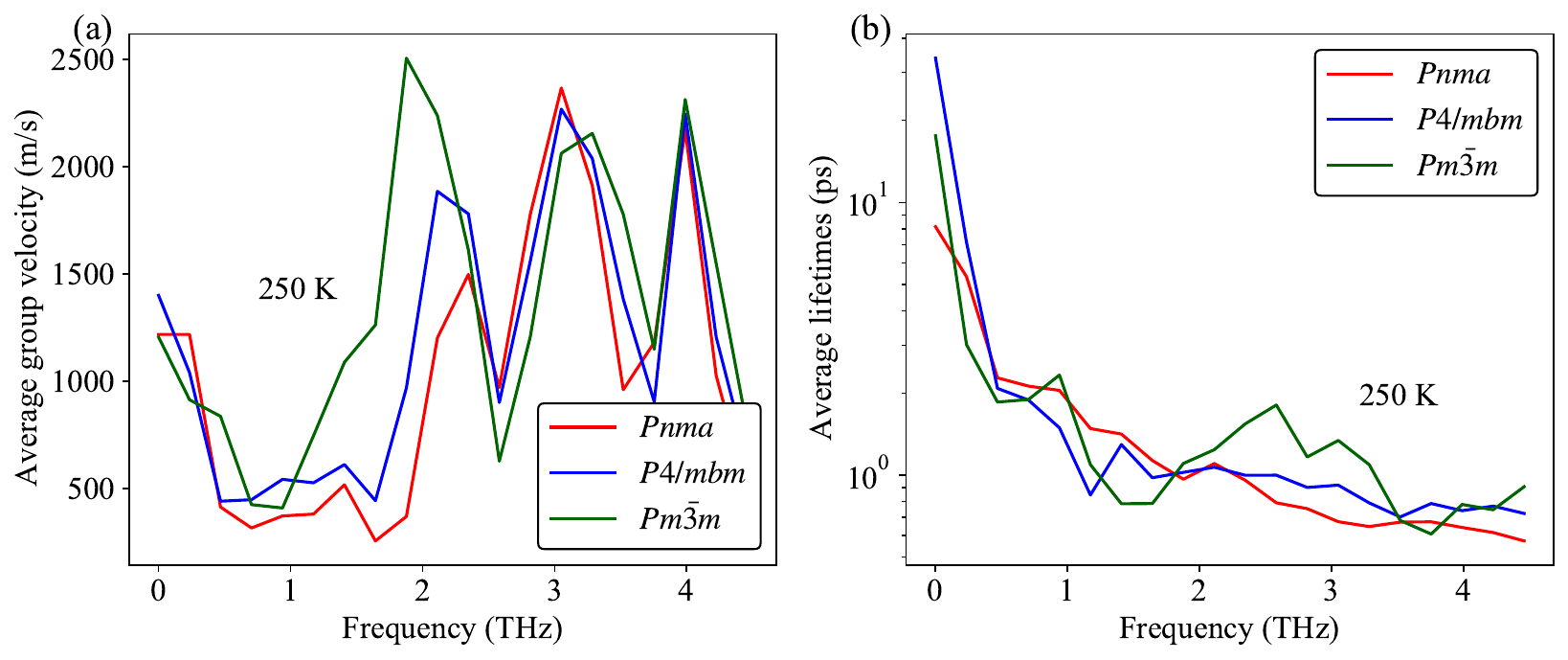}
    \caption{(a) Averaged phonon group velocities by frequency bins in CsPbBr$_3$ for different phases. (b) Averaged phonon lifetimes by frequency bins in CsPbBr$_3$ for different phases.}
    \label{fig:binned}
\end{figure}

In order to elucidate the origin of the hierarchy of the lattice thermal conductivity for different phases at a given temperature, we check phonon properties at 250 K. Although only the orthorhombic phase is stable at this temperature, we can still calculate the lattice thermal conductivity in all phases. This is possible since SSCHA auxiliary phonon frequencies are always positive by definition. The results for the other two, unstable, phases are not physically relevant but will reveal the causes for the different values of $\kappa$ in different phases. In Supp. Fig.~\ref{fig:binned} (a) we show the norm of the phonon group velocities for different phases at 250 K. To make a comparison between different phases easier we bin phonon modes by frequencies in 15 bins and then average the phonon group velocity inside the bin. The largest group velocities are in $Pm\bar{3}m$ cubic phase which is the main reason for the largest lattice thermal conductivity of this phase. However, overall group velocities are quite similar for different phases and the increase in the cubic phase is only visible in a particular frequency range (1.0-2.0 THz).

In Supp. Fig.~\ref{fig:binned} (b) we show the average phonon lifetimes for different phases of CsPbBr$_3$. As for phonon group velocities, we average phonon lifetimes inside distinct frequency bins. In the low-frequency limit (up to 0.5 THz), the $P4/mbm$ tetragonal phase surprisingly has the highest phonon lifetimes. Between 0.5-1.8 THz the $Pnma$ phase has the highest phonon lifetimes, while above 1.8 THz $Pm\bar{3}m$ phase phonon lifetimes dominate. The hierarchy of phonon lifetimes for different phases explains why the $Pnma$ phase has the lowest lattice thermal conductivity at any given temperature.

\begin{figure}
    \centering
    \includegraphics[width=0.9\textwidth]{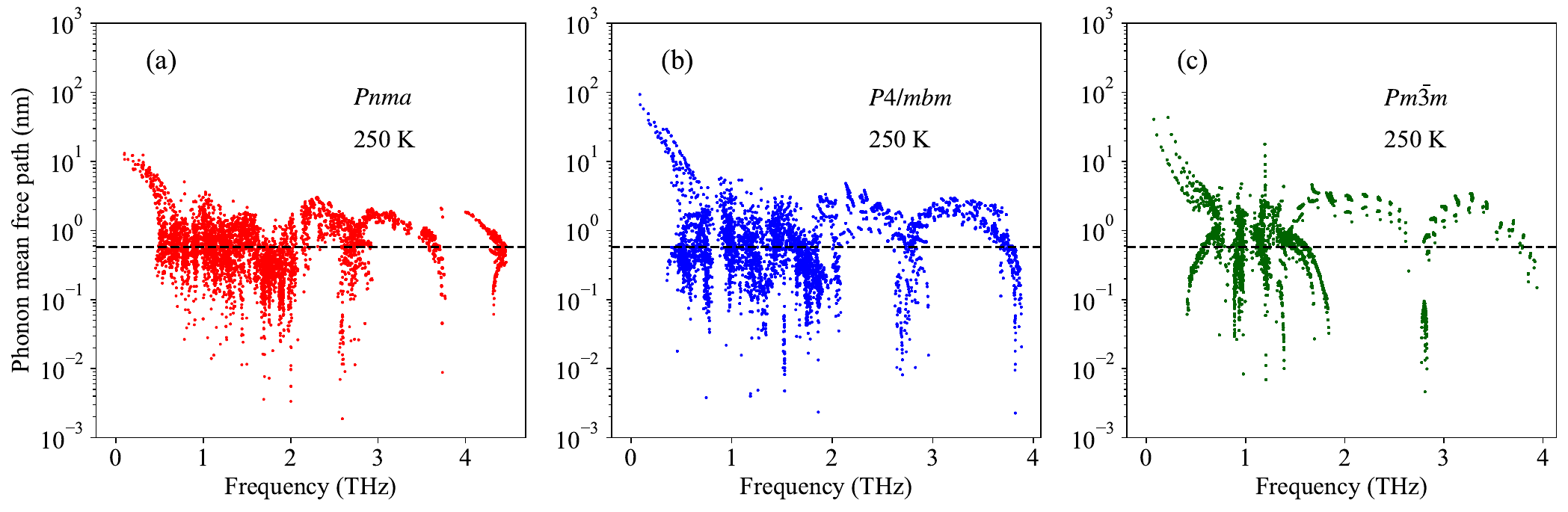}
    \caption{Mean free path of phonons calculated for (a) $Pnma$, (b) $P4/mbm$ and (c) $Pm\bar{3}m$ phase at 250 K. The dashed line represents the characteristic lattice constant at that temperature.} 
    \label{fig:mfps}
\end{figure}

In Supp. Fig.~\ref{fig:mfps} we present the mean free path of phonon for all three phases at 250 K. Almost all phonon modes (except the low-lying acoustic modes in $P4/mbm$ and $Pm\bar{3}m$ phases) have mean free path less than 100 nm which is the size of the nanowires synthesized in experiments that measured the lattice thermal conductivity~\cite{CsPbBr3_kappa1, CsPbBr_kappa2}. A large number of phonon modes in all phases has a mean free path much smaller than the lattice constant at that temperature, indicating the possibility that the coherent contribution to the transport might be important.

\begin{figure}
    \centering
    \includegraphics[width=0.9\textwidth]{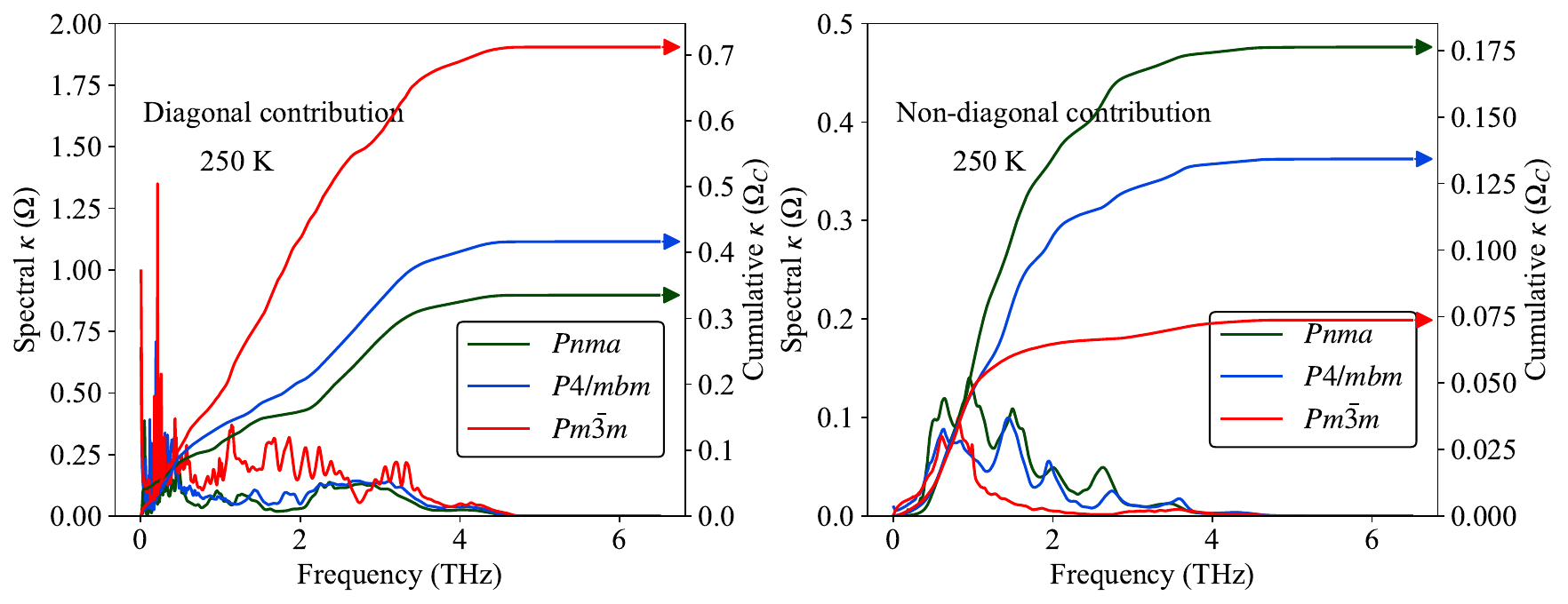}
    \caption{(a) Spectral decomposition of the diagonal part of the lattice thermal conductivity for $xx$ component calculated with Green-Kubo method. (a) Spectral decomposition of the non-diagonal part of the lattice thermal conductivity for $xx$ component calculated with Green-Kubo method.}
    \label{fig:spec_kappa}
\end{figure}

Analyzing the results for the Green-Kubo method is more challenging since we do not have a well-defined transport quantity such as phonon lifetimes. For this reason, we are showing the spectral decomposition of the lattice thermal conductivity in Supp. Fig.~\ref{fig:spec_kappa}. The spectral lattice thermal conductivity is given by:
\begin{align}
    \kappa ^{xy} (\Omega) = \frac{4\pi\beta ^2k_{B}}{NV}\sum _{\mathbf{q},j,j'}v^{x}_{\mathbf{q},j,j'}v^{y*}_{\mathbf{q},j,j'}\omega _{\mathbf{q},j}\omega _{\mathbf{q},j'} \frac{\exp(\beta\Omega)}{\left(\exp(\beta\Omega) - 1\right)^2}\sigma _{\mathbf{q},j}(\Omega)\sigma_{\mathbf{q},j'}(\Omega).
    \label{eq:spec_kappa}
\end{align}
Importantly, this frequency dependence is not the same as in the dynamical lattice thermal conductivity and this quantity can be defined for any $\nu$ of the dynamical lattice thermal conductivity. Integrating $\kappa (\Omega)$ over the positive frequency range will give us the lattice thermal conductivity~\footnote{To obtain the expression Eq.~\ref{eq:spec_kappa}  we used the fact that phonon spectral function is even with respect to $\Omega$.}. Sum over phonon branches $j,j'$ can be split into diagonal $j=j'$ and non-diagonal  $j\neq j'$ parts. Additionally, we can define the cumulative $\kappa$:

\begin{align}
    \kappa ^{xy} (\Omega _C) = \frac{4\pi\beta ^2k_{B}}{NV}\sum _{\mathbf{q},j,j'}v^{x}_{\mathbf{q},j,j'}v^{y*}_{\mathbf{q},j,j'}\omega _{\mathbf{q},j}\omega _{\mathbf{q},j'} \int ^{\Omega _C}_{0}\text{d}\Omega\frac{\exp(\beta\Omega)}{\left(\exp(\beta\Omega) - 1\right)^2}\sigma _{\mathbf{q},j}(\Omega)\sigma_{\mathbf{q},j'}(\Omega). 
\end{align}

The spectral decomposition reveals that all modes in CsPbBr$_3$ contribute to the lattice thermal conductivity, without a frequency range that dominates. Another important fact is that for cubic and tetragonal phases we see a divergence of the spectral $\kappa$ for low frequencies. This comes from the strong softening of phonon modes at this temperature, which makes the phonon spectral functions non-zero for $\Omega$ close to zero. The divergence comes from the exponential function. This situation is unphysical and the numbers we get for these phases at 250 K are not really good estimates for the  lattice thermal conductivity. However, we already know that the tetragonal and cubic phases are not dynamically stable at this temperature and this calculation is purely an exercise. 

\newpage
\subsection{Stability of the orthorhombic $Pnma$ phase}

\begin{figure}
    \centering
    \includegraphics[width=0.9\textwidth]{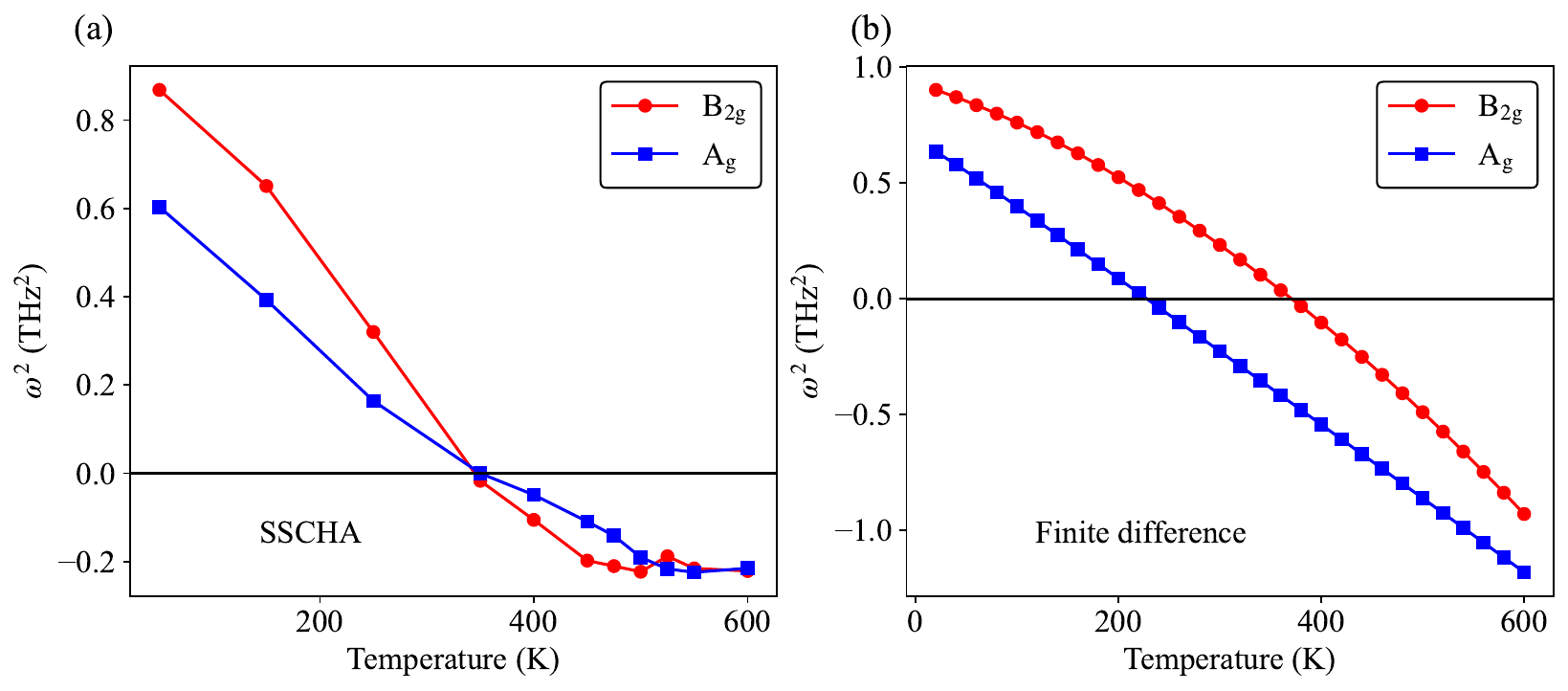}
    \caption{(a) Temperature dependence of the Hessian frequency of soft modes in orthorhombic $Pnma$ phase of CsPbBr$_3$ calculated with SSCHA. (b) Temperature dependence of the Hessian frequency of soft modes in orthorhombic $Pnma$ phase of CsPbBr$_3$ calculated with finite difference force constants.}
    \label{fig:ortho_hess}
\end{figure}

In Supp. Fig.~\ref{fig:ortho_hess} we are showing the temperature dependence of the Hessian frequency of two soft modes for the orthorhombic $Pnma$ phase of CsPbBr$_3$. This calculation includes also the third-order anharmonicity calculated by finite difference with the statical dressed bubble approximation. At temperatures around 300 K, the square of the frequency of these modes becomes negative indicating possible dynamical instability. A$_\mathrm{g}$ mode is symmetry preserving mode, while B$_{2\mathrm{g}}$ mode drives the system to the monoclinic $P2_1/m$ structure. Nevertheless, it can also be a signature of gaining symmetry, meaning that the softening of one of the modes indicates a second-order phase transition to the tetragonal phase.

In Supp. Fig.~\ref{fig:ortho_hess} (a) we are showing results obtained fully with SSCHA. This means we used SSCHA temperature-dependent force constants and changed the structure with temperature. In Supp. Fig.~\ref{fig:ortho_hess} (b) however, we used finite difference force constants obtained for the structure that minimizes Born-Oppenheimer energy surface. Using SSCHA delays the phase transition for about 100 K, but can not ultimately suppress it. Since we know that $Pnma$ should be stable at these temperatures, the reason for this softening of particular phonon modes at unexpectedly low temperatures is the fact that we are not including $\stackrel{(4)}{\boldsymbol{\mathcal{D}}}_{\mathcal{R}}$ for this phase.

\newpage

\subsection{Stability of the cubic $Pm\bar{3}m$ phase}

\begin{figure}
    \centering
    \includegraphics[width=0.6\textwidth]{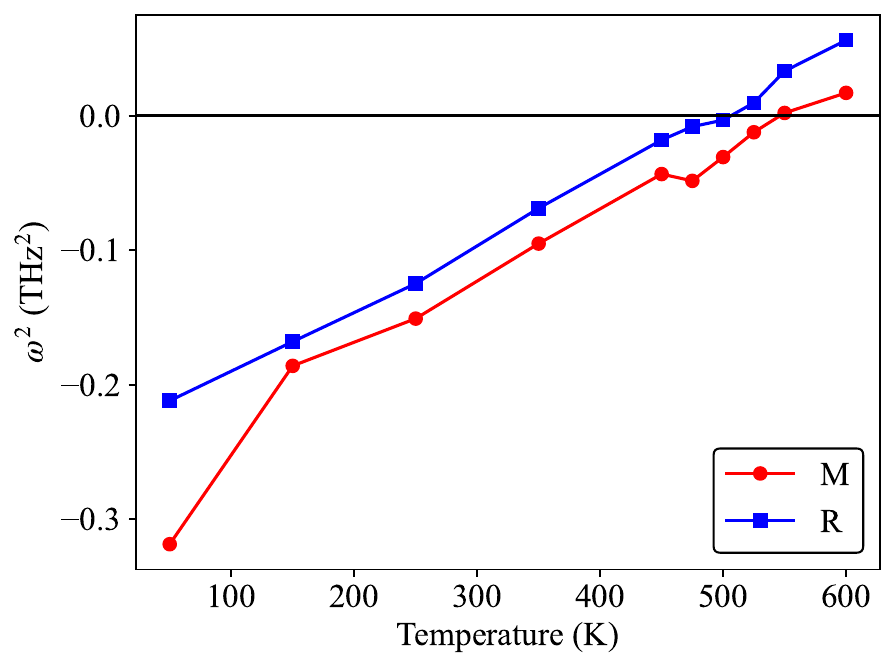}
    \caption{Temperature dependence of the Hessian frequency of soft modes in cubic $Pm\bar{3}m$ phase of CsPbBr$_3$ calculated with SSCHA at $M$ and $R$ point. }
    \label{fig:cub_hess}
\end{figure}

Here we will present results for the Hessian of the cubic phase. The harmonic approximation predicts two main instabilities in the cubic $Pm\bar{3}m$ phase, in $M$ (0.5,0.5,0.0) and $R$ (0.5, 0.5, 0.5) high-symmetry point. The condensation of the unstable $M$ phonon would lead to the tetragonal $P4/mbm$ phase which is the one that we find in CsPbBr$_3$. The $R$ point folds back to the $Z$ point of the tetragonal phase, which is the mode that drives the phase transition between the tetragonal and orthorhombic phases. In Supp. Fig.~\ref{fig:cub_hess} we show results for the Hessian of these two modes including the fourth-order anharmonicity. 
The $R$ mode becomes stable at a lower temperature (500 K). On the other hand, the frequency of the $M$ mode becomes positive at a higher temperature of 550 K, the same temperature at which we have a first-order phase transition. This is an intriguing situation where we fulfill the main requirement for the second-order phase transition (a condensation of the soft mode), but still have the first-order phase transition as evidenced by the discontinuity of the lattice parameters, pointing to a small order-disorder character of the transition.

\newpage
\subsection{Comparing results for different phases}

\begin{figure}
    \centering
    \includegraphics[width=0.99\textwidth]{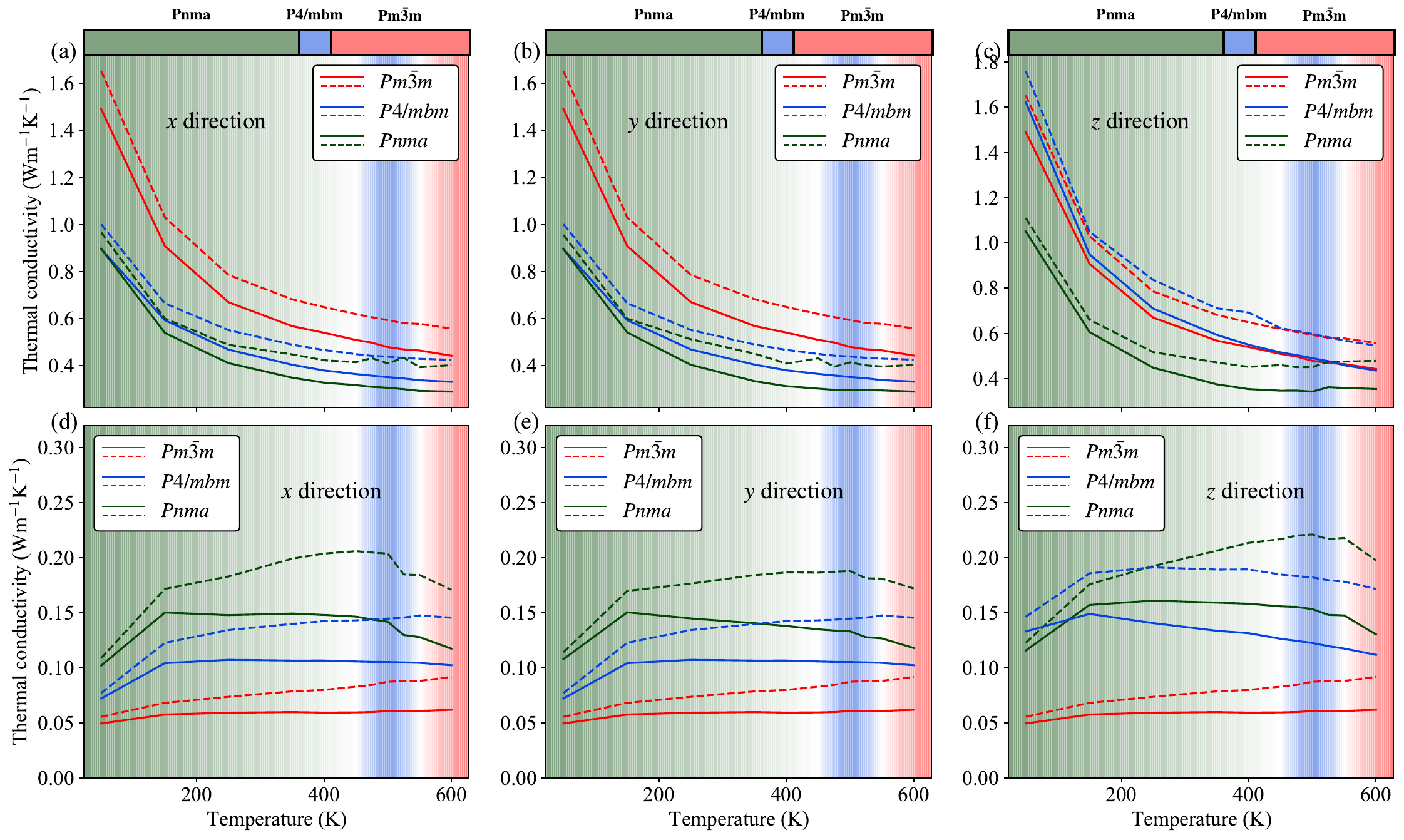}
    \caption{ Total ((a) in $x$ direction, (b) in $y$ direction, (c) in $z$ direction) and coherent ((d) in $x$ direction, (e) in $y$ direction, (f) in $z$ direction) part of the lattice thermal conductivity calculated for three different phases of CsPbBr$_3$. The shaded area of the background shows the calculated phase diagram for CsPbBr$_3$. The bar above the graph shows the experimental phase diagram. The color codes between lines and background are the same ($Pnma$ green, $P4/mbm$ blue, and $Pm\bar{3}m$ red). Full lines represent results obtained with the Boltzmann transport equation, while dashed lines are results obtained with the Green-Kubo method.}
    \label{fig:all_kappas}
\end{figure}

In Supp. Fig.~\ref{fig:all_kappas} we compare the calculated lattice thermal conductivity for different phases. As we have noted in the main text, the $Pnma$ phase has the lowest lattice thermal conductivity, followed by $P4/mbm$, and $Pm\bar{3}m$ has the highest $\kappa$. For the coherent part of the thermal conductivity, we see the opposite trend.

\newpage
\subsection{Drude model for dynamical lattice thermal conductivity}

Drude model for electrical conductivity ($\sigma$) assumes a constant relaxation time ($\tau$) for electrons. The real and imaginary parts of the AC conductivity are then given by:
\begin{align*}
    \operatorname{Re}\sigma (\nu) = \frac{\sigma _0}{1 + \nu^2\tau ^2} && \operatorname{Im}\sigma (\nu) = \nu\tau\frac{\sigma _0}{1 + \nu^2\tau ^2}.
\end{align*}

\begin{figure}
    \centering
    \includegraphics[width=0.9\textwidth]{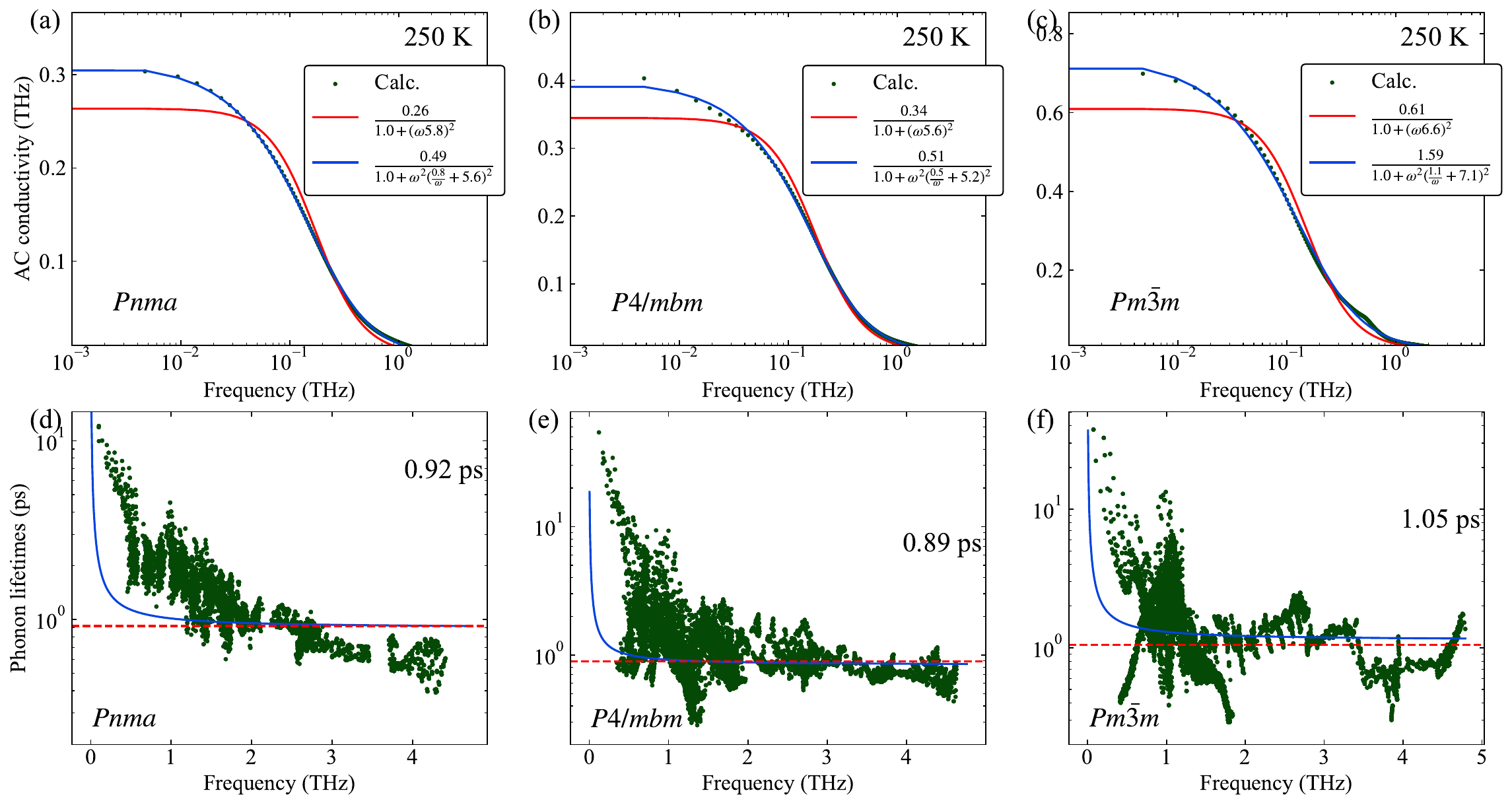}
    \caption{Real part of the dynamical lattice thermal conductivity at 250 K of (a) $Pnma$, (b) $P4/mbm$, and (c) $Pm\bar{3}m$ phase. Points are Green-Kubo calculations, while full lines are fitted with two versions of the Drude model explained in the text. In the bottom panels, we show calculated phonon lifetimes of (d) $Pnma$, (e) $P4/mbm$, and (f) $Pm\bar{3}m$ phase compared to the fitted value of the lifetimes in the Drude model.}
    \label{fig:fit_kappa}
\end{figure}

We fitted this function to the real part of the lattice thermal conductivity, see Supp. Fig.~\ref{fig:fit_kappa} (blue line). The fitting produces a value of phonon lifetime a bit larger than the one obtained from the peak of the imaginary part of the lattice thermal conductivity. However, the fit is quite bad and it does not give a correct value for static lattice thermal conductivity $\sigma _0$. 

We then assumed that phonon lifetimes are inversely proportional to the frequency $\tau = \frac{A}{\nu} + B$ and used this definition in the Drude model. This fit produced much better agreement with the actual calculation, correctly predicting both the tail and the static lattice thermal conductivity. 

To try and disentangle why this is the case let us try to derive the perturbative limit for the diagonal part of the dynamical lattice thermal conductivity starting from Supp. Eq.~\ref{eq:dyn_kappa1}:
\begin{align*}
    \kappa ^{xy} (\nu) = \frac{\pi\beta ^2k_{B}}{NV}\sum _{\mathbf{q},j,j'}v^{x}_{\mathbf{q},j,j'}v^{y*}_{\mathbf{q},j,j'}\omega _{\mathbf{q},j}\omega _{\mathbf{q},j'}\frac{e^{\beta\nu} - 1}{\beta\nu}\int_{-\infty}^{\infty}\textrm{d}\Omega (1+\frac{\Omega}{\Omega + \nu})\frac{e^{\beta\Omega}}{\left(e^{\beta\Omega} - 1\right)\left(e^{\beta(\Omega + \nu)} - 1\right)}\sigma _{\mathbf{q},j'}(\Omega)\sigma_{\mathbf{q},j}(\Omega + \nu).
\end{align*}
The spectral functions are then just Lorentzians centered around $\omega _{\mathbf{q},j}$ and with width $\Gamma _{\mathbf{q}j}$, see Supp. Eq.~\ref{LA_spec_func}. We will then multiply all of the parts of the spectral functions which will give us four separate terms:
\small
\begin{align*}
    \sigma _{\mathbf{q},j'}(\Omega)\sigma_{\mathbf{q},j}(\Omega + \nu) = \frac{1}{4\pi ^2}\left(\frac{-\Gamma _{\mathbf{q},j'}}{(\Omega - \omega _{\mathbf{q},j'})^2 + \Gamma ^2 _{\mathbf{q},j'}} + \frac{\Gamma _{\mathbf{q},j'}}{(\Omega + \omega _{\mathbf{q},j'})^2 + \Gamma ^2 _{\mathbf{q},j'}}\right)\left(\frac{-\Gamma _{\mathbf{q},j}}{(\Omega + \nu - \omega _{\mathbf{q},j})^2 + \Gamma ^2 _{\mathbf{q},j}} + \frac{\Gamma _{\mathbf{q},j}}{(\Omega + \nu + \omega _{\mathbf{q},j})^2 + \Gamma ^2 _{\mathbf{q},j}}\right)
\end{align*}
\normalsize
Then for each of these terms, we will approximate the value of the part of the integrand that does not contain the spectral function ($(1+\frac{\Omega}{\Omega + \nu})\frac{e^{\beta\Omega}}{\left(e^{\beta\Omega} - 1\right)\left(e^{\beta(\Omega + \nu)} - 1\right)}$) with its value at the poles of the spectral function. This allows us then to use the convolution of two Lorentzians to get rid of the integral. The solution for the diagonal part of the lattice thermal conductivity is then:
\begin{align*}\label{eq:dyn_kappa_pert}
    \kappa (\nu) &= \kappa _1(\nu) + \kappa _2(\nu) + \kappa _3(\nu)\numberthis 
\end{align*}
with:
\begin{align*}
    \kappa _1(\nu) &= \frac{1}{NV}\sum _{\mathbf{q},j} \mathcal{C}_{1}(\nu, \omega _{\mathbf{q},j})\frac{\kappa _{\mathbf{q},j}}{\nu ^2\tau ^2 _{\mathbf{q},j} + 1} \\
    \kappa _2(\nu) &= -\frac{1}{NV}\sum _{\mathbf{q},j} \left(\mathcal{C}_{1}(\nu, \omega _{\mathbf{q},j}) + \mathcal{C}_{2}(\nu, \omega _{\mathbf{q},j})\right)\frac{\kappa _{\mathbf{q},j}}{(\nu ^2 + 2\omega _{\mathbf{q},j})\tau ^2 _{\mathbf{q},j} + 1}\\
    \kappa _3(\nu) &= \frac{1}{NV}\sum _{\mathbf{q},j} \mathcal{C}_{2}(\nu, \omega _{\mathbf{q},j})\frac{\kappa _{\mathbf{q},j}}{\nu ^2\tau ^2 _{\mathbf{q},j} + 1}.
\end{align*}
Here $\kappa _{\mathbf{q},j}$ is the contribution to the lattice thermal conductivity by $(\mathbf{q},j)$ mode. The functions $\mathcal{C}_{1}(\nu, \omega _{\mathbf{q},j})/\mathcal{C}_{2}(\nu, \omega _{\mathbf{q},j})$ are coming from the part of the integrand that we approximated by the value at the pole of the spectral function:
\begin{align*}
    \mathcal{C}_{1}(\nu, \omega _{\mathbf{q},j}) &= \frac{1}{4}\frac{e^{\beta \nu} - 1}{\beta\nu}(1 + \frac{\omega _{\mathbf{q},j}}{\omega _{\mathbf{q},j} + \nu})\frac{e^{\beta\omega _{\mathbf{q},j}} - 1}{e^{\beta(\omega _{\mathbf{q},j} + \nu)} - 1} \\
    \mathcal{C}_{2}(\nu, \omega _{\mathbf{q},j}) &= \frac{1}{4}\frac{e^{\beta \nu} - 1}{\beta\nu}(1+\frac{\omega _{\mathbf{q},j} + \nu}{\omega _{\mathbf{q},j}})\frac{e^{-\beta\omega _{\mathbf{q},j}} - 1}{e^{\beta\nu}(e^{-\beta(\omega _{\mathbf{q},j} + \nu)} - 1)}\numberthis
    \label{eq:Cs}
\end{align*}
In the limit $\nu\rightarrow 0$ both of these functions take the value of 0.5. 

Assuming for a second that we have a constant relaxation time and that we take $\mathcal{C}$ functions in the small frequency limits we obtain a Drude model clearly explaining why the fit to Drude model of the results of the full equation~\ref{eq:dyn_kappa1} works. However, even results by this perturbative limit fit better with a frequency-dependent phonon lifetime. Since in this case (Supp. Eq.~\ref{eq:dyn_kappa_pert}) the phonon mode linewidth is a constant we can exclude the frequency dependence of the phonon self-energy as the reason behind better fit for the $\nu$ dependent lifetimes. 

Another source of this apparent frequency dependence of fitted phonon lifetime could be $\mathcal{C}$ functions, see Supp. Eq.~\ref{eq:Cs}. However, results with constant $\mathcal{C} = 0.5$, give virtually the same result which still shows better fitting to the frequency-dependent lifetimes. This leaves a better fit with a frequency-dependent lifetime of dynamical lattice thermal conductivity an open question.

Finally, let us briefly discuss the question of the perturbative limit of the non-diagonal in phonon bands contribution to the dynamical lattice thermal conductivity. Here we would again have three parts contributing to the dynamical $\kappa$:
\begin{align*}
    \kappa _1(\nu) &\sim \frac{(\Gamma _{\mathbf{q}, j} + \Gamma _{\mathbf{q}, j'})}{(\omega _{\mathbf{q},j} - \omega _{\mathbf{q},j'} + \nu)^2 + (\Gamma _{\mathbf{q}, j} + \Gamma _{\mathbf{q}, j'})^2} \\
    \kappa _2(\nu) &\sim \frac{(\Gamma _{\mathbf{q}, j} + \Gamma _{\mathbf{q}, j'})}{(\omega _{\mathbf{q},j} + \omega _{\mathbf{q},j'} + \nu)^2 + (\Gamma _{\mathbf{q}, j} + \Gamma _{\mathbf{q}, j'})^2} \\
    \kappa _3(\nu) &\sim \frac{(\Gamma _{\mathbf{q}, j} + \Gamma _{\mathbf{q}, j'})}{(\omega _{\mathbf{q},j'} - \omega _{\mathbf{q},j} + \nu)^2 + (\Gamma _{\mathbf{q}, j} + \Gamma _{\mathbf{q}, j'})^2}.
\end{align*}
This tells us why there are peaks in the coherent contribution to the dynamical lattice thermal conductivity. Whenever the driving frequency $\nu$ is the same as the frequency spacing between phonon modes, we have a resonance and that shows up as a structure in the dynamical lattice thermal conductivity.

\newpage
\subsection{Comparing finite difference and SSCHA results}

We calculated the harmonic and perturbative third-order force constants for the orthorhombic phase of CsPbBr$_3$ using the finite difference method as implemented in Phonopy~\cite{phonopy-phono3py-JPSJ}. We then calculated the lattice thermal conductivity of CsPbBr$_3$ using these force constants and compared it to SSCHA results (which uses renormalized force constants due to temperature and structure change). The comparison between the two results is given in Figure 5 of the main part.

\begin{figure}
    \centering
    \includegraphics[width=0.9\textwidth]{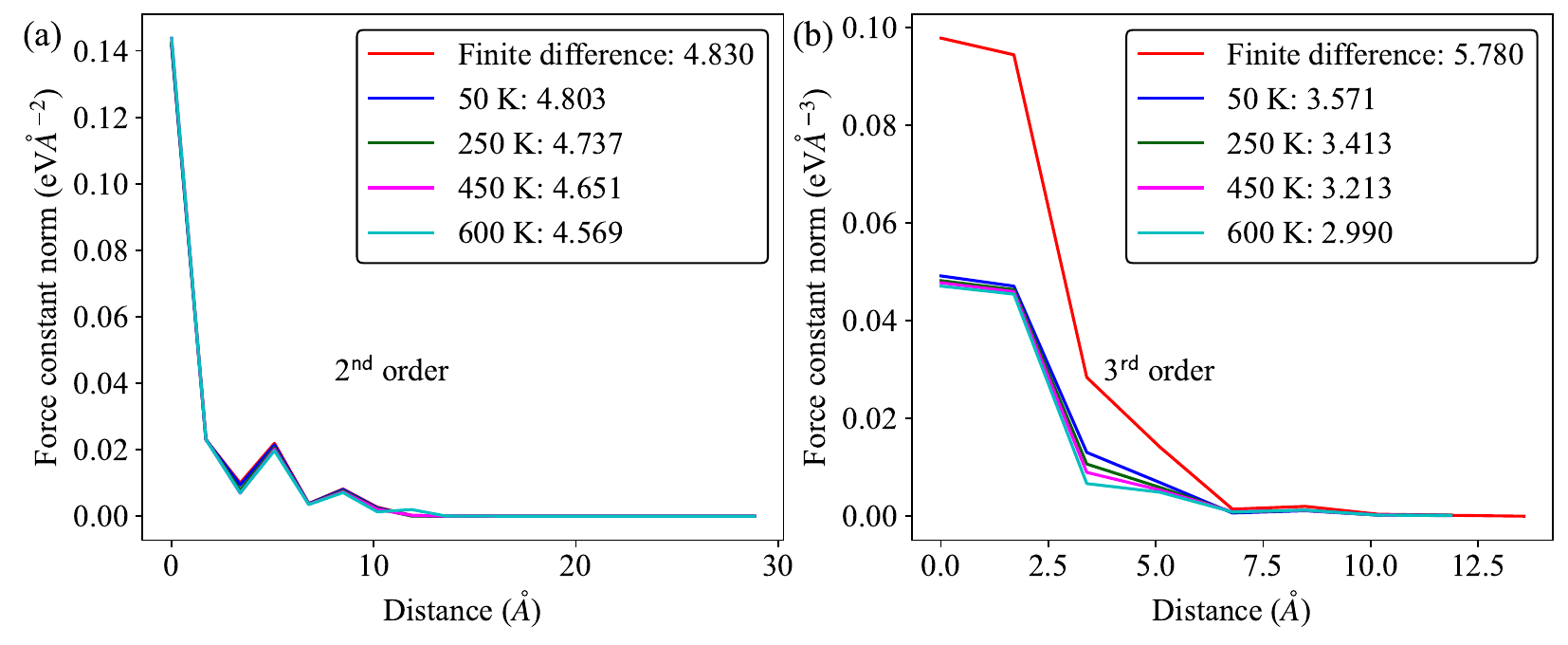}
    \caption{(a) Harmonic second order force constants calculated with finite difference and auxiliary SSCHA force constants at different temperatures. To present results we binned the norms of second-order force constants with respect to the atom-atom distance and plot the maximum norm inside the bin. (b) The same study as in (a) but for third-order force constants. Force constants were binned with respect to the distance between the first and second atoms. The numbers in the legend signify the sum of all norms in the supercell.}
    \label{fig:fcs}
\end{figure}

Here we will explain the lower values of thermal conductivity obtained with finite difference force constants. In Supp. Fig.~\ref{fig:fcs} we show the second and third-order force constants calculated with finite difference method and with SSCHA at different temperatures for the orthorhombic $Pnma$ phase. The force constants in SSCHA are lower, especially the third-order ones, leading to stronger scattering between phonons and lower phonon lifetimes. This situation culminates in lower total thermal conductivity calculated with finite difference force constants. This addresses the important role played by higher-order terms that are captured by the SSCHA force constants.

\end{document}